\documentclass[titlepage,letterpaper,openany,12pt]{book}
\usepackage{amssymb}
\usepackage{amsmath}

\newfont{\cyr}{wncyr10 scaled 1200}
\newfont{\cyrbf}{wncyb10 scaled 1200}

 \newcommand{\CC}{\symbol{'121}}
 \newcommand{\cc}{\symbol{'161}}
 \newcommand{\ch}{\symbol{'17}}
 \newcommand{\dd}{\symbol{'16}}
 \newcommand{\zz}{\symbol{'31}}

\begin{document}

\thispagestyle{empty}

\begin{center}
\vspace{-1cm}{\large UNIVERSIDAD DE SANTIAGO DE CHILE}

{\large DEPARTAMENTO DE F\'{I}SICA, FACULTAD DE CIENCIA}{\Large \vspace{3cm}}

{\LARGE \textbf{Dynamics of Wess-Zumino-Witten} \bigskip }

{\LARGE \textbf{and Chern-Simons Theories} }{\Large \vspace{2cm}}

{\LARGE Olivera Mi\v{s}kovi\'{c}\vspace{1cm}}

{\Large Thesis for obtaining the degree of}{\LARGE \bigskip }

{\Large Doctor of Philosophy\vspace{2cm}}{\LARGE \textbf{\bigskip }}

\begin{tabular}{ll}
{\Large Supervising Professors:}{\LARGE \qquad \medskip } & {\Large Dr.
Ricardo Troncoso} \\
& {\Large Dr. Jorge Zanelli}%
\end{tabular}%
{\Large \vspace{2.5cm}\vspace{2.5cm}}

{\large SANTIAGO -- CHILE}

{\large JANUARY 2004}

\newpage
\end{center}

\thispagestyle{empty}

\begin{center}
\vspace{-1cm}{\large UNIVERSIDAD DE SANTIAGO DE CHILE}

{\large DEPARTAMENTO DE F\'{I}SICA, FACULTAD DE CIENCIA}{\Large \vspace{3cm}}

{\LARGE \textbf{Din\'{a}mica de las Teor\'{\i}as de}}{\huge \bigskip }

{\LARGE \textbf{Wess-Zumino-Witten y Chern-Simons}}{\Large \vspace{2cm}}

{\LARGE Olivera Mi\v{s}kovi\'{c}\vspace{1cm}}

{\Large Tesis para optar el grado de}{\LARGE \bigskip }

{\Large Doctor en Ciencias con Menci\'{o}n en F\'{\i}sica\vspace{2cm}}%
{\LARGE \textbf{\bigskip }}

\begin{tabular}{ll}
{\Large Directores de Tesis:}{\LARGE \qquad \medskip } & {\Large Dr. Ricardo
Troncoso} \\
& {\Large Dr. Jorge Zanelli}%
\end{tabular}%
{\Large \vspace{2.5cm}\vspace{2.5cm}}

{\large SANTIAGO -- CHILE}

{\large ENERO 2004}

\newpage

\thispagestyle{empty}

{\large \textbf{INFORME DE APROBACION}}

{\large \textbf{TESIS DE DOCTORADO}}{\Large \vspace{1.5cm}}
\end{center}

{\large Se in\-for\-ma al Comit\'{e} del Programa de Doctorado en Ci\-en\-ci\-as con
menci\'{o}n en F\'{\i}sica que la Tesis presentada por
el candidato\vspace{1cm}}

\begin{center}
{\large Olivera Mi\v{s}kovi\'{c}\vspace{1cm}}
\end{center}

\noindent
{\large ha sido aprobada por la Comisi\'{o}n Informante de Tesis como
requisito para la obtenci\'{o}n del grado de Doctor en Ciencias con menci\'{o}n
en F\'{\i}sica.\vspace{1.5cm}}

\begin{tabular}{ll}
{\large \textbf{Directores de Tesis:}\vspace{0.7cm}} &  \\
{\large Dr. Ricardo Troncoso\bigskip } & \rule{6cm}{0.02cm} \\
{\large Dr. Jorge Zanelli\vspace{1cm}} & \rule{6cm}{0.02cm} \\
{\large \textbf{Comisi\'{o}n Informante de Tesis:}\vspace{0.7cm}} &  \\
{\large Dr. M\'{a}ximo Ba\~{n}ados\bigskip } & \rule{6cm}{0.02cm} \\
{\large Dr. Jorge Gamboa\bigskip } & \rule{6cm}{0.02cm} \\
{\large Dr. Marc Henneaux\bigskip } & \rule{6cm}{0.02cm} \\
{\large Dr. Mikhail Plyushchay\bigskip } & \rule{6cm}{0.02cm} \\
{\large Dr. Lautaro Vergara (presidente)} & \rule{6cm}{0.02cm}%
\end{tabular}
\bigskip

\begin{center}
\newpage

\pagenumbering{roman}\textbf{Abstract\bigskip }
\end{center}

This thesis is devoted to the study of three problems on the
Wess-Zumino-Witten (WZW) and Chern-Simons (CS) supergravity theories in the
Hamiltonian framework:

1. The two-dimensional super WZW model coupled to supergravity is
constructed. The canonical representation of Kac-Moody algebra is extended
to the super Kac-Moody and Virasoro algebras. Then, the canonical action is
constructed, invariant under local supersymmetry transformations. The metric
tensor and Rarita-Schwinger fields emerge as Lagrange multipliers of the
components of the super energy-momentum tensor.

2. In dimensions $D\geq 5$, CS theories are irregular systems, that
is, they have constraints which are functionally dependent in some sectors
of phase space. In these cases, the standard Dirac procedure is not directly
applicable and must be redefined, as it is shown in the simplified case of
finite number of degrees of freedom. Irregular systems fall into two classes
depending on their behavior in the vicinity of the constraint surface. In
one case, it is possible to regularize the system without ambiguities, while
in the other, regularization is not always possible and the Hamiltonian and
Lagrangian descriptions may be dynamically inequivalent. Irregularities have
important consequences in the linearized approximation of nonlinear theories.

3. The dynamics of CS supergravity theory in $D=5$, based on the
supersymmetric extension of the \emph{AdS} algebra, $su(2,2\left\vert
4\right. ),$ is analyzed. The dynamical fields are the vielbein, the spin
connection, $8$ gravitini, as well as $SU(4)$ and $U(1)$ gauge fields. A
class of backgrounds is found, providing a regular and generic
effective theory. Some of these backgrounds are shown to be BPS states. The
charges for the simplest choice of asymptotic conditions are obtained, and
they satisfy a supersymmetric extension of the classical WZW$_{4}$ algebra,
associated to $su(2,2\left\vert 4\right. )$.

\newpage

\begin{center}
\textbf{Resumen\bigskip }
\end{center}

Esta tesis est\'{a} dedicada al estudio de tres problemas en de teor\'{\i}as
de supergravedad de Wess-Zumino-Witten (WZW) y de Chern-Simons (CS), en el
formalismo hamiltoniano:

1. Se construye un modelo de s\'{u}per WZW en dos dimensiones, acoplado a
supergravedad. La representaci\'{o}n can\'{o}nica del \'{a}lgebra de
Kac-Moody es extendida a las \'{a}lgebras de s\'{u}per Kac-Moody y s\'{u}per
Virasoro. Luego, se construye la acci\'{o}n can\'{o}nica, invariante bajo
transformaciones de supersimetr\'{\i}a locales. El tensor m\'{e}trico y el
campo de Rarita-Schwinger aparecen como multiplicadores de Lagrange de las
componentes del s\'{u}per tensor energ\'{\i}a-momentum.

2. En dimensiones $D\geq 5$, las teor\'{\i}as de CS constituyen
sistemas irregulares, es decir, contienen ligaduras que son funcionalmente
dependientes en algunos sectores del espacio de fase. En estos casos, el
procedimiento de Dirac est\'{a}ndar no es aplicable directamente y debe ser
redefinido, como se muestra en el caso simplificado cuando el sistema tiene
un n\'{u}mero finito de grados de libertad. Los sistemas irregulares pueden
pertenecer a dos clases, dependiendo de su comportamiento en la vecindad de
la superficie de ligadura. En un caso, es posible regularizar el sistema sin
ambig\"{u}edades, mientras que en el otro, la regularizaci\'{o}n no es
siempre posible y las descripciones hamiltoniana y lagrangiana pueden no ser
din\'{a}micamente equivalentes. Estas irregularidades tienen importantes
consecuencias en la aproximaci\'{o}n linealizada de teor\'{\i}as no-lineales.

3. Se analiza la din\'{a}mica de la teor\'{\i}a de supergravedad de
CS en $D=5$, basada en la extensi\'{o}n supersim\'{e}trica del \'{a}lgebra
de \emph{AdS}, $su(2,2\left\vert 4\right. )$. Los campos din\'{a}micos son
el vielbein, la conecci\'{o}n de spin y $8$ gravitini, adem\'{a}s de campos
de gauge para $SU(4)$ y $U(1)$. Se identifica una clase de backgrounds que
da lugar a una teor\'{\i}a efectiva que es regular\textbf{\ }y gen\'{e}rica.
Del mismo modo, se prueba que algunos de estos backgrounds son estados BPS.
Se obtienen las cargas para la elecci\'{o}n m\'{a}s simple de condiciones
asint\'{o}ticas. Estas cargas satisfacen una extensi\'{o}n supersim\'{e}%
trica del \'{a}lgebra cl\'{a}sica de WZW$_{4}$, asociada a $su(2,2\left\vert
4\right. )$.
\newpage

\begin{center}
\cyrbf{Rezime}
\end{center}

\cyr{Ova teza je posve\ch ena prou\cc avanju tri problema
u vezi sa Ves-Zumino-Vitenovim (VZV) i
\CC ern-Sajmonsovim (\CC S) teorijama supergravitacije u okviru
Hamiltonovog formalizma}:

    1. \cyr{Konstruisan je dvodimenzioni super VZV-ov model kuplovan sa
su\-per\-gra\-vi\-ta\-ci\-jom.
Kanonska reprezentacija Kac-Mudijeve algebre je proxirena do
supersimetri\cc ne Kac-Mu\-di\-je\-ve i Virazorove algebre, a zatim je
konstruisano kanonsko dejstvo invarijantno pod transformacijama
lokalne supersimetrije. Metri\cc ki tenzor i Rarita-Xvingerovo polje
su se pojavili kao Lagran\zz evi mno\zz itelji uz komponente
supertenzora energije-impulsa.}

    2. \cyr{\CC S-ove teorije u dimenzijama $D\geq 5$ su iregularni sistemi,
xto zna\cc i da sadr\zz e veze
koje su funkcionalno zavisne u pojedinim oblastima faznog
prostora. U tim slu\cc ajevima ne mo\zz e direktno da se primeni
standardna Dirakova procedura, nego mora da se redefinixe, xto je
ura\dd eno za najjednostavnije sisteme sa kona\cc nim brojem stepeni
slobode. Pokazano je da postoje dve vrste iregularnih sistema, u
zavisnosti od toga kako se ponaxaju u blizini povrxi definisane
vezama. U jednom slu\cc aju sistem mo\zz e da se regularixe, dok u drugom
slu\cc aju regularizacija nije uvek mogu\ch a, poxto Hamiltonov i
Lagran\zz ev formalizam mogu da dovedu do dinami\cc ki neekvivalentnih
rezultata. Iregularnosti imaju zna\cc ajne posledice u linearnoj
aproksimaciji nelinearnih teorija.}

    3. \cyr{Analizirana je dinamika \CC S-ove teorije supergravitacije u $D=5$, bazirane na
supersimetri\cc noj ekstenziji anti-de Siterove algebre,
$su(2,2|4)$. Dinami\cc ka polja te teorije su pentada, spinska
koneksija, $8$ gravitina, kao i gradijentna polja $SU(4)$ i
$U(1)$. Na\dd ena je klasa pozadinskih polja takvih da su
efektivne teorije, definisane u njihovoj okolini, regularne i
generi\cc ke. Tako\dd e je pokazano da neka od tih pozadinskih
polja predstavljaju tzv.\ {\rm BPS}-stanja. Izborom
najjednostavnijih asimptot\-skih uslova su dobijeni o\cc uvani
naboji, \cc ija je algebra supersimetri\cc na ekstenzija klasi\cc
ne {\rm WZW}$_4$ algebre, asocirane sa $su(2,2|4)$.}

\newpage

\begin{center}
\textbf{Acknowledgment}{\large \textbf{\bigskip }}
\end{center}

{\small There are many people who supported me from the moment I left my
home in Belgrade, and came to Chile to work on my Ph.D. Thesis. First of
all, I thank to my parents and my sister, without whose care, support and
positive influence my trip and this work would have not been possible. I
will also be in eternal debt to Rodrigo Olea, who has been sharing with me
his love, his life and his knowledge in physics, and without whom everything
would have been different.}

{\small The person who particularly fills me with the gratitude is my
supervisor Jorge Zanelli, whose love for physics and his profound
understanding of it have been irresestibly inspiring, and whose deep
humanity is overwhelming. Infinite thanks to Ricardo Troncoso, my second
supervisor and friend, for his help, support and cheering up in dark
moments. I would like to thank to all physicists and staff of CECS, with the
director Claudio Teitelboim, who accepted me from the very beginning as a
member of their \textquotedblleft second family\textquotedblright .}

{\small For enlightening and useful discussions, as well as friendly
criticism, thanks to Alexis Am\'{e}zaga, Rodrigo Aros, Eloy Ayon-Beato,
M\'{a}ximo Ba\~{n}ados, Arundhati Dasgupta, Andr\'{e}s Gomberoff, Mokhtar
Hassa\"{\i}ne, Marc Henneaux, Georgios Kofinas, Cristi\'{a}n Mart\'{\i}nez,
Rodrigo Olea, Mikhail Plyushchay, Joel Saavedra, Claudio Teitelboim and
Tatjana Vuka\v{s}inac. I am particularly thankfull to Milutin
Blagojevi\'{c},
Ivana Mi\v{s}kovi\'{c} and Branislav Sazdovi\'{c}, who were regularly
replying my numerous questions by e-mail, and to Milovan Vasili\'{c}, for
his helpful criticism. Many thanks to all physicists and staff of Institute
of Physics in\ Belgrade, Vin\v{c}a\ Institute and the Department of Physics
USACH.}

{\small I will be forever obliged to my friends Aleksandra \'{C}uk, Sne\v{z}%
ana Djurdji\'{c}, D\'{a}fni Fernanda Zenedin Marchioro, Milan Lali\'{c},
Aleksandra Lazovi\'{c}, Ricardo Navarro, Ljubomir Nikoli\'{c}, Ivan Saji\'{c}%
, Jelena To\v{s}i\'{c} and Tatjana Vuka\v{s}inac, for their friendship and
moral support.}

{\small This work is partially funded by Chilean grants JZ 1999, MECESUP USA
9930, FONDECYT 2010017 and Millenium Project. I also want to thank to the
Abdus Salam ICTP, Max Planck-Albert Einstein Institut and Universit\'{e}
Libre de Bruxelles for their hospitality during the completition the parts
of this work, as well as to Sonja and Vlado Mi\v{s}kovi\'{c} for their
financial help. The generous support of Empresas CMPC to CECS is also
acknowledged. CECS is a Millenium Science Institute and is funded in part by
grants from Fundaci\'{o}n Andes and the Tinker Foundation.}\newpage
\rm
\tableofcontents

\chapter{Introduction}
\rm
\pagenumbering{arabic}Wess-Zumino-Witten (WZW) and Chern-Simons (CS) field
theories have been intensively studied in connection with several
applications in physics and mathematics.

The two-dimensional WZW theory\footnote{%
This is also called Wess-Zumino-Novikov-Witten (WZNW) theory.} \cite%
{Wess-Zumino,Novikov,Witten'84}, described by a non-linear sigma model with
non-local interaction, was originally studied by Witten \cite{Witten'84} as
a theory equivalent to non-interacting massless fermions, thus providing
non-Abelian bosonization rules for interacting fermionic theories. The WZW
action is also known as the necessary counter-term for cancelation of
quantum anomalies (the breaking down of a classical symmetry at the quantum
level) \cite{Erlich-Freedman}--\cite{Bogojevic-Miskovic-Sazdovic2}. This
theory is exactly solvable and quantizable, and its action has two
independent (\textquotedblleft left\textquotedblright\ and \textquotedblleft
right\textquotedblright ) chiral symmetries, whose infinite-dimensional
algebras are two copies of the affine, Kac-Moody (KM), algebra. The WZW
theory is also conformally invariant, where the symmetry is described by the
Virasoro algebra. Because of this, the WZW model is relevant in string
theory, as well.

Three-dimensional CS theories have a topological origin, since they can be
defined as CS forms integrated over the boundary of a compact
four-dimensional manifold. These theories have no local degrees of freedom
and they are also exactly solvable and quantizable \cite{WittenCS}. As
topological field theories, they can be used in the classification of
three-dimensional manifolds \cite{WittenTFT}. The quantum CS theories are
known to describe the quantum Hall effect \cite{Zhang-Hansson-Kivelson}. CS
theories can also be defined on three-dimensional manifolds \emph{with} a
boundary. In that case, their transformations under the \textquotedblleft
large\textquotedblright\ gauge transformations are non-trivial and given by
a closed 3-form, that has been used in the context of quantum anomalies. The
fact that both WZW and CS\ theories are related to the quantum anomalies is
not accidental. Their relation reflects a profound connection between them.
For example, the gauge transformations of the CS action give a non-trivial
contribution to the gauged WZW model describing the most general form of a
two-dimensional chiral anomaly \cite{Witten'83}--\cite{Karabali-Schnitzer}.
Any CS theory defined on a three-manifold with a boundary, induces a
two-dimensional WZW model as a topological field theory \cite%
{WittenTFT,Moore-Seiberg,Elitzur-Moore-Schwimmer-Seiberg}.

In general, the dynamics at the boundary is determined by the asymptotic
behavior of the fields. This is essential for a suitable definition of the
global charges of the theory \cite{Regge-Teitelboim}--\cite{Abbott-Deser2}.

The most interesting aspect of these two classes of theories, which will be
further investigated, is their deep connection with lower-dimensional
gravity theories. For example, two-dimensional induced gravity can be
obtained as a gauge extension of the WZW model \cite%
{Polyakov,Knizhnik-Polyakov-Zamolodchikov,Abdalla-Abdalla-Gamboa-Zadra},
while the Liouville theory, describing the asymptotic dynamics of
three-dimensional Einstein-Hilbert gravity with negative cosmological
constant, is equivalent to the two-dimensional induced gravity in the
conformal gauge \cite%
{Forgacs-Wipf-Balog-Feher-Raifeartaigh,Coussaert-Henneaux-Driel}.

On the other hand, three-dimensional gravity, described by the
Einstein-Hilbert action, which is linear in the curvature of space-time, can
be formulated as a CS gauge theory invariant under de Sitter (\emph{dS}),
anti-de Sitter (\emph{AdS})\emph{\ }or Poincar\'{e} groups \cite%
{Achucarro-Townsend,Witten'88,Carlip}. Asymptotically locally \emph{AdS}
gravity, for example, has an infinite-dimensional algebra of asymptotic
symmetries described by the Virasoro algebra, whose realization in terms of
conserved charged requires a non-trivial classical central charge \cite%
{Brown-Henneaux}.

The need to look for alternative gravity theories arises from the fact that
General Relativity, which gives a successful classical description of
gravitational phenomena in four dimensions, does not admit a standard
quantum description yet, while the other three fundamental forces are
consistently unified and described by quantum theories of the Yang-Mills
(YM) type. In this approach, the main obstruction for existence of a quantum
theory of gravity is its nonrenormalizability, \emph{i.e.}, the
impossibility of removing all divergences which appear in the high-energy
sector of the theory, due to the dimension of gravitational constant. While
the renormalizability of YM theories is a consequence of their invariance
under local gauge transformations, since the gauge principle provides a
dimensionless coupling constant, the Einstein-Hilbert theory is invariant
under general coordinate transformations $x\rightarrow x^{\prime }=x^{\prime
}(x)$ (diffeomorphisms). This symmetry, however, does not guarantee the
consistency of the quantum theory, because this symmetry does not have a
fiber-bundle structure as in YM theories (which is sufficient, but not
necessary condition for a theory to be renormalizable).

Supersymmetry is naturally introduced, since supersymmetric theories can
lead to a non-trivial unification of space-time and internal symmetries
within a relativistic quantum field theory (see, \emph{e.g.}, \cite%
{Grisaru(SG)}--\cite{Sohnius}). In these theories, both bosons and fermions
belong to the same representation of the supergroup. The gravitational
interaction emerges naturally from \emph{local supersymmetry}, since the
anticommutator of two supersymmetry generators (supercharges) gives a
generator of local translations. In that way, supergravity theories are
obtained as supersymmetric extensions of the purely gravitional part.

Furthermore, supersymmetric extensions of chiral and conformal symmetries
define supersymmetric WZW models, characterized by super KM and super
Virasoro algebras \cite{Hull-Spence}. In three dimensions, these
superalgebras are obtained as the algebra of the classical charges for \emph{%
AdS} supergravity models, with adequate asymptotic conditions \cite%
{Banados-Bautier-Coussaert-Henneaux-Ortiz,Henneaux-Maoz-Schwimmer}.

The existence of supersymmetry makes possible to construct non-negative
quantities quadratic in the supercharges, which gives rise to the
inequalities known as Bogomol'nyi bounds \cite{Bogomolnyi}. These bounds
guarantee the stability of the ground state (vacuum) in supergravity
theories, so that it remains a state of minimal energy after perturbations.
The Bogomol'nyi bound ensures the positivity of energy in the standard
supergravities \cite{Witten(E),Deser-Teitelboim,Grisaru(E)}, even in
presence of other conserved quantities \cite{Olive-Witten}.

Higher-dimensional theories can be physically meaningful if one supposes
that only four dimensions of space-time are observable, while others are
\textquotedblleft too small\textquotedblright\ to be visible at currently
reachable energies. In that sense, a four-dimensional theory would be an
effective theory. This can be realized by the procedure known as dimensional
reduction, where one assumes that the radius of extra dimensions is
compactified beyond sight (see, \emph{e.g.}, \cite{Witten(KK),Viswanathan}).

It is interesting, thus, to consider higher-dimensional CS theories, which
are defined in all odd dimensions, and have Lagrangians represented by CS
forms \cite{Floreanini-Percacci}--\cite{Nair-Schiff'92}. CS gravity and
supergravity theories are based on the anti-de Sitter $\left[ \emph{AdS}%
\text{, or }SO(D-1,2)\right] $, de Sitter $\left[ \emph{dS}\text{, or }%
SO(D,1)\right] $, and Poincar\'{e} $\left[ ISO(D-1,1)\right] $ gauge groups,
as well as their supersymmetric extensions. They are by construction
invariant under diffeomorphisms and provide a non-standard, consistent
description of gravity as a gauge theory
\cite{Chamseddine'90}--\cite{Hassaine-Olea-Troncoso}.
They are genuine gauge theories which are extensions
of the Einstein-Hilbert action. Their actions are polynomials in the
curvature $R$ and they can also depend explicitly on the torsion $T$.
Furthermore, they possess propagating degrees of freedom \cite%
{Banados-Garay-Henneaux1,Banados-Garay-Henneaux2}, and have a very rich
phase space structure. In CS supergravities, unlike in the standard
supergravity theories, the supersymmetry algebra closes \emph{off shell},
without addition of the auxiliary fields \cite{Troncoso-Zanelli '99a}.

On the other hand, WZW and super WZW theories could be generalized to higher
dimensions as field theories whose symmetries are described by
(supersymmetric) extensions of KM and Virasoro algebras. They are not
studied as much as in the two-dimensional case. For example, the WZW action
is known in four dimensions, with the local symmetry described by a
four-dimensional extension of the KM algebra, or WZW$_{4}$ algebra \cite%
{Nair-Schiff'90,Nair-Schiff'92,Losev-Moore-Nekrasov-Shatashvili}. The
relation between this four-dimensional WZW theory and a CS\ theory in five
dimensions is established only at the level of algebras \cite%
{Banados-Garay-Henneaux2}, while the actions has not been obtained
explicitly. In general, the action of the four-dimensional super WZW model
remains unknown.

Higher-dimensional CS theories have complex configurational space. In
five-di\-men\-si\-o\-nal CS supergravity, for example, it was observed that the
linearized action around an \emph{AdS} background seems to have one more
degree of freedom\emph{\ }than the full nonlinear system \cite%
{Chandia-Troncoso-Zanelli}. This paradoxical behavior arises from the
violation of the \emph{regularity conditions} among the symmetry generators
of the theory, or their functional dependence, in the region of phase space
defined by the selected background. Therefore, CS theories in $D\geq 5$
dimensions are \emph{irregular systems}, where Dirac's standard procedure of
finding local symmetries and physical degrees of freedom \cite{Dirac} fails.%
\footnote{%
A well-known example of an irregular system is a relativistic massless
particle ($p^{\mu }p_{\mu }=0$), which is irregular at the origin of
momentum space ($p_{\mu }=0$).} The problem of the regularity does not
appear in lower-dimensional WZW or CS theories, but can occur in any
physical system, independently of the dimension of space-time.

Constraints satisfying regularity conditions are sometimes referred to as
\emph{effective} constraints \cite{Batlle-Gomis-Pons-RomanRoy}. The issue of
regularity (\emph{effectiveness}) and its relevance for the equivalence
between the Lagrangian and Hamiltonian formalisms has also been discussed in
several references \cite{Garcia-Pons,Di Stefano,Pons-Salisbury-Shepley}.

There are many different areas where the WZW and CS theories find their
applications but, from the point of view of this thesis, the main motivation
to study these theories is: (\emph{i}) CS theories are alternative gravity
and supergravity theories in odd dimensions, (\emph{ii}) every \emph{AdS}-CS
theory induces a conformal field theory at the boundary, a WZW model, and (%
\emph{iii}) WZW models should correspond to (super)gravity theories in even
dimensions.\medskip

Therefore, three problems related to CS theories and to the topics (\emph{i}%
)-(\emph{iii}) are presented in this thesis, where each one can be analyzed
independently.

\begin{itemize}
\item The first problem is the construction of the super WZW model coupled
to supergravity, from the chosen canonical representation of the super
Virasoro algebra. The model is obtained explicitly in two dimensions \cite%
{Miskovic-Sazdovic}, what may provide some insight for finding the unknown
four-dimensional super WZW theory.

\item The second question arises from the need of dealing with irregular CS
systems, and of generalizing the Dirac procedure. The problem is analyzed
for classical mechanical systems with finite number of degrees of freedom
\cite{Miskovic-Zanelli,Miskovic-ZanelliJMP}, and discussed in CS theories as
well \cite{Miskovic-ZanelliJMP}.

\item The third part presents a work currently in progress, based on the
study of the dynamical structure of five-dimensional \emph{AdS}-CS
supergravity, both in the bulk and at the boundary. The physical degrees of
freedom and local symmetries of these theories depend on the symplectic form
(defining the kinetic term). Since the symplectic form is a function of
phase space coordinates whose rank can vary throughout the phase space, CS
theories can be either\emph{\ regular}, or \emph{irregular} \cite%
{Miskovic-ZanelliJMP}. Moreover, they can be \emph{generic}, with a minimal
number of local symmetries \cite{Banados-Garay-Henneaux2}, or
\emph{degenerate} if the symplectic form has a lower rank and additional
symmetries emerge \cite{Saavedra-Troncoso-Zanelli,Saavedra-Troncoso-Zanelli2}.
In the asymptotic sector, the charge algebra of the \emph{AdS}-CS
supergravity is the supersymmetric extension of the WZW$_{4}$ algebra
with a central charge \cite{Miskovic-Troncoso-Zanelli}.
\end{itemize}

The thesis is organized as follows.

In Chapter 2, the two-dimensional WZW model is reviewed. Two independent KM
algebras are obtained as canonical symmetries of the WZW action. The
conformal symmetry in this model is not independent, since the Virasoro
generators can be expressed as bilinears of the KM generators. The super WZW
model is also discussed, using superspace formalism.

In Chapter 3, two-dimensional WZW supergravity is constructed using the
Hamiltonian method. The canonical (first order) action is defined from the
phase space representation of the super Virasoro algebra, and the Lagrange
multipliers corresponding to the super Virasoro generators appear as the
components of the metric field and gravitini.

In Chapter 4, Dirac's procedure is extended to irregular classical
mechanical systems with finite number of degrees of freedom. These systems
are classified, and regularized when possible, by introducing dynamically
equivalent regular Lagrangians. It is shown that a system cannot evolve in
time from a regular phase space configuration into an irregular one, since
regular and irregular configurations always belong to sectors of phase space
that do not intersect.

In Chapter 5, the dynamical structure of the higher-dimensional CS theories
is studied, where the symmetries are analyzed using canonical methods.
Criteria which determine whether the regularity and genericity conditions
are satisfied around the chosen background, are presented.

In Chapter 6, five-dimensional \emph{AdS}-CS supergravity theory, based on $%
SU(2,2\left\vert N\right. )$ group, is studied, as the simplest example of
CS supergravity theories with propagating degrees of freedom. In the
particular case of $N=4$, a class of generic and regular backgrounds is
found, such that all charges can be defined at the boundary. Among them,
there are configurations which are BPS states. The classical charge algebra
is obtained as the supersymmetric extension of the WZW$_{4}$ algebra with a
central extension.

The main results of the Thesis were published in Refs. \cite%
{Miskovic-Sazdovic,Miskovic-Zanelli,Miskovic-ZanelliJMP}, and the manuscript
\cite{Miskovic-Troncoso-Zanelli} is in preparation.

\chapter{Wess-Zumino-Witten model}

Wess-Zumino-Witten (WZW) models are conformal field theories in which an
affine, or Kac-Moody (KM) algebra gives the spectrum of the theory. The
two-dimensional WZW model studied here is a system whose kinetic term is
given by the nonlinear sigma model and the potential is the Wess-Zumino term
\cite{Wess-Zumino,Novikov}. This model was originally studied by Witten in
the context of two-dimensional bosonization, where it provides non-Abelian
bosonization rules describing non-interacting massless fermions \cite%
{Witten'84}. The WZW action was also used as a term cancelling quantum
anomalies \cite{Erlich-Freedman}--\cite{Bogojevic-Miskovic-Sazdovic2}. It is
a chirally and conformally invariant theory, and exactly solvable.

The purpose of this chapter is to present the WZW model and its global and
local symmetries in a systematic way, as well as to introduce the notation.
In the next chapter, a general idea of finding an action, starting only from
the algebra of its local symmetries, will be presented in two dimensions and
for $N=1$ superconformal group. It will be shown that, in that way, the
super Virasoso algebra leads to the super WZW model coupled to supergravity.

\section{The action}

The two-dimensional WZW model has a fundamental field $g$ belonging to a
non-Abelian semi-simple compact Lie group $G$ and the dynamics given by the
action

\begin{equation}
I_{\text{WZW}}\left[ g\right] =I_{0}\left[ g\right] +\Gamma \left[ g\right]
=-a\int\limits_{\mathcal{M}}\left\langle ^{\ast }\mathbf{VV}\right\rangle -%
\frac{k}{3}\int\limits_{\mathcal{B}}\left\langle \mathbf{V}^{3}\right\rangle
\qquad \left( \mathbf{V}\equiv g^{-1}dg\right) \,,  \label{WZW}
\end{equation}%
where $I_{0}\left[ g\right] $ is the action of the non-linear $\sigma $%
-model with a positive dimensionless coupling constant $a$, while $\Gamma %
\left[ g\right] $ is the topological Wess-Zumino term defined over a
three-manifold $\mathcal{B}$ whose boundary is the two-dimensional
space-time, $\partial \mathcal{B}=\mathcal{M},$ and $k$ is a dimensionless
constant. Here $\mathbf{V}$ is a Lie-algebra-valued one-form, $^{\ast }%
\mathbf{V}$ is its Hodge-dual and $\left\langle \cdots \right\rangle $
stands for a trace. The exterior product $\wedge $ between the forms is
understood.

The Wess-Zumino term has a non-local expression, involving integration over
the three-manifold $\mathcal{B}$, but it depends only on its boundary $%
\mathcal{M}$ modulo a constant. Independence from the choice of $\mathcal{B}$
follows from the identity
\begin{equation}
\frac{1}{12\pi }\int\limits_{\mathcal{B}}\left\langle \mathbf{V}%
^{3}\right\rangle -\frac{1}{12\pi }\int\limits_{\mathcal{B}^{\prime
}}\left\langle \mathbf{V}^{3}\right\rangle =-2\pi n\,,\qquad \left( n\in
\mathbb{Z}\right) \,,  \label{index}
\end{equation}%
with $\mathcal{M}=\partial \mathcal{B}=\partial \mathcal{B}^{\prime }$,
which is a consequence of the index theorem. The set $\mathcal{B\cup B}%
^{\prime }$ is a closed oriented manifold, topologically equivalent to a
three-sphere $S^{3}$, provided the boundaries of $\mathcal{B}$ and $\mathcal{%
B}^{\prime }$ have opposite orientations. Thus, the term $2\pi n$ in (\ref%
{index}) arises because of the topologically distinct possibilities to have
the mapping $g(x):$ $S^{3}\rightarrow G$, classified by $\pi _{3}\left(
G\right) \simeq \mathbb{Z}$ (for $G$ compact semi-simple), with the
corresponding winding number $n$. Therefore, all dynamics happens on the
two-dimensional space-time $\mathcal{M}$, provided the constant $k$ is
proportional to an integer,%
\begin{equation}
k=\frac{n}{8\pi }\,.  \label{quant}
\end{equation}%
Then, the quantum amplitude $\int \left[ dA\right] \,e^{iI_{\text{WZW}}\left[
A\right] }$ is independent of the choice of $\mathcal{B}$, provided the
constant $k$ is quantized as in (\ref{quant}).

\subparagraph{Field equations.}

The local coordinates on $\mathcal{M}$ with the signature $\left( -,+\right)
$ are introduced as $x^{\mu }$ $\left( \mu =0,1\right) $. Then the
differential forms can be expressed in the basis of 1-forms on $\mathcal{M}$
as $\mathbf{V}=g^{-1}\partial _{\mu }g\,dx^{\mu }$ and $^{\ast }\mathbf{V}%
=\varepsilon _{\mu }^{\;\;\,\nu }\,g^{-1}\partial _{\nu }g\,dx^{\mu }$ (with
$\varepsilon ^{01}=1$). Under a small variation of the gauge field, $%
g\rightarrow g+\delta g$, the WZW action changes as
\begin{equation}
\delta I_{\text{WZW}}\left[ g\right] =-a\int d^{2}x\,\left\langle
g^{-1}\delta g\,\partial _{\mu }\left( g^{-1}\partial ^{\mu }g\right)
\right\rangle +k\int d^{2}x\,\varepsilon ^{\mu \nu }\left\langle
g^{-1}\delta g\,\partial _{\mu }\left( g^{-1}\partial _{\nu }g\right)
\right\rangle \,.
\end{equation}%
In the \emph{light-cone} coordinates $x^{\pm }=\frac{1}{\sqrt{2}}\left(
x^{0}\pm x^{1}\right) $, it gives the following field equations:\footnote{%
In the \emph{light-cone} coordinates, the antisymmetric tensor becomes $%
\varepsilon ^{+-}=-\varepsilon ^{-+}=1$, and the metric has non-zero
components $\eta _{+-}=\eta _{-+}=-1$.}
\begin{equation}
\,\left( a+k\right) \,\partial _{-}\left( g^{-1}\partial _{+}g\right)
+\left( a-k\right) \partial _{+}\left( g^{-1}\partial _{-}g\right) =0\,.
\label{wzw eom}
\end{equation}%
Therefore, taking into account that $a$ must be positive in order to have
the right sign in the kinetic term of (\ref{WZW}), while the sign of $k=%
\frac{n}{8\pi }$ depends on $n\in \mathbb{Z}$, there are two possible
choices of $a$ which give a theory with chiral symmetry. For $a=-k$ and $n<0$%
, the equation of motion is $\partial _{+}\left( g^{-1}\partial _{-}g\right)
=0$, with the general solution
\begin{equation}
g(x^{+},x^{-})=g_{+}\left( x^{+}\right) g_{-}\left( x^{-}\right) \,,
\label{sol}
\end{equation}%
where $g_{+}$ and $g_{-}$ are elements of $G$. For $a=k$ and $n>0$, (\ref%
{wzw eom}) becomes $\partial _{-}\left( g^{-1}\partial _{+}g\right) =0$ and
the general solution is factorized as $g_{-}\left( x^{-}\right) g_{+}\left(
x^{+}\right) $. Equation (\ref{sol}) is invariant under the \emph{left}
\emph{chiral} transformations $g_{+}\left( x^{+}\right) \rightarrow \Omega
_{+}\left( x^{+}\right) g_{+}\left( x^{+}\right) $ and \emph{right chiral}
transformations $g_{-}\left( x^{-}\right) \rightarrow g_{-}\left(
x^{-}\right) \Omega _{-}^{-1}\left( x^{-}\right) $, or
\begin{equation}
g(x^{+},x^{-})\rightarrow \Omega _{+}\left( x^{+}\right)
g(x^{+},x^{-})\Omega _{-}^{-1}\left( x^{-}\right) \,,\qquad \left( \Omega
_{+},\Omega _{-}\right) \in G\otimes G\,.  \label{ch}
\end{equation}%
This symmetry is related to two independent Kac-Moody algebras, as will be
seen using canonical methods.

\section{Kac-Moody algebra}

The symmetries of the WZW action (\ref{WZW}) with $a=-k>0$ can be found
using the Hamiltonian formalism \cite{Witten'84,Sazdovic'95}. A summary of
the formalism is given in Appendix \ref{HF}.

Let the local coordinates $q^{i}$ parametrize the group manifold, $g=g(q)$,
where the number of coordinates is equal to the dimension of $G$, and the
generators of the group satisfy the Lie algebra $[\mathbf{G}_{a},\mathbf{G}%
_{b}]=f_{ab}^{\;\;c}\,\mathbf{G}_{c}$. Then the Lie-algebra-valued 1-form $%
\mathbf{V}$ can be expanded as
\begin{equation}
\mathbf{V}=g^{-1}dg=dq^{i}E_{i}^{a}\,\mathbf{G}_{a}\,,  \label{def q}
\end{equation}%
where $E_{i}^{a}$ is a vielbein on the group manifold. The Killing metric in
the adjoint representation of the Lie algebra is $g_{ab}=\left\langle
\mathbf{G}_{a}\mathbf{G}_{b}\right\rangle =f_{ad}^{\;\;c}f_{bc}^{\;\;d},$
and in the coordinate basis the metric is
\begin{equation}
\gamma _{ij}\left( q\right) =E_{i}^{a}\left( q\right) E_{j}^{b}\left(
q\right) g_{ab}\,.
\end{equation}%
On the basis of Poincar\'{e}'s lemma, the equation $d\left\langle \mathbf{V}%
\right\rangle ^{3}=0$ can be locally written as exterior derivative of a
two-form,%
\begin{equation}
\left\langle \mathbf{V}\right\rangle ^{3}=-6\,d\rho \,,
\end{equation}%
where $\rho \equiv \frac{1}{2}\,\rho _{ij}\,dq^{i}dq^{j}$. Rewriting this in
components, the following identities are obtained:
\begin{equation}
\frac{1}{2}\,f_{abc}\,E_{i}^{a}E_{j}^{b}E_{k}^{c}=\partial _{i}\rho
_{jk}+\partial _{j}\rho _{ki}+\partial _{k}\rho _{ij}\,.
\end{equation}%
Then the local expression for the action (\ref{WZW}) in the \emph{light-cone}
coordinates becomes:
\begin{equation}
I_{\text{WZW}}\left[ g\left( q\right) \right] =-2k\int d^{2}x\,\left[ \gamma
_{ij}\left( q\right) +2\rho _{ij}\left( q\right) \right] \partial
_{-}q^{i}\partial _{+}q^{j}\,.  \label{local w}
\end{equation}%
Taking $\tau =x^{0}$ as canonical time, one finds that (\ref{local w}) has
no first class constraints, therefore has no local symmetries. But it is
expected to find that this action possesses two types of chiral symmetry
which, unlike standard local symmetries, depend only on one coordinate. The
canonical approach to the chiral symmetries requires to take one of the
\emph{light-cone} coordinates as the time variable. In order to find the
complete chiral invariance of (\ref{local w}), both possibilities $\tau
=x^{-}$ and $\tau =x^{+}$ should be analyzed.

\subsubsection{\emph{a}) Local symmetries}

The space-time coordinates are chosen as $\left( \tau ,\sigma \right)
=\left( x^{-},x^{+}\right) $. Then, the momenta $p_{i}=\frac{\delta I}{%
\delta \dot{q}^{i}}$ canonically conjugated to coordinates $q^{i}$ are not
independent and define the constraints
\begin{equation}
K_{-i}\equiv p_{i}+2k\left( \gamma _{ij}+2\rho _{ij}\right) \partial
_{\sigma }q^{j}\approx 0\,,  \label{current-}
\end{equation}%
where the additional\ subindex \textquotedblleft $-$\textquotedblright\ in $%
K_{-i}$ refers to the choice of $\tau $, while $\approx $ denotes the weak
equality, since the derivatives of $K_{-i}$ do not vanish on the constraint
surface $K_{-i}=0$. An equivalent set of constraints (defining the same
constraint surface) is
\begin{equation}
K_{-a}=-E_{a}^{i}K_{-i}\approx 0\,,
\end{equation}%
where $E_{a}^{i}$ is the inverse vielbein $\left( E_{a}^{i}E_{i}^{b}=\delta
_{a}^{b}\text{ and }E_{i}^{a}E_{a}^{j}=\delta _{i}^{j}\right) $. Then, the
Hamiltonian
\begin{equation}
H=\int d\sigma \,u^{a}K_{-a}
\end{equation}%
depends on the arbitrary multipliers $u^{a}\left( \tau ,\sigma \right) $ and
generates the time evolution of any phase space variable $F\left( q,p\right)
$ by Poisson brackets (PB) as
\begin{equation}
\frac{dF}{d\tau }=\left\{ F,H\right\} \,.
\end{equation}%
The constraints $K_{-a}\approx 0$ are preserved in time if $\partial
_{\sigma }u=0$, thus they do not yield new constraints, and $u=u\left( \tau
\right) $. The PB of constraints give rise to the affine, or KM, algebra
\begin{equation}
\left\{ K_{-a}\left( x\right) ,K_{-b}\left( x^{\prime }\right) \right\}
=f_{ab}^{\;\;c}K_{-c}\left( x\right) \delta \left( \sigma -\sigma ^{\prime
}\right) -4kg_{ab}\partial _{\sigma }\delta \left( \sigma -\sigma ^{\prime
}\right) \,,  \label{KM1}
\end{equation}%
with the \emph{central extension} $-4k$. Here $x=\left( \tau ,\sigma \right)
$ and the PBs are taken at the same $\tau $. The presence of the Schwinger
term proportional to $\partial _{\sigma }\delta $ in (\ref{KM1}) indicates
that not all constraints are first class, however, it is not clear how to
identify first and second class constraints. Assuming that the space is
compact and all variables on $\mathcal{M}$ are periodic functions, with
period $L$, $F\left( \tau ,\sigma +L\right) =F\left( \tau ,\sigma \right) $,
they can be Fourier-expanded as
\begin{equation}
F\left( \tau ,\sigma \right) =\frac{1}{L}\,\sum\limits_{n\in \mathbb{Z}%
}F_{n}\left( \tau \right) e^{-\frac{2\pi i}{L}\,n\sigma }\qquad
\longleftrightarrow \qquad F_{n}\left( \tau \right)
=\int\limits_{0}^{L}d\sigma \,F\left( \tau ,\sigma \right) e^{\frac{2\pi i}{L%
}\,n\sigma }\,.
\end{equation}%
Then, the KM algebra becomes
\begin{equation}
\left\{ K_{-an},K_{-bm}\right\} =f_{ab}^{\;\;c}\,K_{-c\left( n+m\right)
}-8\pi i\,kn\,g_{ab}\delta _{n+m,0}\,.  \label{KM2}
\end{equation}%
The result does not depend on the period $L$, and now it is straightforward
to separate first and second class constraints. From
\begin{equation}
\left\{ K_{-a0},K_{-bn}\right\} =f_{ab}^{\;\;c}K_{-cn}\approx 0\ ,
\end{equation}%
it can be seen that the zero modes $K_{a0}$ are first class constrains and
that they close the PB subalgebra which is isomorphic to the algebra of $G$,
while the non-zero modes $K_{an}$ $\left( n\neq 0\right) $ are second class
constraints,
\begin{equation}
\left\{ K_{-an},K_{-b\left( -n\right) }\right\} \approx -8\pi
i\,kn\,g_{ab}\neq 0\,,\qquad \left( n\neq 0\right) \,.  \label{K second}
\end{equation}%
Therefore, the generator of gauge transformations, containing all first
class constraints, has the form%
\begin{equation}
G_{-}\left[ \lambda \right] \equiv \lambda _{-}^{a}K_{_{-}a0}\,,
\end{equation}%
with a local parameter $\lambda _{-}^{a}\left( \tau \right) $. The dynamical
field $q^{i}$ changes under infinitesimal gauge transformations as
\begin{equation}
\delta _{-}q^{i}=\left\{ q^{i},G\left[ \lambda \right] \right\} =-\lambda
_{-}^{a}E_{a}^{i}\,,
\end{equation}%
which, with the help of the expansion $g^{-1}\delta _{-}g=\delta
_{-}q^{i}E_{i}^{a}\mathbf{G}_{a}$, leads to the transformation law $%
g^{-1}\delta _{-}g=-\mathbf{\lambda }_{-},$ or
\begin{equation}
\delta _{-}g=-g\mathbf{\lambda }_{-}\,.  \label{right inf}
\end{equation}%
The transformations (\ref{right inf}) are the infinitesimal form of the
right chiral gauge transformations%
\begin{equation}
g\rightarrow g\ \Omega _{-}^{-1}\left( x^{-}\right) \,,  \label{right}
\end{equation}%
with a group element defined as $\Omega _{-}\equiv e^{\mathbf{\lambda }%
_{-}}\approx 1+\mathbf{\lambda }_{-}$.

Alternatively, choosing the space-time coordinates as $\left( \tau ,\sigma
\right) =\left( x^{+},-x^{-}\right) $, where the minus sign is adopted to
preserve the orientation between coordinate axes, and with the help of the
identity
\begin{equation}
-g\mathbf{V}g^{-1}=gdg^{-1}\equiv dq^{i}\tilde{E}_{i}^{a}\,\mathbf{G}_{a}\,,
\end{equation}%
the constraints take the form
\begin{equation}
K_{+a}\equiv -\tilde{E}_{a}^{i}\,\left[ p_{i}+2k\left( \gamma _{ij}-2\rho
_{ij}\right) \partial _{\sigma }q^{j}\right] \approx 0\,.  \label{current+}
\end{equation}%
They form an independent KM algebra,%
\begin{equation}
\left\{ K_{+a}\left( x\right) ,K_{+b}\left( x^{\prime }\right) \right\}
=f_{ab}^{\;\;c}K_{+c}\left( x\right) \delta \left( \sigma -\sigma ^{\prime
}\right) +4k\ g_{ab}\partial _{\sigma }\delta \left( \sigma -\sigma ^{\prime
}\right) \,,
\end{equation}%
with the central extension $+4k$, which has the opposite sign to that in (%
\ref{KM1}). The mode expansion gives the algebra (\ref{KM2}) with $%
k\rightarrow -k$ and the corresponding first class gauge generator $G_{+}%
\left[ \lambda \right] \equiv \lambda _{+}^{a}K_{_{+}a0}$ leads to right
chiral transformations $\delta _{+}g=\mathbf{\lambda }_{+}g,$ which are the
infinitesimal form of%
\begin{equation}
g\rightarrow \Omega _{+}\left( x^{+}\right) g\,,  \label{left}
\end{equation}%
where $\Omega _{+}\equiv e^{\mathbf{\lambda }_{+}}\approx 1+\mathbf{\lambda }%
_{+}$. The constraints $K_{+a}$ commute with $K_{-b}$. Both results (\ref%
{right}) and (\ref{left}) are independent gauge symmetries of the action.
This means that the action (\ref{local w}) is invariant under the chiral
transformations (\ref{ch}) generated by $K_{+a0}$ acting from the right, and
$K_{-a0}$ acting from the left. The modes $K_{\pm an},$ $n\neq 0,$ are not
the generators of local symmetries (they are second class constraints), and
they lead to the central extensions $\pm 4k$ in the corresponding KM
algebras.

\subsubsection{\emph{b}) Global symmetries}

Choosing the space-time coordinates as $\tau =x^{-}$ and $\sigma =x^{+}$,
the infinitesimal chiral transformations (\ref{ch}) of $g\in G$ take a form%
\begin{equation}
\delta _{\pm }g=\mathbf{\lambda }_{+}\left( \sigma \right) g-g\mathbf{%
\lambda }_{-}\left( \tau \right) \,,
\end{equation}%
with the Lie-algebra-valued parameters $\mathbf{\lambda }_{\pm }=\lambda
_{\pm }^{a}\mathbf{G}_{a}$ or, in terms of the local fields $q^{i},$ the
transformations become%
\begin{equation}
\delta _{\pm }q^{i}=-\lambda _{+}^{a}\left( \sigma \right) \tilde{E}%
_{a}^{i}-\lambda _{-}^{a}\left( \tau \right) E_{a}^{i}\,.
\end{equation}%
Therefore, \emph{right} chiral transformations, given by the time-dependent
parameter $\mathbf{\lambda }_{-}\left( \tau \right) $, lead to \emph{local
symmetries} of the WZW action. It is already shown that these symmetries are
generated by the first class constraints $K_{-a0}\approx 0$.

On the other hand, \emph{left} chiral transformations correspond to \emph{%
global symmetries} of the action, since they are given by infinite number of
time-independent parameters $\mathbf{\lambda }_{+}\left( \sigma \right) $.
Conserved charges corresponding to these global symmetries are obtained from
Noether currents,%
\begin{equation}
J_{+}^{a}\equiv 4k\,\left( g\partial _{+}g^{-1}\right) ^{a}=4k\,\tilde{E}%
_{i}^{a}\partial _{\sigma }q^{i}\,,
\end{equation}%
and they do not vanish on the constraint surface $K_{-a}=0$.

In order to find the current algebra, it is convenient to introduce a Dirac
bracket (DB) as%
\begin{equation}
\left\{ M,N\right\} ^{\ast }\equiv \left\{ M,N\right\}
-\sum\limits_{m,n\neq 0}\left\{ M,K_{-an}\right\} \Delta _{nm}^{ab}\left\{
K_{-bm},N\right\} \,,
\end{equation}%
where $\Delta _{nm}^{ab}\equiv \frac{1}{8\pi i\,kn}\,g^{ab}\delta _{n+m,0}$ $%
\left( n,m\neq 0\right) $ is the inverse of the PB matrix of second class
constraints $K_{-an}$ $\left( n\neq 0\right) $, given by Eq. (\ref{K second}%
). Then it can be shown that the currents $J_{+}^{a}$ satisfy the KM algebra
with the central charge $4k$,%
\begin{equation}
\left\{ J_{+a}\left( x\right) ,J_{+b}\left( x^{\prime }\right) \right\}
^{\ast }=f_{ab}^{\;\;c}J_{+c}\left( x\right) \delta \left( \sigma -\sigma
^{\prime }\right) +4k\ g_{ab}\partial _{\sigma }\delta \left( \sigma -\sigma
^{\prime }\right) \,.  \label{KM J}
\end{equation}

Similarly, choosing $\tau =x^{+}$ as canonical time, there are Noether
currents $J_{-}=4k\,g^{-1}\partial _{-}g,$ corresponding to the right chiral
symmetries as global symmetries of the WZW model. The currents $J_{-}^{a}$
satisfy a KM current algebra with the central charge $-4k$.

\section{Virasoro algebra}

WZW theory is also invariant under conformal transformations, or
diffeomorphisms $x^{\mu }\rightarrow x^{\prime \mu }(x)$ which change the
line element by a scale factor at each point of space-time,
\begin{equation}
ds^{\prime 2}=g_{\mu \nu }^{\prime }\left( x^{\prime }\right) \,dx^{\prime
\mu }dx^{\prime \nu }=\Lambda \left( x\right) g_{\mu \nu }\left( x\right)
\,dx^{\mu }dx^{\nu }=\Lambda \left( x\right) ds^{2}\,.
\end{equation}%
Conformal transformations have the form of chiral and antichiral mappings $%
x^{+}\rightarrow f_{+}(x^{+})$ and $x^{-}\rightarrow f_{-}(x^{-}),$ and
their generators are the \emph{light-cone }components of the energy-momentum
tensor $T_{++}(x^{+})$ and $T_{--}(x^{-})$, which is traceless ($T_{\;\mu
}^{\mu }=2T_{+-}=0$). They satisfy two independent Virasoro algebras,
\begin{equation}
\left[ T(x),T(y)\right] =-\left[ T(x)+T(y)\right] \delta ^{\prime }(\sigma
_{x}-\sigma _{y})-\frac{c}{12}\delta ^{\prime \prime \prime }(\sigma
_{x}-\sigma _{y})\,,  \label{virasoro}
\end{equation}%
\emph{without central charge} ($c=0$) in a classical theory, or \emph{with}
\emph{central charges }$c=c_{0}$ and $c=-c_{0}$ (for $T_{++}$ and $T_{--}$
respectively) in the quantum case. (The normalization of $c$ is adopted from
string theory.) The appearance of the Schwinger term in the quantum Virasoro
algebra is called a\emph{\ quantum anomaly,} but it can appear in a
classical theory as well \cite{Brown-Henneaux}. The physical meaning of $%
c\neq 0$ in a classical theory is the breaking of conformal symmetry by the
introduction of a macroscopic scale into the system (by boundary conditions,
for example).

The Fourier modes of the Virasoro generators, $L_{n}\,\left( n\in \mathbb{Z}%
\right) $, in a compact space with period $L$, are given by%
\begin{equation}
L_{n}=\frac{L}{2\pi i}\int\limits_{0}^{L}d\sigma \,T\left( \sigma \right)
\,e^{\frac{2\pi i}{L}\,n\sigma },
\end{equation}%
where $L_{n}^{\dagger }=L_{-n}$ for unitary representations. They obey the
well-known commutation rules
\begin{equation}
\left[ L_{n},L_{m}\right] =(n-m)L_{n+m}+\frac{c}{12}\,2\pi i\ n^{3}\delta
_{n+m,0}\,.  \label{Virasoro c}
\end{equation}%
This algebra contains a finite subalgebra, generated by $\{L_{-1},$ $L_{0},$
$L_{1}\},$ associated to global conformal invariance.

Conformal symmetry does not arise in the Hamiltonian analysis because it is
not an independent symmetry. Virasoro generators can be expressed in terms
of KM currents as \cite%
{Blagojevic-Popovic-Sazdovic'98,Blagojevic-Popovic-Sazdovic'99}%
\begin{eqnarray}
T_{--} &=&\frac{1}{4k}\,g^{ab}J_{-a}J_{-b}\,,  \notag \\
T_{++} &=&-\frac{1}{4k}\,g^{ab}J_{+a}J_{+b}\,,  \label{Sugaw}
\end{eqnarray}%
and they represent the \emph{light-cone} components of the energy-momentum
tensor. Conformal invariance leads to $T_{-+}=0$. In terms of the Fourier
modes, the relations (\ref{Sugaw}) become%
\begin{equation}
L_{\pm n}=\mp \frac{1}{8\pi ik}\,\sum\limits_{m\in \mathbb{Z}}g^{ab}J_{\pm
am}J_{\pm b\left( n-m\right) }\,.
\end{equation}

The observation that the energy-momentum tensor is a bilinear in the
currents, is used to construct them using the procedure of Sugawara\emph{\ }%
\cite{Sugawara,Sommerfield}. More precisely, for a given KM algebra, there
is always a Virasoro algebra, such that they form a semi-direct product
\begin{equation}
\left[ L_{n},J_{am}\right] =2m\,J_{a\left( n+m\right) }\,,\qquad \qquad %
\left[ L_{n},K\right] =0\,,  \label{semi}
\end{equation}%
where $K$ is the central extension of the KM algebra. More about Virasoro
and KM algebras in the conformal filed theory can be found in Refs. \cite%
{Goddard-Olive}--\cite{West}.

\section{Gauged WZW model}

The two-dimensional WZW model is closely related to a three-dimensional
Chern-Simons (CS) theory whose dynamical field is a Lie-algebra-valued
1-form, $\mathbf{A}=A^{a}\mathbf{G}_{a}$. Definition of the CS Lagrangian
comes from the Chern character,%
\begin{equation}
P(A)=\left\langle \mathbf{F}^{2}\right\rangle \equiv g_{ab}F^{a}F^{b},
\end{equation}%
where $\mathbf{F}=d\mathbf{A}+\mathbf{A}^{2}$ is the field-strength 2-form
associated with the gauge field $\mathbf{A}$. Since the Chern character is a
closed form ($dP=0$), on the basis of algebraic Poincar\'{e}'s lemma%
\footnote{%
Poincar\'{e} lemma states that any closed form $P\left( A\right) $ can be
locally written as exact form $d\alpha \left( A\right) $. The algebraic
Poincar\'{e} lemma guarantees that a differential form $\alpha \left(
A\right) $ is a local function of $A$ (depending on finite number of
derivatives $\partial A$, $\partial ^{2}A$, \ldots ).}, it can be locally
written as the exterior derivative of a 3-form, called the CS form, which
defines the CS Lagrangian as $dL_{\text{CS}}\left( A\right) =kP\left(
A\right) $. Then the CS action is given by
\begin{equation}
I_{\text{CS}}\left[ A\right] =\int\limits_{\mathcal{B}}L_{\text{CS}}\left(
A\right) =k\int\limits_{\mathcal{B}}\left\langle \mathbf{AF}-\frac{1}{3}\,%
\mathbf{A}^{3}\right\rangle \,,  \label{CS 3}
\end{equation}%
where $\mathcal{B}$ is a three-dimensional manifold (not necessarily without
a boundary).

Under finite gauge transformations
\begin{equation}
\mathbf{A}^{g}=g\left( \mathbf{A}+d\right) g^{-1}=g\left( \mathbf{A}-\mathbf{%
V}\right) g^{-1}\,,\qquad \left( g\in G\right) \,,  \label{large}
\end{equation}%
the field-strength transforms homogeneously $\left( \mathbf{F}^{g}=g\mathbf{F%
}g^{-1}\right) $, the Chern character is invariant and the CS Lagrangian
changes as $L_{\text{CS}}^{g}=L_{\text{CS}}+\omega $, where $\omega $ is a
closed form ($d\omega =0$) which need not be exact for nontrivial topology
of $\mathcal{B}$. Explicitly, under the finite gauge transformations, the CS
action changes as

\begin{equation}
I_{\text{CS}}\left[ A^{g}\right] -I_{\text{CS}}\left[ A\right] =\alpha \left[
A,g\right] \,,  \label{anomaly}
\end{equation}%
where
\begin{equation}
\alpha \left[ A,g\right] =-k\int\limits_{\mathcal{M}=\partial \mathcal{B}%
}\left\langle \mathbf{AV}\right\rangle +\frac{k}{3}\int\limits_{\mathcal{B}%
}\left\langle \mathbf{V}^{3}\right\rangle \,.
\end{equation}%
The last term is recognized as the Wess-Zumino term. The functional $\alpha %
\left[ A,g\right] $ satisfies the so-called \emph{cocycle equation}, or
Polyakov-Wiegmann identity \cite{Polyakov-Wiegmann},%
\begin{equation}
\alpha \left[ A^{g},h\right] -\alpha \left[ A,gh\right] +\alpha \left[ A,g%
\right] =0\,,\qquad \left( g,h\in G\right) \,.  \label{cocycle eq}
\end{equation}%
Any quantity which satisfies the above equation is called Wess-Zumino term,
or \emph{anomaly}, and it describes the non-invariance of the quantum theory
under a classical gauge symmetry.\footnote{%
The CS action (\ref{CS 3}) plays the role of an effective action for a
theory with matter fields $\phi $, whose quantum theory is described by the
functional integral
\begin{equation*}
Z=\int \left[ dA\right] \,e^{iI_{\text{CS}}\left[ A\right] }=\int \left[ dA%
\right] \left[ d\phi \right] \,e^{iI_{0}\left[ \phi ,A\right] }\,,
\end{equation*}%
where the classical action is invariant under gauge transformations, $I_{0}%
\left[ \phi ^{g},A^{g}\right] =I_{0}\left[ \phi ,A\right] ,$ while the
measure has the anomaly $\left[ d\phi \right] ^{g}=\left[ d\phi \right]
\,e^{i\alpha \left[ A,g\right] }.$} Any other object of the form $\beta %
\left[ A^{g}\right] -\beta \left[ A\right] $ also satisfies the cocycle
equation. Since $\left\langle \mathbf{A}^{2}\right\rangle =0$, a natural
nontrivial possibility for $\beta $ is
\begin{equation}
\beta \left[ A\right] =a\int\limits_{\mathcal{M}}\left\langle ^{\ast }%
\mathbf{AA}\right\rangle \qquad \left( a\in \mathbb{R}\right) \,,
\end{equation}%
where $^{\ast }\mathbf{A}$ is a Hodge-dual field. With this choice, the
action which satisfies the cocycle equation (\ref{cocycle eq}), and
therefore describes the anomaly of a classical theory, is
\begin{equation}
I_{\text{GWZW}}\left[ A,g\right] =I_{\text{WZW}}\left[ g\right]
+2a\int\limits_{\mathcal{M}}\left\langle ^{\ast }\mathbf{AV}\right\rangle
+k\int\limits_{\mathcal{M}}\left\langle \mathbf{AV}\right\rangle \,,
\end{equation}%
and is called the \emph{gauged WZW action}. More about this model and its
quantization can be found in Refs. \cite%
{Gawedzki-Kupiainen'88,Gawedzki-Kupiainen'89,Karabali-Schnitzer,SazdovicDanube}%
.

\section{Supersymmetric WZW model}

The supersymmetric generalization of the WZW model is defined in $\left(
1,1\right) $ superspace parametrized by four real coordinates $z^{A}=\left(
x^{\mu },\theta _{\alpha }\right) $, where $x^{\mu }$ $\left( \mu
=0,1\right) $ are local coordinates on a two-dimensional space-time with the
signature $\left( -,+\right) ,$ and $\theta _{\alpha }$ $\left( \alpha
=+,-\right) $ is a Majorana spinor.\footnote{%
The notation $\left( 1,1\right) $ refers to the superspace with two
Grassmann odd variables $\left( \theta _{+},\theta _{-}\right) $.} The super
Wess-Zumino-Witten (SWZW) model is given by the action \cite%
{Vecchia-Knizhnik-Petersen-Rossi,Abdalla-Abdalla},%
\begin{equation}
I_{\text{SWZW}}\left[ g\right] =-k\int d^{4}z\,\left\langle \bar{D}%
S^{\dagger }DS\right\rangle -\frac{2k}{3}\int d^{4}z\,\left\langle
S^{\dagger }\dot{S}\bar{D}S^{\dagger }\gamma _{3}DS\right\rangle \,,
\label{swzw}
\end{equation}%
where $k$ is a positive dimensionless constant, $S$ is a matrix superfield, $%
\left\langle \cdots \right\rangle $ stands for a supertrace and $D_{\alpha }=%
\bar{\partial}_{\alpha }+i\left( \gamma ^{m}\theta \right) _{\alpha
}\partial _{m}$ is the supercovariant derivative defined in the tangent
Minkowski space with coordinates $x^{m}$ $\left( m=0,1\right) $, where $%
\partial _{m}\equiv \partial /\partial x^{m}$ and $\bar{\partial}_{\alpha
}\equiv \partial /\partial \bar{\theta}^{\alpha }\,.$ The $\gamma $-matrices
$\gamma ^{m}$ satisfy the Clifford algebra and $\gamma _{3}\equiv i\gamma
^{0}\gamma ^{1}.$ Integration is carried out over the superspace variables,
where $d^{4}z\equiv d^{2}x\,d^{2}\theta $ and basic integrals for Grassman
odd numbers are $\int d\theta =0$ and $\int d\theta \,\theta =1$. All
conventions and representations are given in Appendix \ref{super}.

The supermatrix $S$ is expanded in the superspace as
\begin{eqnarray}
S &=&g+i\theta \psi +\frac{i}{2}\,\bar{\theta}\theta \,F\,,  \notag \\
S^{\dagger } &=&g^{\dagger }+i\theta \psi ^{\dagger }+\frac{i}{2}\,\bar{%
\theta}\theta \,F^{\dagger }\,.
\end{eqnarray}%
Supposing, for simplicity, that $S$ is unitary ($SS^{\dagger }=1$), one
obtains
\begin{eqnarray}
g^{\dagger } &=&g^{-1}\,,  \notag \\
\psi ^{\dagger } &=&-g^{\dagger }\psi g^{\dagger }\,, \\
F^{\dagger } &=&-g^{\dagger }Fg^{\dagger }-ig^{\dagger }\bar{\psi}g^{\dagger
}\psi g^{\dagger }\,,  \notag
\end{eqnarray}%
where the identity $\bar{\theta}\psi \ \bar{\theta}\psi ^{\dagger }=-\frac{1%
}{2}\ \bar{\theta}\theta \ \bar{\psi}\psi ^{\dagger },$ valid for Majorana
fermions, is used. The equations of motion obtained from the action (\ref%
{swzw}) are
\begin{equation}
\bar{D}\left( S^{\dagger }\gamma ^{+}DS\right) =0\,,
\end{equation}%
where $\gamma ^{\pm }=\frac{1}{2}\,\left( 1\pm \gamma _{3}\right) $ are
projective $\gamma $-matrices.

In order to be convinced that this model is indeed a supersymmetric
extension of the WZW model (\ref{WZW}), the action (\ref{swzw}) can be
written in components and the Berezin integrals over Grassman variables are
performed. Then,
\begin{equation}
I_{\text{SWZW}}\left[ S\right] =I_{\text{WZW}}\left[ g\right] ++I_{\text{f}}%
\left[ g,\psi \right] +I_{F}\left[ g,\psi ,F\right] \,,  \label{terms}
\end{equation}%
where the bosonic sector $I_{\text{WZW}}\left[ g\right] $ is the WZW action (%
\ref{WZW}) with $a=k$, and the additional terms demanded by the
supersymmetry are
\begin{equation}
\begin{tabular}{lll}
$I_{\text{f}}\left[ g,\psi \right] $ & $=$ & $-ik\int d^{2}x\,\left\langle
\bar{\psi}^{\dagger }\left( \partial \hspace{-0.2cm}/+\frac{1}{2}\,\gamma
_{3}\partial \hspace{-0.2cm}/gg^{\dagger }\right) \psi \right\rangle \,,$ \\
$I_{F}\left[ g,\psi ,F\right] $ & $=$ & $k\int d^{2}x\,\left\langle
F^{\dagger }F-\frac{i}{4}\,\bar{\psi}^{\dagger }\gamma _{3}\psi \left(
F^{\dagger }g-g^{\dagger }F\right) \right\rangle \,,$%
\end{tabular}
\label{susy}
\end{equation}%
where $\partial \hspace{-0.2cm}/\equiv \gamma ^{\mu }\partial _{\mu }$. The
matrix field $F$ is auxiliary (it does not have a kinetic term in $I_{F}$)
and it can be integrated out by means of its equations of motion,%
\begin{equation}
F=\frac{i}{2}\,g\left( \bar{\psi}^{\dagger }\psi +\frac{1}{2}\,\bar{\psi}%
^{\dagger }\gamma _{3}\psi \right) \ .  \label{gF}
\end{equation}%
Putting the solution for $F,$ given by (\ref{gF}), into the action $I_{F}%
\left[ g,\psi ,F\right] ,$ and after using the Fierz identity for Majorana
fermions,
\begin{equation}
\left\langle \left( \bar{\psi}^{\dagger }\psi \right) ^{2}+\left( \bar{\psi}%
^{\dagger }\gamma _{3}\psi \right) ^{2}\right\rangle =0\,,
\end{equation}%
one obtains that this action vanishes,
\begin{equation}
I_{F}\left[ g,\psi ,F\left( g,\psi \right) \right] =0\,.
\end{equation}%
Although the fermions $\psi $ in $I_{\text{f}}$ are, in general,
interacting, after making the following reparametrization
\begin{equation}
\chi =g^{\dagger }\gamma ^{+}\psi +\gamma ^{-}\psi g^{\dagger }\,,
\label{rep}
\end{equation}%
the fermionic action reduces to the action of a free fermion
\begin{equation}
I_{0}\left[ \chi \right] =I_{\text{f}}\left[ g,\psi \right] =-\frac{ik}{2}%
\int d^{2}x\,\left\langle \bar{\chi}\partial \hspace{-0.2cm}/\chi
\right\rangle \,,  \label{free}
\end{equation}%
and the fermions are completely decoupled from the WZW term. With the
reparametrization (\ref{rep}), the superfield $S$ is factorized as:
\begin{equation}
S=\left( 1+i\theta _{+}\chi _{-}\right) g\left( 1-i\theta _{-}\chi
_{+}\right) \,.
\end{equation}

\subparagraph{Equations of motion.}

The classical equations of motion following from the action
\begin{equation}
I_{\text{SWZW}}\left[ S\right] =I_{\text{WZW}}\left[ g\right] +I_{0}\left[
\chi \right]
\end{equation}%
lead to the general solution in \emph{light-cone} coordinates:
\begin{equation}
g\left( x^{+},x^{-}\right) =g_{-}\left( x^{-}\right) g_{+}\left(
x^{+}\right) \,,  \label{sol 1}
\end{equation}%
and for fermions
\begin{equation}
\chi _{+}=\chi _{+}\left( x^{+}\right) \,,\qquad \qquad \chi _{-}=\chi
_{-}\left( x^{-}\right) \,,
\end{equation}%
giving the factorization of the superfield as
\begin{equation}
S\left( x^{+},x^{-},\theta _{+},\theta _{-}\right) =S_{-}\left( x^{-},\theta
_{+}\right) S_{+}\left( x^{+},\theta _{-}\right) \,,
\end{equation}%
with the factors:
\begin{equation}
S_{-}=\left( 1+i\theta _{+}\chi \right) \,g_{-}\,,\qquad \qquad
S_{+}=g_{+}\,\left( 1-i\theta _{-}\chi _{+}\right) \,.  \label{sol 3}
\end{equation}

\subparagraph{Local symmetries.}

The general form of solutions (\ref{sol 1} - \ref{sol 3}) implies that the
SWZW model possesses:

(\emph{i}) KM or chiral symmetries $G\otimes G$, where the components of the
superfield transform under $\Omega _{-}\left( x^{-}\right) $ and $\Omega
_{+}\left( x^{-}\right) $, with $\Omega _{-}\Omega _{-}^{\dagger }=\Omega
_{+}\Omega _{+}^{\dagger }=1$, as
\begin{equation}
\begin{tabular}{lll}
$g$ & $\rightarrow $ & $\Omega _{-}g\Omega _{+}^{-1}\,,\medskip $ \\
$\chi _{-}$ & $\rightarrow $ & $\Omega _{-}\chi _{-}\Omega
_{-}^{-1}\,,\medskip $ \\
$\chi _{+}$ & $\rightarrow $ & $\Omega _{+}\chi _{+}\Omega _{+}^{-1}\,.$%
\end{tabular}%
\end{equation}

(\emph{ii}) The supersymmetry partner of the local KM transformations is an
additional invariance, with local parameters Majorana spinors $\eta
_{+}\left( x^{+}\right) $ and $\eta _{-}\left( x^{-}\right) $, where the
fields transform as
\begin{equation}
\delta g=0\,,\qquad \qquad \delta \chi _{\pm }=\eta _{\pm }\,.
\end{equation}

(\emph{iii}) The action is, by construction, also invariant under the \emph{%
conformal supersymmetry transformations }which change the line element of
the superpace $ds^{2}=d\ell ^{+}d\ell ^{-}\,$(where $d\ell ^{\pm }\equiv
dx^{\pm }-d\theta ^{\pm }\theta ^{\pm }$) by a scale factor: $ds^{\prime
2}=\Omega \,ds^{2}$. The fields change under superconformal transformations
as
\begin{eqnarray}
\delta _{\epsilon }g &=&i\bar{\epsilon}\psi \,,  \notag \\
\delta _{\epsilon }\psi &=&\left( \partial \hspace{-0.2cm}/g+ig\bar{\psi}%
\gamma ^{+}\psi \right) \epsilon \,,  \label{super c}
\end{eqnarray}%
where the local parameter $\epsilon $ is a Majorana spinor satisfying the
constraint $\partial \hspace{-0.2cm}/\gamma ^{\mu }\epsilon =0,$ that has
the solution $\epsilon ^{\pm }\equiv \gamma ^{\pm }\epsilon =\epsilon ^{\pm
}\left( x^{\pm }\right) \,.$ The supersymmetry transformations (\ref{super c}%
) imply the following transformations of $g$ and $\chi $ in components:
\begin{equation}
\begin{tabular}{lll}
$\delta _{\epsilon }g$ & $=$ & $i\left( \epsilon ^{+}\chi ^{-}g-\epsilon
^{-}g\chi ^{+}\right) \,,\medskip $ \\
$\delta _{\epsilon }\chi ^{+}$ & $=$ & $\left( g^{\dagger }\partial
_{+}g+i\chi _{+}\chi _{+}\right) \epsilon ^{-}\,,\medskip $ \\
$\delta _{\epsilon }\chi ^{-}$ & $=$ & $\left( g\partial _{-}g^{\dagger
}+i\chi _{-}\chi _{-}\right) \epsilon ^{+}\,.$%
\end{tabular}%
\end{equation}

Similarly to the non-supersymmetric case, it is possible to find the chiral
\emph{supercurrents} which close two independent super KM algebras. The
generators of the superconformal transformations, \emph{i.e.}, the group
invariants of the supercurrents which can be obtained by a generalized
Sugawara construction, close two independent super Virasoro algebras\emph{\ }%
without central charges. The details of this analysis can be found in Ref.
\cite{Vecchia-Knizhnik-Petersen-Rossi}. Those algebras will be constructed
in the next chapter following a different approach, in the context of the
SWZW model coupled to supergravity.

The general form of a super Virasoro algebra\emph{\ }has two central charges
$c$ and $\hat{c}$ which commute with all generators. The infinite set of
super Virasoro generators $L_{n}$ $\left( n\in \mathbb{Z}\right) $ and $%
G_{r} $ (where $r\in \mathbb{Z}$ for the \emph{Ramond sector} \cite{Ramond},
while $r\in \mathbb{Z+}\frac{1}{2}$ for the \emph{Neveu-Schwarz sector} \cite%
{Neveu-Schwarz1,Neveu-Schwarz2}), obey the following (anti)commutation
relations:
\begin{eqnarray}
\left[ L_{n},L_{m}\right] &=&\left( n-m\right) L_{n+m}+\frac{c}{12}\,n\left(
n^{2}-1\right) \delta _{n+m,0}\,,  \notag \\
\left[ G_{r},L_{n}\right] &=&\left( r-\frac{n}{2}\right) G_{n+r}\,, \\
\left\{ G_{r},G_{s}\right\} &=&2L_{r+s}+\frac{\hat{c}}{3}\,\left( r^{2}-%
\frac{1}{4}\right) \delta _{r+s,0}\,.  \notag
\end{eqnarray}%
The Ramond and Neveu-Schwarz sectors correspond to two different periodic
conditions of fermions, $\theta \left( e^{2\pi i}z\right) =\theta \left(
z\right) $ or $\theta \left( e^{2\pi i}z\right) =-\theta \left( z\right) .$
In the Neveu-Schwarz sector, the five generators $\left\{
L_{-1},L_{0},L_{1},G_{-1/2},G_{1/2}\right\} $ form a closed subalgebra $%
osp\left( 1\left\vert 2\right. \right) $, while in the Ramond sector a
closed subalgebra does not exist.

\chapter{Supersymmetric WZW model coupled to supergravity}

The extension of the WZW model has been studied with rigid supersymmetry
\cite{Vecchia-Knizhnik-Petersen-Rossi,Abdalla-Abdalla}, and with local
supersymmetry \cite%
{Bergshoeff-Daemi-Salam-Sarmadi-Sezgin,Gates-Nishino,Ketov}. The models were
constructed and analyzed using Lagrangian formalism in both superfield and
component notations. In the previous chapter, it is shown that in a case of
global supersymmetry it is possible to choose the fermionic field such that
fermions are completely decoupled.

In this chapter, supersymmetric WZW model coupled to two-dimensional
supergravity will be constructed \cite{Miskovic-Sazdovic} as a theory which
appears as a Lagrangian realization of the super Virasoro algebra. In the
Hamiltonian formalism, the components of the metric tensor and
Rarita-Schwinger field appear naturally as Lagrange multipliers
corresponding to the constraints satisfying the super Virasoro PB algebra.
Similar approach has been used to find a diffeomorphisms invariant action
for the spinning string \cite{Brink-Deser-Zumino-Vecchia-Howe}--\cite%
{Deser-Zumino}.

\section{Super Virasoro generators}

The Hamiltonian formalism (see Appendix \ref{ET}) will be used to construct
an action invariant under the gauge transformations for a given algebra of
the group $G$. Consider a PB \emph{representation} of the algebra in the
form
\begin{equation}
\left\{ \phi _{r},\phi _{s}\right\} =f_{rs}^{\;\;\;p}\phi _{p}\,,
\label{general1}
\end{equation}%
whose elements $\phi _{r}$ are functions of the coordinates $q^{i}$ and
their conjugate momenta $p_{i}$. By definition, $\phi _{r}$ are \emph{first
class} constraints, and the canonical Hamiltonian is assumed to be zero
(there are no local degrees of freedom). Then, the canonical action
\begin{equation}
I\left[ q,p,u\right] =\int dt\,\left( p_{i}\,\dot{q}^{i}-u^{r}\phi
_{r}\right)  \label{action E}
\end{equation}%
is invariant under the gauge transformations generated by the first class
constraints $\phi _{r}$. Any phase-space function $F(q,p)$ changes as
\begin{equation}
\delta _{\varepsilon }F=\{F,\varepsilon ^{r}\phi _{r}\}\,,  \label{general2}
\end{equation}%
and the Lagrange multipliers $u^{r}$ transform as
\begin{equation}
\delta _{\varepsilon }u^{r}=\dot{\varepsilon}^{r}+f_{ps}^{\;\;\;r}\,u^{s}%
\varepsilon ^{p}\,.  \label{general3}
\end{equation}%
The multipliers will be identified as gauge fields, later.

A similar approach has been used for the construction of the action for
W-strings propagating on a group manifold and on curved backgrounds \cite%
{Mikovic-Sazdovic'95,Mikovic-Sazdovic'97}, and also for two-dimensional
induced gravity \cite{Popovic-Sazdovic}. The covariant extension of the WZW
model with respect to an arbitrary internal group has been obtained in \cite%
{Sazdovic'00} and with respect to the $SL(2,\mathbb{R})$ internal group and
diffeomorphisms in \cite%
{Blagojevic-Popovic-Sazdovic'98,Blagojevic-Popovic-Sazdovic'99}, by the same
method. Here, the last approach will be supersymmetrized.

\subparagraph{Super Kac-Moody algebra.}

The representation of the supersymmetric KM algebra will be constructed
starting from the known bosonic KM sector. The field $g\in G$ is a mapping
from a two-dimensional Riemannian space-time $\mathcal{M}$ to a semi-simple
Lie group $G$, parametrized by local coordinates $q^{i}$, $g=g(q)$, and
generated by the anti-Hermitean generators $\mathbf{G}_{a}$ closing a Lie
algebra with structure constants $f_{ab}^{\;\;c}$. Two Maurer-Cartan (Lie
algebra valued) one-forms can be defined, $\mathbf{A}_{+}=g^{-1}dg$ and $%
\mathbf{A}_{-}=-g\mathbf{A}_{+}g^{-1}=gdg^{-1}$, whose expansions
\begin{equation}
\mathbf{A}_{+}=dq^{i}E_{+i}^{a}\,\mathbf{G}_{a}\,,\qquad \qquad \mathbf{A}%
_{-}\equiv dq^{i}E_{-i}^{a}\,\mathbf{G}_{a}\,,
\end{equation}%
define vielbeins $E_{\pm i}^{a}\left( q\right) $ on the group manifold, with
inverses $E_{\pm a}^{i}\left( q\right) $ ($E_{\pm i}^{a}E_{\pm a}^{j}=\delta
_{i}^{j}$ and $E_{\pm i}^{a}E_{\pm b}^{i}=\delta _{b}^{a}$). The Killing
metric is $g_{ab}=\frac{1}{2}\,\left\langle \mathbf{G}_{a}\mathbf{G}%
_{b}\right\rangle =\frac{1}{2}\,f_{ac}^{\;\;d}f_{bd}^{\;\;c}.$ The metric $%
\gamma _{ij}\left( q\right) $ in the coordinate basis does not depend on the
choice of the Maurer-Cartan form,
\begin{equation}
\gamma _{ij}=E_{+i}^{a}\,E_{+j}^{b}\,g_{ab}=E_{-i}^{a}\,E_{-j}^{b}\,g_{ab}\,,
\end{equation}%
and has the inverse $\gamma ^{ij}\left( q\right) $. The forms $\left\langle
\left( \mathbf{A}_{+}\right) ^{3}\right\rangle $ and $\left\langle \left(
\mathbf{A}_{-}\right) ^{3}\right\rangle $ are closed, and can be locally
written as
\begin{equation}
\frac{1}{2}\,\left\langle \left( \mathbf{A}_{\pm }\right) ^{3}\right\rangle
=-6\,d\rho \,,  \label{exact}
\end{equation}%
where $\rho \equiv \frac{1}{2}\,\rho _{ij}\,dq^{i}dq^{j}\,$ is a two-form,
independent on the choice of $\mathbf{A}_{+}$ or $\mathbf{A}_{-}$. Rewriting
in components, the expressions (\ref{exact}) leads to the identities
\begin{equation}
\pm \frac{1}{2}\,f_{abc}\,E_{\mp i}^{a}E_{\mp j}^{b}E_{\mp k}^{c}=\partial
_{i}\rho _{jk}+\partial _{j}\rho _{ki}+\partial _{k}\rho _{ij}\,.
\end{equation}

Choosing the canonical representation of (bosonic) KM currents in 2D
Minkowski space-time $x^{\mu }=(\tau ,\sigma )$ $\left( \mu =0,1\right) $ is
the form (\ref{current-}) and (\ref{current+}) \cite%
{Sazdovic'95,Sazdovic'00,Sazdovic'02}, one has
\begin{equation}
j_{\pm i}\equiv p_{i}+k\,\omega _{\pm ij}\,\partial _{\sigma }q^{j}\,,
\label{current}
\end{equation}%
where the momentum-independent part is
\begin{equation}
\omega _{\pm ij}\equiv \rho _{ij}\pm \frac{1}{2}\,\gamma _{ij}\,.  \label{p}
\end{equation}%
The basic PB are $\left\{ q^{i}\left( x\right) ,p_{j}\left( x^{\prime
}\right) \right\} =\delta _{j}^{i}\,\delta \left( \sigma -\sigma ^{\prime
}\right) ,$ so that PB of currents (\ref{current}) close two independent KM
algebras of the group $G$, with the central charges $\pm k$,
\begin{equation}
\{j_{\pm a}\left( x\right) ,j_{\pm b}\left( x^{\prime }\right)
\}=f_{ab}^{\;\;c}\,j_{\pm c}\left( x\right) \,\delta \left( \sigma -\sigma
^{\prime }\right) \pm k\,g_{ab}\partial _{\sigma }\delta \left( \sigma
-\sigma ^{\prime }\right) \,,  \label{KMalgebra}
\end{equation}%
where $j_{\pm a}=-E_{\pm a}^{i}\,j_{\pm i}$ and $\{j_{+a}\left( x\right)
,j_{-b}\left( x^{\prime }\right) \}=0.$ From now on, the $\delta $-function
will be denoted as $\delta \equiv \delta \left( \sigma -\sigma ^{\prime
}\right) ,$ always when its argument cannot be confused.

In order to supersymmetrize this algebra, the fermionic fields $\hat{\chi}%
_{\pm a}$ are introduced. Because the fermionic part of the Lagrangian
should be linear in the time derivative, there always exist second class
constraints $S_{\pm a}\equiv \pi _{\pm a}-ik\,\hat{\chi}_{\pm a}\approx 0,$
linear in the coordinate $\hat{\chi}_{\pm a}$ and in the corresponding
canonical momenta $\pi _{\pm a}$. The Dirac brackets for the fermionic
fields are $\{\hat{\chi}_{\pm a},\hat{\chi}_{\pm b}\}^{\ast }=-\frac{i}{2k}%
\,g_{ab}\,\delta $, while for the bosonic currents $j_{\pm a},$ they remain
the same as the PB. So, one can start from the relation (\ref{KMalgebra})
and
\begin{equation}
\{\hat{\chi}_{\pm a},\hat{\chi}_{\pm b}\}=-\frac{i}{2k}\,g_{ab}\,\delta \,,
\label{db}
\end{equation}%
where, from now, the star can be omitted for the sake of simplicity. Note
that in both bosonic and fermionic cases all quantities of opposite
chiralities commute.

It is easy to check that bilinears in the fermionic fields
\begin{equation}
\tilde{\jmath}_{\pm a}\equiv -ik\,f_{abc}\,\hat{\chi}_{\pm }^{b}\hat{\chi}%
_{\pm }^{c}
\end{equation}%
satisfy the KM algebra without central charges and have nontrivial brackets
with $\hat{\chi}_{\pm b}$,
\begin{eqnarray}
\{\tilde{\jmath}_{\pm a},\tilde{\jmath}_{\pm b}\} &=&f_{ab}^{\;\;c}\,\tilde{%
\jmath}_{\pm c}\delta \,,  \notag \\
\{\tilde{\jmath}_{\pm a},\hat{\chi}_{\pm b}\} &=&f_{ab}^{\;\;c}\,\hat{\chi}%
_{\pm c}\delta \,.
\end{eqnarray}%
Therefore, using $\tilde{\jmath}_{\pm a}$, the new currents can be
introduced,
\begin{equation}
J_{\pm a}\equiv j_{\pm a}+\tilde{\jmath}_{\pm a}\,,
\end{equation}%
such that KM algebras remain unchanged, and which, with its supersymmetric
partners $\hat{\chi}_{\pm a},$\ satisfy two independent super KM algebras:
\begin{eqnarray}
\{J_{\pm a},J_{\pm b}\} &=&f_{ab}^{\;\;c}\,J_{\pm c}\,\delta \pm
k\,g_{ab}\,\partial _{\sigma }\delta \,,  \notag \\
\{J_{\pm a},\hat{\chi}_{\pm b}\} &=&f_{ab}^{\;\;c}\,\hat{\chi}_{\pm
c}\,\delta \,,  \label{SKM} \\
\{\hat{\chi}_{\pm a},\hat{\chi}_{\pm b}\} &=&-\frac{i}{2k}\,g_{ab}\,\delta
\,.  \notag
\end{eqnarray}%
A super KM current is defined as

\begin{equation}
I_{\pm a}(z)\equiv \sqrt{2}k\hat{\chi}_{\pm a}(x)+\theta _{\mp }J_{\pm
a}(x)\,,  \label{Scurrent}
\end{equation}
where $\theta _\alpha $ $\left( \alpha =+,-\right) $ is a Majorana spinor,
and four real local coordinates $z^M=\left( x^\mu ,\theta _\alpha \right) $
parametrize $\left( 1,1\right) $ superspace. Then the algebra (\ref{SKM})
can be rewritten in the form
\begin{equation}
\{I_{\pm a}(z_1),I_{\pm b}(z_2)\}=\delta _{\pm 12}f_{ab}^{\;\;c}\,I_{\pm
c}(z_1)-ik\,g_{ab}\,D^{\pm }\delta _{\pm 12}\,,  \label{sKMalgebra2}
\end{equation}
where $D^{\pm }\equiv \frac \partial {\partial \theta _{\mp }}\,\mp i\theta
_{\mp }\frac \partial {\partial \sigma }$ is the super covariant derivative,
while $\delta _{\pm 12}=(\theta _{\pm 1}-\theta _{\pm 2})\delta (\sigma
_1-\sigma _2)$ is a generalization of the Dirac $\delta $-function to the
super $\delta $-function. Derivatives are always taken over the first
argument of $\delta $-functions. Superspace notation is given in Appendix %
\ref{super}.

\subparagraph{Super Virasoro algebra.}

The next step is to construct a super energy-momentum tensor as a function
of the super KM currents, which is a group invariant. Up to the third power
of $I_{\pm a}$, there are only two group invariants
\begin{equation}
\mathcal{I}_{\pm 1}\equiv g_{ab}\,D^{\pm }I_{\pm }^{a}I_{\pm }^{b}\,,\qquad
\qquad \mathcal{I}_{\pm 2}\equiv if_{abc}\,I_{\pm }^{a}I_{\pm }^{b}I_{\pm
}^{c}\,,
\end{equation}%
because $g_{ab}\,I_{\pm }^{a}I_{\pm }^{b}$ is identically equal to zero
(since super currents are odd variables). Therefore, the super
energy-momentum tensor is looked for in the form
\begin{equation}
T_{\pm }\equiv \alpha _{\pm }\mathcal{I}_{\pm 1}+\beta _{\pm }\mathcal{I}%
_{\pm 2}\,,  \label{T}
\end{equation}%
where the requirement for the superalgebra closure determines the ratio of
coefficients $\alpha _{\pm }$ and $\beta _{\pm }$. In components notation,
one has
\begin{equation}
T_{\pm }=\mp G_{\pm }+\theta _{\mp }L_{\pm }\,,  \label{Ctei}
\end{equation}%
where $L_{\pm }$ is the bosonic part which closes in a (bosonic) Virasoro
algebra, while $G_{\pm }$ is its supersymmetric partner. The explicit
expressions can be found from the expansions of the invariants:
\begin{eqnarray}
\mathcal{I}_{\pm 1} &=&\sqrt{2}kJ_{\pm }^{a}\hat{\chi}_{a}+\theta _{\mp
}\left( J_{\pm }^{a}J_{\pm a}\mp 2ik^{2}\partial _{\sigma }\hat{\chi}_{\pm
}^{a}\hat{\chi}_{\pm a}\right) \,,  \notag \\
\mathcal{I}_{\pm 2} &=&2ik^{2}\left( \sqrt{2}k\,f_{abc}\,\hat{\chi}_{\pm
}^{a}\hat{\chi}_{\pm }^{b}\hat{\chi}_{\pm }^{c}+3\theta _{\mp }f_{abc}\,\hat{%
\chi}_{\pm }^{a}\hat{\chi}_{\pm }^{b}J_{\pm }^{c}\right) \,.
\end{eqnarray}%
The simplest way to find the coefficients in (\ref{T}) is to use the fact
that for every super KM algebra there is a super Virasoro algebra such that
they form a semi-direct product, Eq. (\ref{semi}) or, in another words, that
PB\ of a Virasoro generator and a current, gives a current. Therefore, it is
easy to see that
\begin{equation}
\left\{ G_{\pm },\hat{\chi}_{\pm a}\right\} =\pm \frac{i\alpha _{\pm }}{%
\sqrt{2}}J_{\pm a}\delta \pm \sqrt{2}k\,\left( \alpha _{\pm }-3i\beta _{\pm
}k\right) \,,  \label{close}
\end{equation}%
and the \emph{r.h.s.} of (\ref{close}) closes on currents only if the last
term vanishes. Therefore
\begin{equation}
\frac{\beta _{\pm }}{\alpha _{\pm }}=-3ik\,,
\end{equation}%
and the super energy-momentum tensor is normalized as
\begin{equation}
T_{\pm }\equiv \mp \frac{1}{2k}\,\left( g_{ab}\,D^{\pm }I_{\pm }^{a}I_{\pm
}^{b}+\frac{i}{3k}\,f_{abc}\,I_{\pm }^{a}I_{\pm }^{b}I_{\pm }^{c}\right) \,,
\label{Stei}
\end{equation}%
where the bosonic part is
\begin{equation}
L_{\pm }=\mp \frac{1}{2k}\left( J_{\pm }^{a}J_{\pm a}\pm 2ik^{2}\,\hat{\chi}%
_{\pm }^{a}\,\partial _{\sigma }\hat{\chi}_{\pm a}+2ik\,f_{abc}\,\hat{\chi}%
_{\pm }^{a}\hat{\chi}_{\pm }^{b}J_{\pm }^{c}\right) \,,  \label{components1}
\end{equation}%
while its supersymmetric partner is
\begin{equation}
G_{\pm }=\frac{1}{\sqrt{2}}J_{\pm }^{a}\hat{\chi}_{\pm a}+\frac{i\sqrt{2}k}{3%
}f_{abc}\,\hat{\chi}_{\pm }^{a}\hat{\chi}_{\pm }^{b}\hat{\chi}_{\pm }^{c}\,.
\label{components2}
\end{equation}%
It is useful to express (\ref{components1}) and (\ref{components2}) in terms
of commuting quantities $j_{\pm a}$ and $\hat{\chi}_{\pm a}$:
\begin{eqnarray}
L_{\pm } &=&\mp \frac{1}{2k}\,j_{\pm }^{a}j_{\pm a}-ik\,\hat{\chi}_{\pm
}^{a}\partial _{\sigma }\hat{\chi}_{\pm a}\,,  \notag  \label{components3} \\
G_{\pm } &=&\frac{1}{\sqrt{2}}\,j_{\pm }^{a}\hat{\chi}_{\pm a}-\frac{ik}{3%
\sqrt{2}}\,f_{abc}\,\hat{\chi}_{\pm }^{a}\hat{\chi}_{\pm }^{b}\hat{\chi}%
_{\pm }^{c}\,.
\end{eqnarray}

Using the super KM algebra (\ref{SKM}), the following brackets between the
components of energy-momentum tensor and currents are obtained,
\begin{equation}
\begin{array}{ll}
\{L_{\pm },J_{\pm a}\}=-J_{\pm a}\,\partial _{\sigma }\delta \,, & \{G_{\pm
},J_{\pm a}\}=\pm \frac{k}{\sqrt{2}}\,\hat{\chi}_{\pm a}\,\partial _{\sigma
}\delta \,,\medskip \\
\{L_{\pm },\hat{\chi}_{\pm a}\}=\frac{1}{2}\,(\partial _{\sigma }\hat{\chi}%
_{\pm a}\,\delta -\hat{\chi}_{\pm a}\partial _{\sigma }\delta )\,,\qquad &
\{G_{\pm },\hat{\chi}_{\pm a}\}=-\frac{i}{2\sqrt{2}k}\,J_{\pm a}\,\delta \,,%
\end{array}
\label{T,J}
\end{equation}%
as well as the brackets between the components of energy-momentum tensor
themselves
\begin{eqnarray}
\{L_{\pm },L_{\pm }\} &=&-(\partial _{\sigma }L_{\pm }\delta +2L_{\pm
}\partial _{\sigma }\delta )=-\left[ L_{\pm }(x)+L_{\pm }(x^{\prime })\right]
\partial _{\sigma }\delta \,,  \notag  \label{virasoro1} \\
\{G_{\pm },G_{\pm }\} &=&\pm \frac{i}{2}\,L_{\pm }\delta \,,  \notag \\
\{L_{\pm },G_{\pm }\} &=&-\frac{1}{2}\,\left( \partial _{\sigma }G_{\pm
}\delta +3G_{\pm }\partial _{\sigma }\delta \right) \,,  \notag \\
\{G_{\pm },L_{\pm }\} &=&-\frac{1}{2}\,\left( 2\partial _{\sigma }G_{\pm
}\delta +3G_{\pm }\partial _{\sigma }\delta \right) \,.
\end{eqnarray}%
In terms of the super fields, instead of (\ref{T,J}) one has
\begin{equation}
\{T_{\pm }(z_{1}),I_{\pm a}(z_{2})\}=\pm \frac{i}{2}\,\left( D^{\pm }I_{\pm
a}\,D^{\pm }\delta _{\pm 12}+I_{\pm a}\,{D^{\pm }}^{2}\delta _{\pm
12}\right) \,,  \label{TI}
\end{equation}%
and instead of (\ref{virasoro1})
\begin{equation}
\{T_{\pm }(z_{1}),T_{\pm }(z_{2})\}=\pm \frac{i}{2}\,\left( 2{D^{\pm }}%
^{2}T_{\pm }\,\delta _{\pm 12}+D^{\pm }T_{\pm }\,D^{\pm }\delta _{\pm
12}+3T_{\pm }\,{D^{\pm }}^{2}\delta _{\pm 12}\right) \,.  \label{virasoro2}
\end{equation}%
This is a supersymmetric extension of the Virasoro algebra \emph{without
central charge}.

Since $L_{\pm }$ and $G_{\pm }$ are the first class constrains, one can
apply the general canonical method to construct a theory invariant under
\emph{diffeomorphisms} generated by $L_{\pm }$ and under \emph{local
supersymmetry} generated by $G_{\pm }$. It is known that, in the bosonic
case, the similar approach leads to a covariant extension of the WZW theory
with respect to diffeomorphisms \cite%
{Blagojevic-Popovic-Sazdovic'98,Blagojevic-Popovic-Sazdovic'99}, thus here
one expects to obtain covariant extension of WZW theory with respect to
local supersymmetry.

\section{Effective Lagrangian and gauge transformations}

In order to construct a covariant theory, one uses the generators
\begin{equation}
\phi _{r}=(L_{-},L_{+},G_{-},G_{+})\,,  \label{case}
\end{equation}%
as first class constraints, with the explicit expressions given in equations
(\ref{components3}), and the PB algebra (\ref{virasoro1}) instead of the
equation (\ref{general1}).

\subparagraph{The action.}

According to (\ref{action}), the canonical Lagrangian is introduced as
\begin{equation}
\hat{L}=\dot{q}^{i}p_{i}+ik\,\overset{.}{\hat{\chi}_{+}^{a}}\hat{\chi}%
_{+a}+ik\,\overset{.}{\hat{\chi}_{-}^{a}}\hat{\chi}%
_{-a}-h^{-}L_{-}-h^{+}L_{+}-i\psi ^{-}G_{-}-i\psi ^{+}G_{+}\,,  \label{Clag}
\end{equation}%
with multipliers $u^{r}=(h^{-},h^{+},\psi ^{-},\psi ^{+})$. Note that, on
Dirac brackets, the second class constrains are zero $(S_{\pm a}=0)$, giving
the Grassmann odd momenta as $\pi _{\pm a}=ik\,\hat{\chi}_{\pm a}$. The
remaining momentum variables can be eliminated by means of their equations
of motion,%
\begin{equation}
p^{i}=\frac{k}{h^{-}-h^{+}}\,\left[ \dot{q}^{i}+\left( h^{+}{{\omega _{+}}%
^{i}}_{j}-h^{-}\omega {{_{-}}^{i}}_{j}\right) \partial _{\sigma }{q}^{j}+%
\frac{i}{\sqrt{2}}\,\left( \psi ^{-}\hat{\chi}_{-}^{i}+\psi ^{+}\hat{\chi}%
_{+}^{i}\right) \right] \,,  \label{momentum}
\end{equation}%
where $\hat{\chi}_{\pm }^{i}=E_{\pm a}^{i}\hat{\chi}_{\pm }^{a}$. On the
equations of motion (\ref{momentum}), the currents (\ref{current}) become
\begin{equation}
j_{\pm }^{i}=\frac{k}{2}\,\left[ \hat{\partial}_{\pm }q^{i}+\frac{i}{\sqrt{-2%
\hat{g}}}\,\left( \psi ^{-}\hat{\chi}_{-}^{i}+\psi ^{+}\hat{\chi}%
_{+}^{i}\right) \right] \,,  \label{j.motion}
\end{equation}%
so that the Lagrangian (\ref{Clag}) can be written as a sum of three terms
\begin{equation}
\hat{L}=\hat{L}_{\text{WZW}}+\hat{L}_{\text{f}}+\hat{L}_{\text{int}}\,,
\label{lagrangian}
\end{equation}%
which have the form:
\begin{equation}
\begin{tabular}{lll}
$\hat{L}_{\text{WZW}}$ & $=$ & $-\frac{k}{2}\sqrt{-\hat{g}}\,\omega _{-ij}\,%
\hat{\partial}_{-}q^{i}\hat{\partial}_{+}q^{j}\,,\medskip $ \\
$\hat{L}_{\text{f}}$ & $=$ & $-ik\sqrt{-\hat{g}}\,\left( \hat{\chi}_{+}^{a}%
\hat{D}_{-}\hat{\chi}_{+a}+\hat{\chi}_{-}^{a}\hat{D}_{+}\hat{\chi}%
_{-a}\right) \,,\medskip $ \\
$\hat{L}_{\text{int}}$ & $=$ & $\frac{ik}{2\sqrt{2}}\,\left( \psi ^{+}\hat{%
\chi}_{+i}\hat{\partial}_{+}q^{i}+\psi ^{-}\hat{\chi}_{-i}\hat{\partial}%
_{-}q^{i}\right) \,.$%
\end{tabular}%
\end{equation}%
Here, $\hat{L}_{\text{WZW}}$ and $\hat{L}_{\text{f}}$ are the WZW and
fermion Lagrangians respectively, covariantized in super gravitational
fields $\hat{g}_{\mu \nu }$ and $\psi ^{\pm }$, while $\hat{L}_{\text{int}}$
describes the interaction between bosonic fields $q^{i}$ and fermionic
fields $\hat{\chi}_{\pm i}$. The tensor $\hat{g}_{\mu \nu }$ is introduced
instead of variables $(h^{+},h^{-})$ (see Appendix \ref{cone})
\begin{equation}
\hat{g}_{\mu \nu }\equiv -\frac{1}{2}\,\left(
\begin{array}{cc}
-2h^{+}h^{-} & h^{+}+h^{-} \\
h^{+}+h^{-} & -2%
\end{array}%
\right) \,.
\end{equation}%
Covariant derivatives, acting on fermionic fields $\hat{\chi}_{\pm }^{a}$,
are defined by
\begin{equation}
\hat{D}_{\mp }\hat{\chi}_{\pm }^{a}\equiv \hat{\partial}_{\mp }\hat{\chi}%
_{\pm }^{a}+\frac{i}{3\sqrt{-2\hat{g}}}\,f_{bc}^{\;\;a}\psi ^{\pm }\hat{\chi}%
_{\pm }^{b}\hat{\chi}_{\pm }^{c}\pm \frac{i}{8\hat{g}}\,\psi ^{-}\psi ^{+}\,%
\hat{\chi}_{\mp }^{a}\,,  \label{act}
\end{equation}%
where $\hat{\partial}_{\pm }={{\hat{e}}^{\mu }}_{\;\;\pm }\partial _{\mu }$,
and ${{\hat{e}}^{\mu }}_{\;\;\pm }$ are also given in Appendix.

\subparagraph{Local symmetries.}

The general canonical method provides a mechanism to write out gauge
symmetries of the Lagrangian (\ref{lagrangian}). Instead of relations (\ref%
{general2}), with the help of (\ref{components3}), the following gauge
transformations of the fields are found,
\begin{eqnarray}
\delta q^{i} &=&\frac{1}{k}\,(\varepsilon ^{-}j_{-}^{i}-\varepsilon
^{+}j_{+}^{i})-\frac{i}{\sqrt{2}}\,(\eta ^{+}\hat{\chi}_{+}^{i}+\eta ^{-}%
\hat{\chi}_{-}^{i})\,,  \notag  \label{fields} \\
\delta \hat{\chi}_{\pm }^{a} &=&-\varepsilon ^{\pm }\partial _{1}\hat{\chi}%
_{\pm }^{a}-\frac{1}{2}\,\left( \partial _{1}\varepsilon ^{\pm }\right) \hat{%
\chi}_{\pm }^{a}-\frac{1}{2k\sqrt{2}}\,\eta ^{\pm }\,J_{\pm }^{a}\,,
\end{eqnarray}%
and instead of (\ref{general3}) using (\ref{virasoro1}), one obtains the
gauge transformations of the multipliers,
\begin{eqnarray}
\delta h^{\pm } &=&\partial _{0}\varepsilon ^{\pm }+h^{\pm }\partial
_{1}\varepsilon ^{\pm }-\varepsilon ^{\pm }\partial _{1}h^{\pm }\pm \frac{i}{%
2}\,\psi ^{\pm }\eta ^{\pm }\,,  \notag  \label{multipliers} \\
\delta \psi ^{\pm } &=&\frac{1}{2}\,\psi ^{\pm }\partial _{1}\varepsilon
^{\pm }-\left( \partial _{1}\psi ^{\pm }\right) \varepsilon ^{\pm }+\partial
_{0}\eta ^{\pm }+h^{\pm }\partial _{1}\eta ^{\pm }-\frac{1}{2}\,\left(
\partial _{1}h^{\pm }\right) \eta ^{\pm }\,.
\end{eqnarray}%
The bosonic fields $\varepsilon ^{\pm }$ and the fermionic fields $\eta
^{\pm }$ are the parameters of diffeomorphisms and local supersymmetry
transformations respectively.

\section{Lagrangian formulation}

It turns out that the Lagrangian (\ref{lagrangian}) is invariant under the
following rescaling of fields by two arbitrary parameters $F(x)$ and $f(x)$:
\begin{equation}
\begin{array}{lll}
{{\hat{e}}^{\pm }}_{\;\;\mu } & \rightarrow & {e^{\pm }}_{\mu }\equiv
e^{F\pm f}\,{{\hat{e}}^{\pm }}_{\;\;\mu }\,,\medskip \\
\psi ^{\pm } & \rightarrow & \psi _{\mp (\mp )}\equiv \frac{1}{2\sqrt{-\hat{g%
}}}\,e^{-\frac{1}{2}\,(F\mp 3f)}\,\psi ^{\pm }\,,\medskip \\
\hat{\chi}_{\pm }^{a} & \rightarrow & \chi _{\pm }^{a}\equiv e^{-\frac{1}{2}%
\,(F\pm f)}\,\hat{\chi}_{\pm }^{a}\,.%
\end{array}
\label{rescaling1}
\end{equation}%
As a consequence, one has:
\begin{equation}
\begin{array}{lll}
\sqrt{-\hat{g}} & \rightarrow & \sqrt{-g}=e^{2F}\,\sqrt{-\hat{g}}\,,\medskip
\\
\hat{\partial}_{\pm } & \rightarrow & \partial _{\pm }\equiv e^{-(F\pm f)}\,%
\hat{\partial}_{\pm }\,.%
\end{array}%
\end{equation}%
In terms of the rescaled fields, the rescaled Lagrangian has the same form
as the original one (\ref{lagrangian}),
\begin{equation}
\begin{array}{lll}
L & = & L_{\text{WZW}}+L_{\text{f}}+L_{\text{int}}\,,\medskip \\
L_{\text{WZW}} & = & -\frac{k}{2}\sqrt{-g}\,\omega _{-ij}\,\partial
_{-}q^{i}\partial _{+}q^{j}\,,\medskip \\
L_{\text{f}} & = & -ik\sqrt{-g}\,\left( \chi _{+}^{a}D_{-}\chi _{+a}+\chi
_{-}^{a}D_{+}\chi _{-a}\right) \,,\medskip \\
L_{\text{int}} & = & \frac{ik}{\sqrt{2}}\,\sqrt{-g}\,\left[ \psi _{-(-)}\chi
_{+i}\partial _{+}q^{i}+\psi _{+(+)}\chi _{-i}\partial _{-}q^{i}\right] \,,%
\end{array}
\label{newL}
\end{equation}%
where the covariant derivative acts on $\chi _{\pm }^{a}$ as
\begin{equation}
D_{\mp }\chi _{\pm }^{a}\equiv \partial _{\mp }\chi _{\pm }^{a}+\frac{i\sqrt{%
2}}{3}\,f_{ab}^{\;\;c}\psi _{\mp (\mp )}\chi _{\pm }^{b}\chi _{\pm }^{c}\mp
\frac{i}{2}\,\psi _{+(+)}\psi _{-(-)}\,\chi _{\mp }^{a}\,.
\end{equation}%
Note that the term with derivatives over $F$ and $f$ vanishes because of
nilpotency of the field $\hat{\chi}_{\pm }^{i}$.

The introduction of the new fields $F$ and $f$ gives additional gauge
freedom to the Lagrangian (\ref{newL}). As a consequence of the
transformation $\delta _{\Lambda }F=\Lambda $, the Lagrangian becomes
invariant under the \emph{local Weyl transformations }
\begin{equation}
\begin{array}{lll}
\delta _{\Lambda }{e^{\pm }}_{\mu } & = & \Lambda {e^{\pm }}_{\mu
}\,,\medskip \\
\delta _{\Lambda }\psi _{\pm (\pm )} & = & -\frac{1}{2}\,\Lambda \psi _{\pm
(\pm )}\,,\medskip \\
\delta _{\Lambda }\chi _{\pm }^{a} & = & -\frac{1}{2}\,\Lambda \chi _{\pm
}^{a}\,,%
\end{array}
\label{weyl}
\end{equation}%
while all $F$ independent fields remain Weyl invariant. Furthermore, the
Lagrangian (\ref{newL}) does not change under the \emph{local Lorentz
transformations}
\begin{equation}
\begin{array}{lll}
\delta _{\ell }{e^{\pm }}_{\mu } & = & \mp \ell {e^{\pm }}_{\mu }\,,\medskip
\\
\delta _{\ell }\psi _{\pm (\pm )} & = & \pm \frac{3}{2}\,\ell \,\psi _{\pm
(\pm )}\,,\medskip \\
\delta _{\ell }\chi _{\pm }^{a} & = & \pm \frac{1}{2}\,\ell \,\chi _{\pm
}^{a}\,,%
\end{array}
\label{lorentz}
\end{equation}%
which are generated by the transformation $\delta _{\ell }f=-\ell $. The
vielbein ${e^{\pm }}_{\mu }$ transforms as a Lorentz vector, the fields $%
\psi _{\pm (\pm )}$ and $\chi _{\pm }^{a}$ transform like components of a
spinor field with the spin $\frac{3}{2}$ and $\frac{1}{2}$ respectively,
while all $f$ independent fields are Lorentz scalars.

The Lagrangian (\ref{newL}) depends only on two components $\psi _{\pm (\pm
)}$ of the Rarita-Schwinger spinor field $\psi _{\mu \alpha }={e^{a}}_{\mu
}\psi _{a(\alpha )}$. It means that there is also the additional\emph{\
local super Weyl symmetry}
\begin{equation}
\delta _{\xi }\psi _{\pm (\mp )}=\pm \xi _{\pm }\,,  \label{Sweyl}
\end{equation}%
where $\xi _{\alpha }$ is an odd parameter. All other fields are super Weyl
invariant. Transformations (\ref{weyl}) -- (\ref{Sweyl}) can be written in a
covariant form,
\begin{equation}
\begin{array}{llll}
\text{Lorentz:} & \delta _{\ell }{e^{a}}_{\mu }=-\ell \,{\varepsilon ^{a}}%
_{b}\,{e^{b}}_{\mu }\,, & \delta _{\ell }\psi _{\mu }=\frac{1}{2}\,\ell
\,\gamma _{5}\,\psi _{\mu }\,, & \delta _{\ell }\chi ^{a}=\frac{1}{2}\,\ell
\gamma _{5}\,\chi ^{a}\,,\medskip \\
\text{Weyl:} & \delta _{\Lambda }{e^{a}}_{\mu }=\Lambda {e^{a}}_{\mu }\,, &
\delta _{\Lambda }\psi _{\mu }=\frac{1}{2}\Lambda \psi _{\mu }\,, & \delta
_{\Lambda }\chi ^{a}=\frac{1}{2}\,\Lambda \chi ^{a}\,,\medskip \\
\text{super Weyl:\quad } & \delta _{\xi }{e^{a}}_{\mu }=0\,, & \delta _{\xi
}\psi _{\mu }=\gamma _{\mu }\xi \,, & \delta _{\xi }\chi ^{a}=0\,,%
\end{array}
\label{all}
\end{equation}%
where $\gamma _{\mu }\equiv {e^{a}}_{\mu }\gamma _{a}$. The representation
of the $\gamma $-matrices is given in Appendix \ref{cone}. Note that the
fields ${{\hat{e}}^{\pm }}_{\;\;\mu }$, $\psi ^{\pm }$ and $\hat{\chi}_{\pm
}^{a}$, introduced in the Hamiltonian approach are Lorentz, Weyl and super
Weyl invariants.

The fields $F$ and $f$ do not enter the original Lagrangian (\ref{lagrangian}%
) but are introduced by the rescaling of fields (\ref{rescaling1}). Thus
their change under diffeomorphisms and supersymmetry transformations cannot
be found just applying the general Hamiltonian rules (\ref{general2}), (\ref%
{general3}). They are introduced in such a way that the new fields ${e^{a}}%
_{\mu }$ and $\psi _{\mu \alpha }$ have proper Lorentz, Weyl and super Weyl
transformations. Now, one demands that ${e^{a}}_{\mu }$ transforms like a
vector and $\psi _{\mu \alpha }$ transforms like a Rarita-Schwinger field
under general coordinate transformations with a local parameter $\varepsilon
^{\mu }(x)$ and under $\mathcal{N}=1$ supersymmetric transformations with a
local spinor parameter $\zeta _{\alpha }(x)$. It means that (see for example
\cite{Hoker-Phong})
\begin{eqnarray}
\delta {e^{a}}_{\mu } &=&-\varepsilon ^{\nu }\partial _{\nu }{e^{a}}_{\mu }-{%
e^{a}}_{\nu }\partial _{\mu }\varepsilon ^{\nu }-\frac{i}{2}\,\bar{\zeta}%
\gamma ^{a}\psi _{\mu }\,,  \notag  \label{diffSUSY1} \\
\delta \psi _{\mu } &=&-\varepsilon ^{\nu }\partial _{\nu }\psi _{\mu }-\psi
_{\nu }\,\partial _{\mu }\varepsilon ^{\nu }+\frac{1}{2}\,\triangledown
_{\mu }\zeta \,,
\end{eqnarray}%
where $\triangledown _{\mu }\zeta ={e^{a}}_{\mu }\triangledown _{a}\zeta $
and $\triangledown _{a}\zeta =\left( \partial _{a}+\frac{1}{2}\gamma
_{5}\omega _{a}\right) \,\zeta $. Writing out in components, it gives
\begin{equation}
\begin{array}{lll}
\delta (F\pm f) & = & h^{\pm }\partial _{1}\varepsilon ^{0}-\partial
_{1}\varepsilon ^{1}-\varepsilon ^{\mu }\partial _{\mu }(F\pm f)\pm
ie^{-(F\pm f)}\zeta _{\mp }\psi _{1\mp }\,,\medskip \\
\quad \delta h^{\pm } & = & \partial _{0}\varepsilon ^{1}-h^{\pm }(\partial
_{0}\varepsilon ^{0}-\partial _{1}\varepsilon ^{1})-(h^{\pm })^{2}\partial
_{1}\varepsilon ^{0}-\varepsilon ^{\mu }\partial _{\mu }h^{\pm }\mp \frac{i}{%
2}\,e^{-\frac{1}{2}(F\pm f)}\zeta _{\mp }\psi ^{\pm }\,,\medskip \\
\quad \delta \psi ^{\pm } & = & \left[ \frac{1}{2}\,\partial
_{1}(\varepsilon ^{1}-h^{\pm }\varepsilon ^{0})+\frac{1}{2}\,(\partial
_{1}h^{\pm })\varepsilon ^{0}-\partial _{0}\varepsilon ^{0}-h^{\pm }\partial
_{1}\varepsilon ^{0}\right] \psi ^{\pm }-\varepsilon ^{\mu }\partial _{\mu
}\psi ^{\pm }\,,\medskip \\
&  & +\left( \partial _{0}+h^{\pm }\partial _{1}-\frac{1}{2}\,\partial
_{1}h^{\pm }\right) \left( \zeta _{\mp }e^{-\frac{1}{2}\,(F\pm f)}\right) +%
\frac{i}{4}\,\zeta _{\mp }\psi _{\pm (\mp )}\psi ^{\pm }\,,%
\end{array}
\label{diffSUSY2}
\end{equation}%
where $\psi ^{\mp }\equiv 2e^{-\frac{1}{2}\,(F\mp f)}\left( \psi _{0\pm
}+h^{\mp }\psi _{1\pm }\right) $ in according with (\ref{rescaling1}).

In order to establish the relation between the Hamiltonian and Lagrangian
transformations one should compare the Hamiltonian transformations (\ref%
{multipliers}) with the Lagrangian one, the last two equations of (\ref%
{diffSUSY2}). One finds that these gauge transformations can be identified
by choosing the following relation between the gauge parameters
\begin{equation*}
\varepsilon ^{\pm }\equiv \varepsilon ^{1}-h^{\pm }\varepsilon ^{0}\,,\qquad
\eta ^{\pm }\equiv 2\zeta _{\mp }e^{-\frac{1}{2}\,(F\pm f)}-\varepsilon
^{0}\psi ^{\pm }\,,
\end{equation*}%
and imposing the gauge fixing $\psi _{\pm (\mp )}=0$, because in the
Hamiltonian approach all quantities are super Weyl invariant. Substituting
the equations of motion (\ref{j.motion}) in (\ref{fields}), the momentum
independent formulation of transformation law of matter variables $q^{i}$, $%
\hat{\chi}_{\pm }^{a}$ is obtained. In terms of the Lagrangian variables $%
\varepsilon ^{\mu }$ and $\zeta _{\pm }$, they stand

\begin{eqnarray}
\delta q^{i} &=&-\varepsilon ^{\mu }\,\partial _{\mu }q^{i}-i\sqrt{2}%
\,\left( \zeta _{-}\chi _{+}^{i}+\zeta _{+}\chi _{-}^{i}\right) \,,  \notag
\\
\delta \hat{\chi}_{\pm }^{a} &=&-\varepsilon ^{\mu }\,\partial _{\mu }\hat{%
\chi}_{\pm }^{a}+\varepsilon ^{0}\sqrt{-\hat{g}}\hat{\triangledown}_{\mp }%
\hat{\chi}_{\pm }^{a}+\frac{1}{2}\,\left( h^{\pm }\partial _{1}\varepsilon
^{0}-\partial _{1}\varepsilon ^{1}\right) \hat{\chi}_{\pm }^{a}+
\label{density} \\
&+&\frac{1}{2k\sqrt{2}}\,\left( \varepsilon ^{0}\psi ^{\pm }-2\zeta _{\mp
}e^{-\frac{1}{2}(F\pm f)}\right) J_{\pm }^{a}\,.  \notag
\end{eqnarray}

In the flat space limit ${e^{a}}_{\mu }\rightarrow \delta _{\mu }^{a}$ and $%
\psi _{\mu }\rightarrow 0$, the Lagrangian (\ref{newL}) becomes
\begin{equation}
L_{0}=-\frac{k}{2}\,\omega _{-ij}\,\partial _{-}q^{i}\partial
_{+}q^{j}-ik\,\left( \chi _{+}^{a}\partial _{-}\chi _{+a}+\chi
_{-}^{a}\partial _{+}\chi _{-a}\right) \,,  \label{Flagrangian}
\end{equation}%
and bosonic and fermionic parts are decoupled. The first term is the bosonic
WZW action, while the second one is the Lagrangian of free spinor fields $%
\chi _{\pm }^{a}$. The Lagrangian (\ref{Flagrangian}) describes the $%
\mathcal{N}=1$ supersymmetric WZW theory discussed in the previous chapter
\cite{Vecchia-Knizhnik-Petersen-Rossi,Abdalla-Abdalla}.

\section{Conclusions}

Using the general canonical method, the Lagrangian for the super WZW model
coupled to two-dimensional supergravity is constructed. The basic
ingredients of this approach are the symmetry generators, which are
functions of the coordinates and momenta and satisfy the super KM and super
Virasoro algebras. The application of the Hamiltonian method naturally
incorporates gauge fields (the metric tensor and Rarita-Schwinger fields) as
Lagrange multipliers of the symmetry generators. This method also gives a
prescription for finding gauge transformations for both the matter and gauge
fields.

The Hamiltonian formalism deals with the Lagrangian multipliers $h^{\pm }$
and $\psi ^{\pm }$, which are just the part of the gauge fields necessary to
represent the symmetry of the algebra. In the Lagrangian formulation, the
covariant description of the fields is needed. In order to complete the
vielbeins ${e^{a}}_{\mu }$, it was necessary to introduce the new bosonic
components $F$ and $f$, while for completing the Rarita-Schwinger fields $%
\psi _{\mu \alpha },$ the new fermionic fields $\psi _{\pm (\mp )}$ are
needed. The new components are not physical because they do not appear in
the Lagrangian, but they give additional gauge freedom corresponding to the
additional gauge symmetries. These are local Weyl and local Lorentz
symmetries for the bosonic fields $F$ and $f$ respectively, and local super
Weyl symmetry for the fermionic field $\psi _{\mp (\pm )}$. The fields $F$
and $f$ are not parts of the Hamiltonian formalism, so their transformation
laws under reparametrizations and supersymmetry is found requiring that
vielbeins ${e^{a}}_{\mu }$ transform as vectors, and $\psi _{\mu \alpha }$
transform as a Rarita-Schwinger field. Consequently, the complete relation
between Hamiltonian and Lagrangian formulations is established. The
connection between the corresponding fields, gauge parameters and gauge
transformations is found. In the flat superspace limit, the result of Ref.
\cite{Vecchia-Knizhnik-Petersen-Rossi} is reproduced.

\chapter{Irregular constrained systems}

In the previous chapters, the Hamiltonian analysis was used to find all
local symmetries of the WZW model and its supersymmetric extension. The
canonical method was applied to construct the super WZW action coupled to
supergravity, starting from the representation of the super Virasoro algebra.

In general, Dirac's Hamiltonian analysis provides a systematic technique for
finding the gauge symmetries and the physical degrees of freedom of
constrained systems like gauge theories and gravity \cite{Dirac}. These
theories contain more dynamical variables than is minimally requested by
physics and, consequently, they are not independent. Therefore, in those
theories constraints arise as functions of local coordinates in phase space,
and they are usually functionally independent, at least in the most common
cases of physical interest. However, there are some exceptional cases in
which functional independence is violated, when it is not always clear how
to identify local symmetries and physical degrees of freedom. In this
chapter, these \emph{irregular} systems are analyzed. For the review of the
Hamiltonian formalism see Appendix \ref{HF}.

\section{Regularity conditions}

Consider a constrained system on phase space $\Gamma $ with local
coordinates $z^{n}=\left( q,p\right) $ $\left( n=1,\ldots ,2N\right) $ and a
complete set of constraints
\begin{equation}
\phi _{r}\left( z\right) \approx 0\,,\qquad \left( r=1,\ldots ,R\right) \,,
\end{equation}%
which define a constraint surface\footnote{%
This terminology is standard in Hamiltonian analysis although, in general,
the set $\Sigma $ is not a manifold since it can contain discontinuities, be
non-differentiable, etc.}
\begin{equation}
\Sigma =\left\{ \bar{z}\in \Gamma \mid \phi _{r}(\bar{z})=0\;\left(
r=1,\ldots ,R\right) \;\left( R\leq 2N\right) \right\} \,.
\end{equation}%
Dirac's procedure guarantees that the system remains on the constraint
surface during its evolution.

Choosing different coordinates on $\Gamma $ may lead to different forms for
the constraints whose functional independence is not obvious. The \emph{%
regularity conditions} (RCs) were introduced by Dirac to test this \cite%
{DiracCJM}:\emph{\medskip }

\emph{The constraints }$\phi _{r}\approx 0$\emph{\ are regular\ if and only
if their small variations }$\delta \phi _{r}\ $\emph{evaluated on }$\Sigma $%
\emph{\ define }$R$\emph{\ linearly independent functions of }$\delta z^{n}$%
\emph{.\medskip }

To first order in $\delta z^{n}$, the variations of the constraints have the
form
\begin{equation}
\delta \phi _{r}=J_{rn}\,\delta z^{n}\,,\qquad (r=1,...,R)\,,
\label{linearized constraints}
\end{equation}%
where $J_{rn}\equiv \left. \partial \phi _{r}/\partial z^{n}\right\vert
_{\Sigma }$ is the Jacobian evaluated on the constraint surface. An
equivalent definition of the RCs is \cite{Henneaux-Teitelboim}:\medskip

\emph{The set of constraints} $\phi _{r}\approx 0$ \emph{is regular if and
only if the Jacobian} $J_{rn}=\left. \partial \phi _{r}/\partial
z^{n}\right\vert _{\Sigma }$ \emph{has\ maximal rank,} $\Re (\mathbf{J}%
)=R\,. $\medskip

A system of just one constraint can also fail the test of regularity, as for
the constraint $\phi =q^{2}\approx 0$ in a $2$-dimensional phase space $%
\left( q,p\right) $. In this case, the Jacobian $\mathbf{J}=\left(
2q,0\right) _{q^{2}=0}=0$ has zero rank. The same problem occurs with the
constraint $q^{k}\approx 0$, for $k>1$, which has a zero of $k$-th order on
the constraint surface. Thus one \emph{constraint} may be dependent on
itself, while one \emph{function} is, by definition, always functionally
independent.

\subparagraph{Equivalent constraints.}

Different sets of constraints are said to be \emph{equivalent} if they
define the same constraint surface. This definition refers to the locus of
constraints, not to the equivalence of the resulting dynamics. Since the
surface $\Sigma $ is defined by the zeros of the constraints, while the
regularity conditions depend on their derivatives, it is possible to replace
a set of irregular $\phi $'s by an \emph{equivalent} set of \emph{regular}
constraints $\tilde{\phi}$.

In the following sections, the \emph{nature} of the constraints that give
rise to irregularity is analyzed, and \emph{where} the irregularities can
occur.

\section{Basic types of irregular constraints}

Irregular constraints can be classified according to their behavior in the
vicinity of the surface $\Sigma $. For example, linearly dependent
constraints have Jacobian with constant rank $R^{\prime }$ throughout $%
\Sigma $, and
\begin{equation}
\phi _{r}\equiv J_{rn}(\bar{z})\left( z^{n}-\bar{z}^{n}\right) \approx
0\,,\qquad \Re (\mathbf{J})=R^{\prime }<R\,.  \label{linear}
\end{equation}
These constraints are regular systems in disguise simply because $%
R-R^{\prime }$ constraints are redundant and should be discarded. The subset
with $R^{\prime }$ linearly independent constraints gives the correct
description. For example, linearly dependent constraints are clearly in this
category.\ Apart from this trivial case, two main types of truly irregular
constraints, which do not possess a linear approximation in the vicinity of
some points of $\Sigma $\textbf{,} can be distinguished:

\subparagraph{Type I. Multilinear constraints.}

Consider the constraint
\begin{equation}
\phi \equiv \prod\limits_{i=1}^{M}f_{i}(z)\approx 0\,,  \label{multilinear}
\end{equation}%
where the functions $f_{i}$ have simple zeros. Each factor defines a surface
of codimension 1,
\begin{equation}
\Sigma _{i}\equiv \left\{ \bar{z}\in \Gamma \mid f_{i}(\bar{z})=0\right\} \,,
\label{codimension 1}
\end{equation}%
and $\Sigma $ is the collection of all surfaces, $\Sigma =\bigcup \Sigma
_{i}.$ The rank of the Jacobian of $\phi $ is reduced at intersections
\begin{equation}
\Sigma _{ij}\equiv \Sigma _{i}\bigcap \Sigma _{j}\,.  \label{intersections}
\end{equation}%
Thus, the RCs hold everywhere on $\Sigma $, except at the intersections $%
\Sigma _{ij}$, where $\phi $ has zeros of higher order. Note that the
intersections (\ref{intersections}) also include the points where more than
two $\Sigma $'s overlap.

\subparagraph{Type II. Nonlinear constraints.}

Consider the constraint of the form
\begin{equation}
\phi \equiv \left[ f(z)\right] ^{k}\approx 0\,,\qquad \left( k>1\right) \,,
\label{nonlinear}
\end{equation}%
where the function $f(z)$\ has a simple zero. This constraint has a zero of
order $k$ in the vicinity of $\Sigma $, its Jacobian vanishes on the
constraint surface and, therefore, the RCs fail\footnote{%
Here it is assumed $k>1,$ in order to avoid infinite values for $\frac{%
\partial \phi }{\partial z^{i}}$ on $\Sigma $.}. It could seem harmless to
replace $\phi $ by the equivalent regular constraint $f(z)\approx 0$, but it
is allowed to do only if it does not change the dynamics of original system,
what is discussed below.

Types I and II are the two fundamental generic classes of irregular
constraints. In general, there can be combinations of them occurring
simultaneously near a constraint surface, as in constraints of the form $%
\phi =\left[ f_{1}(z)\right] ^{2}f_{2}(z)\approx 0$, etc.

\section{Classification of constraint surfaces}

The previous classification refers to the way in which $\phi $ approaches
zero. Now it is going to be discussed \emph{where} regularity can be
violated.

The example of $q^{2}\approx 0$ showed that only one constraint can be
irregular, even if, as a function, it is functionally independent. This is
possible since functional independence of the \emph{functions }$\phi ^{r}(z)$
permits the Jacobian $\partial \phi ^{r}/\partial z^{i}$ to have rank lower
than maximal on a submanifold of measure zero,
\begin{equation}
\Xi =\left\{ w\in \Gamma \left\vert \Re \left( \frac{\partial \phi ^{r}}{%
\partial z^{i}}\right) _{z=w}<R\right. \right\} \,,  \label{R(J)}
\end{equation}%
while the \emph{constraints }$\phi ^{r}\approx 0$ have the Jacobian
evaluated at the particular surface $\Sigma $. Their intersection
\begin{equation}
\Sigma _{0}=\Sigma \cap \Xi
\end{equation}%
defines a submanifold on $\Sigma $ where the RCs are violated, while on the
rest of $\Sigma $ the RCs are satisfied.

Thus, barring accidental degeneracies such as linearly dependent
constraints, one of these three situation may present themselves:

\begin{description}
\item \textbf{A.} \emph{The} \emph{RCs are satisfied everywhere on the
constraint surface}: The surfaces $\Sigma $ and $\Xi $ do not intersect and $%
\mathbf{J}$ has maximal rank throughout $\Sigma $. These are regular systems.

\item \textbf{B.} \emph{The RCs fail everywhere on the constraint surface:} $%
\Sigma _{0}=\Sigma $ is a submanifold of $\Xi $ and $\mathbf{J}$ has
constant rank $R^{\prime }<R$ on\emph{\ }$\Sigma _{0}.$ This irregular
system contains nonlinear (type II) constraints.

\item \textbf{C.} \emph{The RCs fail on }$\Sigma _{0}$: The intersection $%
\Sigma _{0}$ is a measure zero submanifold on $\Sigma ,$ so that $\Re $ $%
\left( \mathbf{J}\right) =R^{\prime }<R$\ on $\Sigma _{0}$, while $\Re $ $%
\left( \mathbf{J}\right) =R$ elsewhere on $\Sigma $. This irregular system
contains multilinear (type I) constraints.
\end{description}

In the last case, the constraint surface can be decomposed into two non
overlapping sets $\Sigma _{0}$ and $\Sigma _{R}$. Then, the rank of the
Jacobian jumps from $\Re \left( \mathbf{J}\right) =R$ on $\Sigma _{R}$, to $%
\Re \left( \mathbf{J}\right) =R^{\prime }$ on $\Sigma _{0}$ and the manifold
$\Sigma $ is not differentiable at $\Sigma _{0}$. Although the functions $%
\phi _{r}$ are continuous and differentiable, this is not sufficient for
regularity.

For example, a massless relativistic particle\ in Minkowski space has phase
space $(q^{\mu },p_{\nu })$ with both regular and irregular sectors. The
constraint $\phi \equiv p^{\mu }p_{\mu }\approx 0\;$has Jacobian $\mathbf{J}%
=(0,2p^{\mu })_{\phi =0}$, and its rank is one everywhere, except at the
apex of the cone, $p^{\mu }=0$, where the light-cone is not differentiable
and the Jacobian has rank zero. From the viewpoint of irreducible
representations of the Poincar\'{e} group, the orbits with $p^{\mu }=0$
correspond to the trivial representation of the group, and this point is
excluded from the phase space of the massless particle (see, \emph{e.g.},
\cite{Leiva-Plyushchay}).

The lack of regularity, however, is not necessarily due to the absence of a
well defined smooth tangent space for $\Sigma $. Consider for example the
multilinear constraint
\begin{equation}
\phi (x,y,z)=(x-1)(x^{2}+y^{2}-1)\approx 0\,.
\end{equation}%
Here the constraint surface $\Sigma $ is composed of two sub-manifolds: the
plane $\Pi =\{(x,y,z)\mid $ $x-1\approx 0\}$, and the cylinder $%
C=\{(x,y,z)\mid $ $x^{2}+y^{2}-1\approx 0\}$, which are tangent to each
other along the line $L=\left\{ (x,y,z)\mid x=1,y=0,z\in \mathbb{R}\right\} $%
. The Jacobian on $\Sigma $ is
\begin{equation}
\mathbf{J}=\left( 3x^{2}+y^{2}-2x-1,2y\left( x-1\right) ,0\right) _{\phi
=0}\,,
\end{equation}%
and its rank is one everywhere, except on $L$, where it is zero. The
constraint $\phi $ is irregular on this line. However, the tangent vectors
to $\Sigma $ are well defined there. The irregularity arises because $\phi $
is a multilinear constraint of the type described by (\ref{multilinear}) and
has two simple zeros overlapping on $L$. The equivalent set of regular
constraints on $L$ is $\{\phi _{\Pi }=x-1\approx 0,\,\phi
_{C}=x^{2}+y^{2}-1\approx 0\}$.

\section{Treatment of systems with multilinear constraints}

In what follows regular systems and linearly dependent constraints will not
be discussed. They are either treated in standard texts, or they can be
trivially reduced to the regular case.

Consider a system of type I, as in Eq. (\ref{multilinear}). In the vicinity
of an irregular point where only two surfaces (\ref{codimension 1})
intersect, say $\Sigma _{1}$ and $\Sigma _{2}$, the constraint $\phi \approx
0$ is equivalently described by the set of regular constraints

\begin{equation}
f_{1}\approx 0\,,\qquad f_{2}\approx 0\,.  \label{regular set}
\end{equation}
This replacement generically changes the Lagrangian of the system, and the
orbits, as well. Suppose the original canonical Lagrangian is

\begin{equation}
L(q,\dot{q},u)=p_{i}\dot{q}^{i}-H(q,p)-u\phi (q,p)\,,  \label{LC old}
\end{equation}
where $H$ is the Hamiltonian containing all regular constraints. Replacing $%
\phi $ by (\ref{regular set}), gives rise to an effective extended
Lagrangian
\begin{equation}
L_{12}(q,\dot{q},v)=p_{i}\dot{q}^{i}-H(q,p)-v^{1}f_{1}(q,p)-v^{2}f_{2}(q,p)%
\,.  \label{LC new}
\end{equation}
defined on $\Sigma _{12}$. Thus, instead of the \emph{irregular} Lagrangian (%
\ref{LC old}) defined on the whole $\Sigma $, there is a collection of \emph{%
regularized }effective Lagrangians defined in the neighborhood of the
different intersections of $\Sigma _{i}$s. For each of these regularized
Lagrangians, the Dirac procedure can be carried out to the end.

This can be illustrated with the example of a Lagrangian in a $\left(
2+N\right) $-dimensional configuration space $(x,y,q^{1},\ldots ,q^{N})$,
\begin{equation}
L=\frac{1}{2}\,\sum\limits_{k=1}^{N}\left( \dot{q}^{k}\right) ^{2}+\frac{1}{2%
}\,\left( \dot{x}^{2}+\dot{y}^{2}\right) -\lambda xy\,.  \label{example}
\end{equation}%
This Lagrangian describes a free particle moving\ on the set $\left\{ \left.
\left( x,y,q^{k}\right) \in \mathbb{R}^{N+2}\right\vert xy=0\right\} \subset
\mathbb{R}^{N+2},$ which is the union of two $(N+1)$-dimensional planes
where $x$ and $y$ vanish respectively. The constraint surface defined by $%
xy=0$ can be divided into the following sets:
\begin{eqnarray}
\Sigma _{1} &=&\left\{ \left. \left( x,0,q^{k};p_{x},p_{y},p_{k}\right)
\right\vert x\neq 0\right\} \,,  \notag \\
\Sigma _{2} &=&\left\{ \left. \left( 0,y,q^{k};p_{x},p_{y},p_{k}\right)
\right\vert y\neq 0\right\} \,,  \label{sigma-y} \\
\Sigma _{12} &=&\left\{ \left( 0,0,q^{k};p_{x},p_{y},p_{k}\right) \right\}
\,.  \notag
\end{eqnarray}%
The constraint is regular on $\Sigma _{1}\bigcup \Sigma _{2}$, while on $%
\Sigma _{12}$ it is irregular and can be exchanged by $\left\{ \phi
_{1}=x\approx 0\right. $, $\left. \phi _{2}=y\approx 0\right\} $. The
corresponding regularized Lagrangians are
\begin{eqnarray}
L_{1} &=&\frac{1}{2}\,\sum\limits_{k=1}^{N}\left( \dot{q}^{k}\right) ^{2}+%
\frac{1}{2}\,\dot{x}^{2}\,,  \notag \\
L_{2} &=&\frac{1}{2}\,\sum\limits_{k=1}^{N}\left( \dot{q}^{k}\right) ^{2}+%
\frac{1}{2}\,\dot{y}^{2}\,,  \label{L2} \\
L_{12} &=&\frac{1}{2}\,\sum\limits_{k=1}^{N}\left( \dot{q}^{k}\right) ^{2}\,,
\notag
\end{eqnarray}%
and the Lagrange multipliers have dropped out, so the regularized
Lagrangians describe physical degrees of freedom only -- as expected.

The corresponding regularized Hamiltonians are
\begin{eqnarray}
H_{1} &=&\frac{1}{2}\,\sum\limits_{k=1}^{N}p_{k}^{2}+\frac{1}{2}%
\,p_{x}^{2}\,,  \notag \\
H_{2} &=&\frac{1}{2}\,\sum\limits_{k=1}^{N}p_{k}^{2}+\frac{1}{2}%
\,p_{y}^{2}\,,  \label{H2} \\
H_{12} &=&\frac{1}{2}\,\sum\limits_{k=1}^{N}p_{k}^{2}\,,  \notag
\end{eqnarray}%
which are defined in the corresponding reduced manifolds of phase space
(obtained after completing the Dirac-Bergman procedure):
\begin{eqnarray}
\tilde{\Sigma}_{1} &=&\left\{ \left. \left( x,0,q^{k};p_{x},0,p_{k}\right)
\right\vert x\neq 0\right\} \,,  \notag \\
\tilde{\Sigma}_{2} &=&\left\{ \left. \left( 0,y,q^{k};0,p_{y},p_{k}\right)
\right\vert y\neq 0\right\} \,,  \label{sigma total} \\
\tilde{\Sigma}_{12} &=&\left\{ \left( 0,0,q^{k};0,0,p_{k}\right) \right\} \,.
\notag
\end{eqnarray}

It is straightforward to generalize the proposed treatment when more than
two surfaces $\Sigma _{i}$ overlap.

\subparagraph{Evolution of a multilinearly constrained system.}

In the presence of a multilinear constraint, there are regions of the phase
space where the Jacobian has different rank. A question is, whether the
system can evolve from a generic configuration in a region of maximal rank,
to a configuration of lower rank, in finite time, and in the case that that
were possible, what happens with the system afterwards.

To answer this question, consider the example (\ref{example}) for $N=1$,
\begin{equation}
L=\frac{1}{2}\,\left( \dot{x}^{2}+\dot{y}^{2}+\dot{z}^{2}\right) -\lambda
xy\,.  \label{N=1}
\end{equation}%
In the regions $\Sigma _{1}$ and $\Sigma _{2}$ [see Eqs. (\ref{sigma-y})],
the rank is maximal and the free particle can move freely along the $x$- or $%
y$-axis, respectively.

Suppose that the initial state is
\begin{equation}
x(0)=a>0\,,\quad y(0)=0\,,\quad z(0)=0\,,\quad \dot{x}(0)=-v<0\,,\quad \dot{y%
}(0)=0\,,\quad \dot{z}(0)=0\,,  \label{t=0}
\end{equation}
so that the particle is moving on $\Sigma _{1}$, with finite speed along the
$x$-axis towards $x=0$ on $\Sigma _{12}$. The evolution is given by $\bar{x}%
(t)=a-vt$, $\bar{y}(t)=0$, $\bar{z}(t)=0$ and the particle clearly reaches $%
x=0$ in a finite time ($T=a/v$). What happens then? According to the
evolution equation, for $x<0$ the trajectory takes the form $\bar{x}%
(t)=a^{\prime }-v^{\prime }t$, $\bar{y}(t)=0$, $\bar{z}(t)=0$, however the
action would be infinite unless $a=a^{\prime }$ and $v=v^{\prime }$.
Therefore, the particle continues unperturbed past beyond the point where
the RCs fail. So, the irregular surface is not only reachable in a finite
time, but it is crossed without any observable effect on the trajectory.

From the point of view of the trajectory in phase space, it is clear that
the initial state $\left( a,0,0;-v,0,0\right) $ lies on the surface $\tilde{%
\Sigma}_{1}$, and at $t=T$ the system reaches the point $\left(
0,0,0;-v,0,0\right) $, which \emph{does not lie on} the surface $\tilde{%
\Sigma}_{12}=\left\{ \left( 0,0,z;0,0,p_{z}\right) \right\} $.

While it is true that at $t=T$ the Jacobian changes rank, it would be
incorrect to conclude that the evolution suffers a jump since the dynamical
equations are perfectly valid there. In order to have a significant change
in the dynamics, the Jacobian should change its rank in an open set.

\subparagraph{Degenerate systems.}

The problem of evolution in irregular systems should not be confused with
the issues arising in \emph{degenerate systems}, in which the symplectic
matrix of \emph{regular} constraints $\left\{ \phi _{r},\phi _{s}\right\}
=\Omega _{rs}\left( z\right) $ is a phase space function which changes its
rank at $t=\tau $. In those systems, it \emph{is} possible for a system in a
generic initial configuration, to reach (in a finite time) a configuration
where the symplectic form $\Omega _{rs}\left( z\left( \tau \right) \right) $
has lower rank, leading to a \emph{new} gauge invariance which cancels a
number of degrees of freedom
\cite{Saavedra-Troncoso-Zanelli,Saavedra-Troncoso-Zanelli2}.

That problem is unrelated to the one discussed in \emph{irregular }systems
and can be treated independently. In the irregular case it is the functional
independence of the constraints that fails; in the degenerate dynamical
systems it is the symplectic structure that breaks down.

\section{Systems with nonlinear constraints}

Consider the case of irregular systems of type II. It will be shown that a
nonlinear irregular constraint can be replaced by an equivalent linear one
without changing the dynamical contents of the theory, provided the linear
constraint is second class. Otherwise, the resulting Hamiltonian dynamics
will be, in general, inequivalent to that of the original Lagrangian system.

In order to illustrate this point, consider a system given by the Lagrangian
\begin{equation}
L(q,\dot{q},u)=\frac{1}{2}\,\gamma _{ij}\,\dot{q}^{i}\dot{q}^{j}-u\left[ f(q)%
\right] ^{k}\,,  \label{ex-1}
\end{equation}
where $k>1$\ and

\begin{equation}
f(q)\equiv c_{i}q^{i}\neq 0,\quad i=1,...,N\,.
\end{equation}
Here it is assumed that the metric $\gamma _{ij}$ to be constant and
invertible, and the coefficients $c_{i}$ are also constant. The
Euler-Lagrange equations describe a free particle in an $N$-dimensional
space, with time evolution $\bar{q}^{i}(t)=v_{0}^{i}\,t+q_{0}^{i}$, where $%
u(t)$ is a Lagrange multiplier. This solution is determined by $2N$ initial
conditions, $q^{i}(0)=q_{0}^{i}$ and $\dot{q}^{i}(0)=v_{0}^{i}$ subject to
the constraints $c_{i}q_{0}^{i}=0$ and $c_{i}v_{0}^{i}=0$. Thus, the system
possesses $N-1$ physical degrees of freedom.

In the Hamiltonian approach this system has a primary constraint $\pi \equiv
\frac{\partial L}{\partial \dot{u}}\approx 0$\ whose preservation in time
leads to the secondary constraint
\begin{equation}
\phi \equiv \left[ f(q)\right] ^{k}\approx 0\,.  \label{phi}
\end{equation}%
According to (\ref{nonlinear}), this is a nonlinear constraint and there are
no further constraints. As a consequence, the system has only two first
class constraints $\left\{ \pi \approx 0,\;f^{k}\approx 0\right\} $, and $%
N-1 $ degrees of freedom, as found in the Lagrangian approach.

On the other hand, if one chooses instead of (\ref{phi}), the equivalent
linear constraint
\begin{equation}
f(q)=c_{i}q^{i}\approx 0\,,  \label{root}
\end{equation}
then its time evolution yields a \emph{new} constraint,
\begin{equation}
\chi (p)\equiv \gamma ^{ij}c_{i}\,p_{j}\approx 0\,.
\end{equation}
Now, since
\begin{equation}
\left\{ f,\chi \right\} =\gamma ^{ij}c_{i}\,c_{j}\equiv \left\Vert
c\right\Vert ^{2}\,,
\end{equation}
two cases can be distinguished:

\begin{itemize}
\item If $\left\Vert c\right\Vert =0$, there are three first class
constraints, $\pi \approx 0$, $f\approx 0$ and $\chi \approx 0$, which means
that the system has $N-2$\ physical degrees of freedom. In this case,
substitution of (\ref{phi}) by the equivalent linear constraint (\ref{root}%
), yields a \emph{dynamically inequivalent} system\footnote{%
The expression \textquotedblleft substitution of
constraints\textquotedblright\ always refers to two steps: (\emph{i}) the
exchange of a set of constraints by an equivalent set, and (\emph{ii}) the
introduction of a corresponding effective Lagrangian of type (\ref{LC new}).}%
.

\item If $\left\| c\right\| \neq 0$, then $f\approx 0$ and $\chi \approx 0$
are second class, while $\pi \approx 0$ is first class, which leaves $N-1$
physical degrees of freedom\ and the substitution does not change the
dynamics of the system.
\end{itemize}

Thus, if $f^{k}\approx 0$ is irregular, replacing it by the regular
constraint $f\approx 0$ changes the dynamics if $f$ is a first class
function, but it gives the correct result if it is a second class function.

Note that in the Lagrangian description there is no room to distinguish
first and second class constraints, so it would seem like the value of $%
\mathbf{||}c||$ didn't matter. However, the inequivalence of the
substitution can be understood in the Lagrangian analysis as well. Suppose
that it were permissible to exchange the constraint $f^{k}\approx 0$ by $%
f\approx 0$ in the Lagrangian. Then, instead of (\ref{ex-1}), one would have
\begin{equation}
\tilde{L}(q,u)=\frac{1}{2}\,\gamma _{ij}\,\dot{q}^{i}\dot{q}^{j}-uf(q)\,.
\label{ex-2}
\end{equation}
It can be easily checked that (\ref{ex-2}) yields $N-2$ degrees of freedom
when $\left\Vert c\right\Vert =0$, and $N-1$ degrees of freedom when $%
\left\Vert c\right\Vert \neq 0$, which agrees with the results obtained in
the Hamiltonian analysis. Note that the substitution of $f^{k}$ by $f$
modifies the dynamics only if $\gamma ^{ij}c_{i}\,c_{j}=0$, but this can
happen nontrivially only if the metric $\gamma _{ij}$ is not positive
definite.

In general, a nonlinear irregular constraint $\phi \approx 0$\ has a
multiple zero on the constraint surface $\Sigma $, which means that its
gradient vanishes on $\Sigma $ as well. An immediate consequence of $\frac{%
\partial \phi }{\partial z^{i}}\approx $ $0$, is that $\phi $ commutes with
all \emph{finite} functions on $\Gamma $,
\begin{equation}
\left\{ \phi ,F(z)\right\} \approx 0\,.  \label{pathological 1}
\end{equation}
As a consequence, $\phi \approx 0$ is first class and is always preserved in
time, $\dot{\phi}\approx 0\,.$ On the other hand, a nonlinear constraint
cannot be viewed as a symmetry generator simply because it does not generate
any transformation,
\begin{equation}
\delta _{\varepsilon }z^{i}=\left\{ z^{i},\varepsilon \phi \right\} \approx
0\,.  \label{trivial}
\end{equation}
Consistently with this, $\phi $ cannot be gauge-fixed, as there is no finite
function $\mathcal{P}$ on $\Gamma $ such that $\left\{ \phi ,\mathcal{P}%
\right\} \neq 0\,.$

In this sense, a nonlinear first class constraint that cannot be
gauge-fixed, cancels only half a degree of freedom. The other half degree of
freedom cannot be cancelled because the gauge-fixing function does not exist
and, in particular, it cannot appear in the Hamiltonian. Although it allow
counting the degrees of freedom in a theory, these systems are pathological
and their physical relevance is questionable since their Lagrangians cannot
be regularized.

When a nonlinear constraint $\phi \approx 0$ can be exchanged by a regular
one, the Lagrangian is regularized as in the case of multilinear
constraints. For example, the system (\ref{ex-1}) with $\left\| c\right\|
\neq 0$ has Hamiltonian
\begin{equation}
H=\frac{1}{2}\,\gamma ^{ij}p_{i}p_{j}+\lambda \pi +uf(q)\,,
\end{equation}
where $f=c_{i}q^{i}$ will turn out to be a second class constraint. The
corresponding regularized Lagrangian coincides with $\tilde{L}$, Eq. (\ref%
{ex-2}), as expected.

In Refs. \cite{Garcia-Pons,Pons-Salisbury-Shepley} irregular systems of the
type II were discussed. It was pointed out that there was a possible loss of
dynamical information in some cases. From our point of view, it is clear
that this would occur when $f$ is a first class function.

\section{Some implications of the irregularity}

\subsubsection{\emph{a}) Linearization}

It has been observed in five dimensional Chern-Simons theory, that\textbf{\ }%
the effective action for the linearized perturbations of the system around
certain backgrounds seems to have more degrees of freedom than the fully
nonlinear theory \cite{Chandia-Troncoso-Zanelli}. This is puzzling since the
heuristic picture is that the degrees of freedom of a system correspond to
the small perturbations around a local minimum of the action, and therefore
the number of degrees of freedom should not change when the linearized
approximation is used.

In view of the discussion in the previous section, it is clear that a
possible solution of the puzzle lies in the fact that substituting a
nonlinear constraint by a linear one may change the dynamical features of
the theory. But the problem with linear approximations is more serious: the
linearized approximation retains only up to quadratic and bilinear terms in
the Lagrangian, which give linear equations for the perturbations. Thus,
irregular constraints in the vicinity of the constraint surface are erased
in the linearized action. The smaller number of constraints in the effective
theory can lead to the wrong conclusion that the effective system possess
more degrees of freedom than the unperturbed theory. The lesson to be
learned is that the linear approximation is not valid in the part of the
phase space where the RCs fail.

This is illustrated by the same example discussed earlier (\ref{ex-1}). One
can choose as a background the solution $(\bar{q}^{1},\ldots ,\bar{q}^{N},%
\bar{u})$, where $\bar{q}^{i}(t)=q_{0}^{i}+v_{0}^{i}t$ satisfies the
constraint
\begin{equation}
c_{i}\bar{q}^{i}=0\,,  \label{fq}
\end{equation}%
and $\bar{u}(t)$ is an arbitrarily given function. This describes a free
particle moving in the ($N-1$)-dimensional plane defined by (\ref{fq}). The
linearized effective Lagrangian, to second order in the small perturbations $%
s^{i}=q^{i}-\bar{q}^{i}(t)$ and $w=u-\bar{u}(t)$, has the form
\begin{equation}
L_{\mathtt{eff}}\left( s,w\right) =\frac{1}{2}\,\gamma _{ij}\left( v_{0}^{i}+%
\dot{s}^{i}\right) \left( v_{0}^{j}+\dot{s}^{j}\right) -\bar{u}\left(
c_{i}s^{i}\right) ^{2}\,,  \label{Leff}
\end{equation}%
and the equations of motion are
\begin{equation}
\ddot{s}^{i}+\Gamma _{j}^{i}(t)\,s^{j}=0\,,\qquad \left( i=1,\ldots
,N\right) \,,  \label{Lineq}
\end{equation}%
where $\Gamma _{j}^{i}\equiv 2\bar{u}\,\gamma ^{ik}c_{k}c_{j}$ is the
eigen-frequency matrix. Since $\bar{u}$\ is not a dynamical variable, it is
not varied and the nonlinear constraint $\left( c_{i}s^{i}\right) ^{2}=0$ is
absent from the linearized equations. The system described by (\ref{Lineq})
possesses $N$ physical degrees of freedom, that is, one degree of freedom
more than the original nonlinear theory (\ref{ex-1}).

The only indication that one of these degrees of freedom has a nonphysical
origin is the following: If $\Vert c\Vert \neq 0$, splitting the components
of $s^{i}$ along $c_{i}$ and orthogonal to $c_{i}$\ as
\begin{equation}
s^{i}(t)\equiv s(t)\gamma ^{ij}c_{j}+s_{\perp }^{i}(t)\,,
\end{equation}%
gives rise to the projected equations
\begin{eqnarray}
\ddot{s}_{\perp }^{i} &=&0\,,  \label{s-perp} \\
\ddot{s}+2\bar{u}(t)\Vert c\Vert ^{2}s &=&0\,.  \label{s}
\end{eqnarray}%
The $N-1$ components of $s_{\perp }^{i}(t)$ obey a deterministic second
order equation, whereas $s(t)$ satisfies an equation which depends on an
indeterminate arbitrary function $\bar{u}(t)$. The dependence of $s=\bar{s}%
(t,\bar{u}(t))$ on the background Lagrange multiplier $\bar{u}$ is an
indication that $s$ is a nonphysical degree of freedom, since $u$ was an
arbitrary function to begin with. This is not manifest in Eq. (\ref{s}),
where $\bar{u}$\ is a fixed function and, from a naive point of view, $s(t)$
is determined by the same equation, regardless of the physically obscure
origin of the function $\bar{u}$. It is this naive analysis that leads to
the wrong conclusion indicated above.

Note that a linearized theory may be consistent by itself, but it is not
necessarily a faithful approximation of a nonlinear theory.

One way to avoid the inconsistencies between the original theory and the
linearized one would be to first regularize the constraints (if possible)
and then linearize the corresponding regular Lagrangian.\footnote{%
There may be also other problems in linear approximation (\emph{e.g.}, when
a topology is non-trivial), but these cases are not discussed here.}

\subsubsection{\emph{b}) Dirac conjecture}

Dirac conjectured that \emph{all} first class constraints generate gauge
symmetries. It was shown that Dirac's conjecture is not true for first class
constraints of the form $f^{k}$ $(k>1)$, and following from $\dot{f}\approx
0 $ \cite{Castellani,Blagojevic}. Therefore, for systems with nonlinear
constraints, the conjecture does not work and they generically provide
counterexamples of it \cite{Henneaux-Teitelboim,Allcock,Cawley}.

From the point of view of irregular systems, it is clear that Dirac's
conjecture fails for nonlinear constraints because they do not generate any
local transformation, c.f. Eq. (\ref{trivial}). In Refs. \cite%
{Garcia-Pons,Pons-Salisbury-Shepley} it was observed that Dirac's conjecture
may not hold in the presence of irregular constraints of type II.

In the case of multilinear constraints, however, Dirac's conjecture holds.
The fact that at irregular points the constraints do not generate any
transformation only means that these are fixed points of the gauge
transformation.

\subsubsection{\emph{c}) Quantization}

Although it is possible to deal systematically with classical theories
containing irregular constraints, there may be severe problems in their
quantum description. Consider a path integral of the form
\begin{equation}
Z\sim \int [dq][dp][du]\exp i\left[ p\dot{q}-H(q,p)-u\phi (q,p)\right] \,,
\end{equation}%
where $\phi =\left[ f(q,p)\right] ^{k}$ is a nonlinear constraint.
Integration on $u$ yields to $\delta \left( f^{k}\right) $, which is not
well-defined for a zero of order $k>1$, making the quantum theory ill
defined. Only if the nonlinear constraint could be exchanged by the regular
one, $f(q,p)\approx 0$, the quantum theory could be saved. An example of
this occurs in the standard approach for QED, where it is usual practice to
introduce the nonlinear (Coulomb) gauge fixing term $u(\partial
_{i}A^{i})^{2}$ in order to fix the Gauss law $\phi =\partial _{i}\pi
^{i}\approx 0$ (where $\pi ^{i}\equiv \frac{\delta I_{ED}}{\delta \dot{A}_{i}%
}$). Since the function $f(A)=\partial _{i}A^{i}$ is second class, the
substitution of $f^{2}\left( A\right) \approx 0$ by a regular constraint $%
f\left( A\right) \approx 0$ does not change its dynamical structure.

The other possibility to quantize a system with a nonlinear constraint is to
modify the original Lagrangian so that it becomes regular, leading to the
the change in dynamics as well. This possibility is considered in the Siegel
model of the chiral boson \cite{Siegel}, whose quantum theory is analyzed in
\cite{Henneaux-TeitelboimCB}.

There are other examples of irregular systems whose quantum theories are
discussed in the literature. For example, it is shown that for the models of
relativistic particles with higher spin ($s>\frac{1}{2}$), called \emph{%
systems admitting no gauge conditions}, since they contain irregular first
class constraints, different quantization methods can lead to different
physical results \cite{Plyushchay-Razumov1,Plyushchay-Razumov2}. Other
examples of quantum irregular systems are planar gauge field theories \cite%
{Grignani-Plyushchay-Sodano} and topologically massive gauge fields \cite%
{Nirov-Plyushchay}.

\section{Summary}

The dynamics of a system possessing constraints which may violate the
regularity conditions (functional independence) on some subsets of the
constraint surface $\Sigma $, was discussed. These so-called irregular
systems are seen to arise generically because of nonlinearities in the
constraints and can be classified into two families: multilinear (type I)
and nonlinear (type II).

\begin{itemize}
\item Type I constraints are of the form $\phi =\prod f_i(z)$, where $f_i$
possess simple zeros. These constraints violate the regularity conditions
(RCs) on sets of measure zero on the constraint surface $\Sigma $.

\item Type I constraints can be exchanged by equivalent constraints which
are regular giving an equivalent dynamical system.

\item Type II constraints are of the form $\phi =f^k$ $(k>1)$ where $f$ has
a simple zero. They violate the RCs on sets of nonzero measure on $\Sigma $.

\item A type II constraint can be replaced by an equivalent linear one only
if the latter is second class; if the equivalent linear constraint is first
class, substituting it for the original constraint would change the system.

\item In general, the orbits can cross the configurations where the RCs are
violated without any catastrophic effect for the system. If the symplectic
form degenerates at the irregular points, additional analysis is required.

\item The naive linearized approximation of an irregular constrained system
generically changes it by erasing the irregular constraints. In order to
study the perturbations around a classical orbit in an irregular system, it
would be necessary to first regularize it (if possible) and only then do the
linearized approximation.
\end{itemize}

\chapter{Higher-dimensional Chern-Simons theories as irregular systems}

In this chapter the Hamiltonian dynamics of Chern-Simons (CS) theories in $%
D\geq 5$ is analyzed. It is known that regular and generic CS theories are
invariant under the gauge symmetries and diffeomorphisms \cite%
{Banados-Garay-Henneaux1,Banados-Garay-Henneaux2}, but these theories also
possess sectors in phase space which are irregular \cite{Miskovic-ZanelliJMP}%
. Here a criterion for the choice of a regular background in CS theories is
found, and the implications of the irregularity in CS theories are discussed.

The dynamics arising at the boundary is not considered here.

\section{Chern-Simons action \label{CSI}}

A CS theory is one where the fundamental field is a Lie-algebra-valued
connection one-form $\mathbf{A}=A^{a}\mathbf{G}_{a}$, where the
corresponding anti-Hermitean generators close a $N$-di\-men\-si\-o\-nal algebra $%
\left[ \mathbf{G}_{a},\mathbf{G}_{b}\right] =f_{ab}^{\;\;c}\,\mathbf{G}_{a}$%
. The connection field defines a covariant derivative $D$ that acts on a $p$%
-form $\omega $ as $D\omega =d\omega +[\mathbf{A},\omega $\/$]$.\footnote{%
A $p$-form is the object $\omega =\frac{1}{p!}\,\omega _{n_{1}\cdots
n_{p}}\,dx^{n_{1}}\cdots dx^{n_{p}},$ and the commutator of $p$-form $\omega
$ and $q$-form $\Omega $ is given by $\left[ \omega ,\Omega \right] =\omega
\Omega -\left( -\right) ^{pq}\Omega \omega $.}

The CS Lagrangian is a $(2n+1)$ - form such that its exterior derivative (in
$2n+2$ dimensions) is
\begin{equation}
dL_{\text{CS}}=k\,\,g_{a_{1}\cdots a_{n+1}}\,F^{a_{1}}\cdots F^{a_{n+1}}\,.
\label{dL_CS}
\end{equation}
Here $k$ is a dimensionless constant, $g_{a_{_{1}}\ldots a_{n+1}}$ is a
completely symmetric, invariant tensor $(Dg_{a_{1}\cdots a_{n+1}}=0)$ and $%
\mathbf{F=}F^{a}\mathbf{G}_{a}=d\mathbf{A}+\mathbf{A}^{2}$ is the
field-strength 2-form which satisfies the \emph{Bianchi identity} $D\mathbf{F%
}=0$.

The definition (\ref{dL_CS}) is based on the fact that the form on the \emph{%
r.h.s.} is \emph{closed }(its exterior derivative vanishes due to the
Bianchi identity) and therefore can \emph{locally} be written as $dL_{\text{%
CS}}$. Applying Stokes' theorem to (\ref{dL_CS}), the integral of the total
derivative becomes the integration over the $D$-dimensional manifold $%
\mathcal{M}_{D}=\partial \mathcal{M}_{D+1}$ without boundary and one obtains
the integral identity $\int_{\mathcal{M}_{D+1}}dL_{\text{CS}}(A)=\int_{%
\mathcal{M}_{D}}L_{\text{CS}}(A)$. In general, a CS theory is defined on an
arbitrary (not necessary closed) manifold $\mathcal{M}$,
\begin{equation}
I_{\text{CS}}[A]=\int\limits_{\mathcal{M}}L_{\text{CS}}(A)\,.  \label{I_CS}
\end{equation}
Variation of the expression (\ref{dL_CS}) with respect to the gauge field $%
A^{a}$ gives (up to an exact form)
\begin{equation}
\delta L_{\text{CS}}=k\left( n+1\right) \,\,g_{aa_{1}\cdots
a_{n}}\,F^{a_{1}}\cdots F^{a_{n}}\delta A^{a}\,,  \label{var L}
\end{equation}
from where, supposing the suitable boundary conditions which give a well
defined extremum for $I_{\text{CS}}$, the Euler-Lagrange equations are
obtained,
\begin{equation}
g_{aa_{1}\cdots a_{n}}\,F^{a_{1}}\cdots F^{a_{n}}=0\,.
\end{equation}
Written with all indices, the equations of motion are
\begin{equation}
\varepsilon ^{\mu \mu _{1}\nu _{1}\cdots \mu _{n}\nu _{n}}g_{aa_{1}\cdots
a_{n}}\,F_{\mu _{1}\nu _{1}}^{a_{1}}\cdots F_{\mu _{n}\nu _{n}}^{a_{n}}=0\,,
\label{e.o.m.}
\end{equation}
where $x^{\mu }$ $\left( \mu =0,1,\ldots ,2n\right) $ are local coordinates
at $\mathcal{M}$.

The CS action possesses the following local symmetries:

\begin{itemize}
\item By construction, it is invariant under general coordinate
transformations or \emph{diffeomorphisms}
\begin{eqnarray}
x^{\mu } &\rightarrow &x^{\prime \mu }=x^{\mu }+\xi ^{\mu }(x)\,,  \notag \\
\mathbf{A}(x) &\rightarrow &\mathbf{A}^{\prime }(x^{\prime })=\mathbf{A}%
(x)\,,\qquad \qquad \left( \delta _{\xi }\mathbf{A}=-\pounds _{\xi }\mathbf{A%
}\right) \,,  \label{diff}
\end{eqnarray}%
where $\pounds _{\xi }A_{\mu }^{a}\equiv \partial _{\mu }\xi ^{\nu }A_{\nu
}^{a}+\xi ^{\nu }\partial _{\nu }A_{\mu }^{a}$ stands for a Lie derivative;

\item It has an infinitesimal \emph{gauge symmetry},
\begin{equation}
\delta _{\lambda }\mathbf{A}=-D\mathbf{\lambda }\,,  \label{gauge transf.}
\end{equation}%
since the \emph{r.h.s.} of (\ref{dL_CS}) is an explicitly gauge invariant
expression.
\end{itemize}

Under \emph{large} gauge transformations, a CS action changes for a closed
form. In this chapter the local dynamics is analyzed, and all effects
arising at the boundary are neglected.

\section{Hamiltonian analysis of Chern-Simons theories}

Here the features of the sectors with \emph{regular} dynamics of $(2n+1)$%
-dimensional CS theories are reviewed, where the dimension is larger than
three \cite{Banados-Garay-Henneaux2}.

In the Hamiltonian approach, one supposes that the space-time manifold $%
\mathcal{M}$ has topology $\mathbb{R}\times \sigma $, where $\sigma $ is a $%
2n$-dimensional spatial manifold parametrized by local coordinates $x^{i}$.
The CS action $I_{\text{CS}},$ defined by (\ref{dL_CS}) and (\ref{I_CS}), is
linear in velocities (first order formalism) and can be written as
\begin{equation}
I_{\text{CS}}\left[ A\right] =\int d^{2n+1}x\,\left( \mathcal{L}_{a}^{i}\,%
\dot{A}_{i}^{a}-A_{0}^{a}\chi _{a}\right) \,,  \label{linear in V}
\end{equation}%
where\footnote{%
The expression for $\chi _{a}$ can be obtained directly from (\ref{var L})
for a special choice of variations $\delta A^{a}=\delta A_{0}^{a}\,dt$, and
the definition $\chi _{a}=-\frac{\delta I_{\text{CS}}}{\delta A_{0}^{a}}$
following from (\ref{linear in V}).}
\begin{equation}
\chi _{a}\equiv -\frac{k}{2^{n}}\,\left( n+1\right) \,\varepsilon
^{i_{1}j_{1}\cdots i_{n}j_{n}}\,g_{aa_{1}\cdots
a_{n}}F_{i_{1}j_{1}}^{a_{1}}\cdots F_{i_{n}j_{n}}^{a_{n}}\,,
\end{equation}%
and $\varepsilon ^{i_{1}\cdots i_{2n}}\equiv \varepsilon ^{0i_{1}\cdots
i_{2n}}$. The explicit form of $\mathcal{L}_{a}^{i}\,$is not necessary since
it defines the kinetic term through the \emph{symplectic form }$\Omega
_{ab}^{ij}:$%
\begin{eqnarray}
\Omega _{ab}^{ij}\left( x,x^{\prime }\right) &=&\frac{\delta \mathcal{L}%
_{b}^{j}\left( x^{\prime }\right) }{\delta A_{i}^{a}\left( x\right) }-\frac{%
\delta \mathcal{L}_{a}^{i}\left( x\right) }{\delta A_{j}^{b}\left( x^{\prime
}\right) }  \notag \\
&=&-\frac{kn}{2^{n-1}}\,\left( n+1\right) \,\varepsilon ^{iji_{2}j_{2}\cdots
i_{n}j_{n}}g_{aba_{2}\cdots a_{n}}F_{i_{2}j_{2}}^{a_{1}}\cdots
F_{i_{n}j_{n}}^{a_{n}}\,\delta \left( x-x^{\prime }\right) \,.
\label{sympl. m.}
\end{eqnarray}%
(For derivation see Appendix \ref{sm}.) This matrix plays a central role in
the dynamics of CS theories, as it will be seen later. The gauge field $%
A_{i}^{a}$ and its canonically conjugated momenta $\pi _{a}^{i}\equiv \delta
I_{\text{CS}}/\delta \dot{A}_{i}^{a}$ defines the phase space $\Gamma $,
with the basic PB
\begin{equation}
\left\{ A_{i}^{a}\left( x\right) ,\pi _{b}^{j}\left( x^{\prime }\right)
\right\} =\delta _{i}^{j}\delta _{b}^{a}\delta \left( x-x^{\prime }\right)
\end{equation}%
taken at the same time $x^{0}=x^{\prime 0}=t$, where $\delta \left(
x-x^{\prime }\right) $ stands for a Dirac $\delta $-function at the spatial
section. The field $A_{0}^{a}$ is a Lagrange multiplier.

\subparagraph{Constraints.}

Constraints in the theory are

\begin{eqnarray}
\phi _{a}^{i} &\equiv &\pi _{a}^{i}-\mathcal{L}_{a}^{i}(A)\approx 0\,,
\notag \\
\mathcal{G}_{a} &\equiv &-\chi _{a}+D_{i}\phi _{a}^{i}\approx 0\,,
\end{eqnarray}%
where the covariant derivative acts on $\phi _{a}^{i}$ as $D_{i}\phi
_{a}^{i}\equiv \partial _{i}\phi _{a}^{i}+f_{ab}^{\;\;c}A_{i}^{b}\phi
_{c}^{i}.$ The total Hamiltonian is
\begin{equation}
H=\int d^{2n}x\,\left( -A_{0}^{a}\mathcal{G}_{a}+u_{i}^{a}\phi
_{a}^{i}\right) \,,
\end{equation}%
and it depends on $N$ arbitrary functions $A_{0}^{a}$ and $2nN$ arbitrary
functions $u_{i}^{a}$. Constraints define the constraint surface $\Sigma
\subset \Gamma $ and they close the following PB algebra:
\begin{equation}
\begin{array}{ll}
\left\{ \pi _{a}^{0},\;\text{all\ }\right\} =0\,,\medskip & \left\{ \mathcal{%
G}_{a},\mathcal{G}_{b}\right\} =f_{ab}^{\;\;c}\phi _{c}^{i}\delta \,, \\
\left\{ \phi _{a}^{i},\phi _{b}^{j}\right\} =\Omega _{ab}^{ij}\delta
\,,\qquad \qquad & \left\{ \mathcal{G}_{a},\phi _{b}^{i}\right\}
=f_{ab}^{\;\;c}\phi _{c}^{i}\delta \,,%
\end{array}
\label{PB}
\end{equation}%
with $\Omega _{ab}^{ij}(x,x^{\prime })=\Omega _{ab}^{ij}\left( x\right)
\delta \left( x-x^{\prime }\right) $. The constraints $\phi _{a}^{i}$ and $%
\mathcal{G}_{a}$ evolve as
\begin{eqnarray}
\dot{\phi}_{a}^{i} &\approx &\left\{ \phi _{a}^{i},H\right\} =\Omega
_{ab}^{ij}u_{j}^{b}-f_{ab}^{\;\;c}\phi _{c}^{i}\approx \Omega
_{ab}^{ij}u_{j}^{b}=0\,,  \label{evolution phi} \\
\overset{\cdot }{\mathcal{G}}_{a} &\approx &\left\{ \pi _{a}^{0},H\right\}
=f_{ab}^{\;\;c}\left( A_{0}^{b}\mathcal{G}_{c}-u_{i}^{b}\phi _{c}^{i}\right)
\approx 0\,,  \label{evolution G}
\end{eqnarray}%
and, therefore, there are no new constraints. The equation (\ref{evolution
phi}) gives restrictions on a number of multipliers $u_{a}^{i}$, depending
on the rank of $\Omega _{ab}^{ij}$.

\subparagraph{Generic theories.}

The dynamics of CS theories basically depends on the symplectic matrix $%
\Omega _{ab}^{ij}$. This matrix is degenerate for any CS theory, due to the
identity (Appendix \ref{sm})
\begin{equation}
\Omega _{ab}^{ik}F_{kj}^{b}=-\delta _{j}^{i}\chi _{a}\approx 0\,.
\label{modes}
\end{equation}%
Therefore, there are at least $2n$ zero modes of the symplectic matrix, $%
\left( \mathbf{V}_{i}\right) _{j}^{a}=F_{ij}^{a}$.

Among all CS theories based on a gauge group of dimension $N$, there is a
family of \emph{generic} theories for which:

\begin{description}
\item (\emph{a}) the zero modes of $\Omega _{ab}^{ij}$ are the $2n$ \emph{%
linearly independent} vectors $\left( \mathbf{V}_{i}\right)
_{j}^{a}=F_{ij}^{a}$;

\item (\emph{b}) there are no other zero modes and the rank of symplectic
matrix is the largest,
\begin{equation}
\Re \left( \mathbf{\Omega }\right) =2n\left( N-1\right) \,.
\end{equation}
\end{description}

For $D\geq 5$, a CS theory cannot be generic on the whole phase space. For
example, any CS theory possess a \emph{pure gauge} solution ($F=0$) which
does not satisfy generic conditions (\emph{a}) and (\emph{b}). Therefore, a
generic CS theory is determined by (\emph{i}) an invariant tensor $%
g_{a_{1}\cdots a_{n+1}}$ and (\emph{ii}) a domain, or a sector of the phase
space on which the generic conditions (\emph{a}) and (\emph{b}) are
satisfied. This sector can be chosen as an open set around a solution $\bar{A%
}$ of the constraints $\chi (\bar{A})=0$ such that (\emph{a}) and (\emph{b})
are fulfilled for $\bar{F}$ and $\bar{\Omega}$.

\subparagraph{First and second class constraints.}

In order to separate first from second class constraints, a (non-singular)
transformation of $\phi _{a}^{i}$ has to be made, which diagonalize the
symplectic form in (\ref{PB}):
\begin{equation}
\mathcal{H}_{i}\equiv F_{ij}^{a}\phi _{a}^{j}\approx 0\,,\qquad \qquad
\theta _{\alpha }\equiv S_{\alpha i}^{a}\phi _{a}^{i}\approx 0\,,
\label{diffeomorphisms}
\end{equation}%
where $\alpha =1,\ldots ,2n\left( N-1\right) .$ The constraints $\mathcal{H}%
_{i}$ correspond to the zero modes of $\Omega _{ab}^{ij}$ and satisfy the
algebra
\begin{equation}
\left\{ \mathcal{H}_{i}\left( x\right) ,\mathcal{H}_{j}\left( x^{\prime
}\right) \right\} =\left[ \mathcal{H}_{i}\left( x^{\prime }\right) \partial
_{j}+\mathcal{H}_{j}\left( x\right) \partial _{i}-F_{ij}^{a}\mathcal{G}%
_{a}\left( x\right) \right] \delta \left( x-x^{\prime }\right) ,
\end{equation}%
thus they are first class. Constraints $\theta _{\alpha }$ are second class.
The explicit form of the tensor $S_{\alpha j}^{a}$ is not always possible to
find, and the condition
\begin{equation}
\Re \left( \Delta _{\alpha \beta }\right) =2n(N-1)\,,\qquad \qquad \Delta
_{\alpha \beta }\equiv S_{\alpha _{i}}^{a}\Omega _{ab}^{ij}S_{\beta j}^{b}\,,
\end{equation}%
provides that the only non-vanishing brackets on $\Sigma $ are
\begin{equation}
\left\{ \theta _{\alpha },\theta _{\beta }\right\} \approx \Delta _{\alpha
\beta }\,\delta \,.
\end{equation}

Therefore, in a generic CS theory with $2nN$ dynamical fields $A_{a}^{i},$ $%
N $ first class constraints $\mathcal{G}_{a}$, $2n$ first class constraints $%
\mathcal{H}_{i}\,,$ and $2n\left( N-1\right) $ second class constraints $%
\theta _{\alpha }$, the number of physical degrees of freedom is:
\begin{equation}
f_{2n+1}(N)=nN-n-N\,,\quad \quad (n,N>1)\,.  \label{degrees of freedom}
\end{equation}

\subparagraph{Local symmetries.}

A CS theory is invariant under the following local transformations:

\begin{itemize}
\item \emph{Gauge transformations,} generated by
\begin{equation}
G\left[ \lambda \right] \equiv \int d^{2n}x\,\lambda ^{a}\mathcal{G}_{a}\,,
\end{equation}%
which act on gauge fields as
\begin{equation}
\delta _{\lambda }A_{i}^{a}=\left\{ A_{i}^{a},G\left[ \lambda \right]
\right\} =-D_{i}\lambda ^{a}\,;
\end{equation}

\item \emph{Improved spatial diffeomorphisms,} generated by
\begin{equation}
\mathcal{H}\left[ \varepsilon \right] \equiv \int d^{2n}x\,\varepsilon ^{i}%
\mathcal{H}_{i}\,,
\end{equation}%
which change the gauge field as
\begin{equation}
\delta _{\varepsilon }A_{i}^{a}=\left\{ A_{i}^{a},\mathcal{H}\left[
\varepsilon \right] \right\} =\varepsilon ^{j}F_{ji}^{a}\,.  \label{improved}
\end{equation}
\end{itemize}

Local symmetries which are not independent from the above ones, are \emph{%
spatial diffeomorphisms,} which differ from improved spatial diffeomorphisms
by a gauge transformation,
\begin{equation}
\delta _{\xi }A_{i}^{a}=-\pounds _{\xi }A_{i}^{a}=-\xi
^{j}F_{ji}^{a}-D_{i}\left( \xi ^{i}A_{i}^{a}\right) \,,
\end{equation}%
and\emph{\ (improved) time reparametrizations}, which change the gauge
fields as
\begin{equation}
\delta _{\varepsilon }A_{i}^{a}=\varepsilon ^{0}F_{0i}^{a}\,.
\end{equation}%
Due to the equations of motion $\Omega _{ab}^{ij}F_{0j}^{b}=0$, the vector $%
F_{0j}^{a}$ must be a linear combination of zero modes of $\Omega _{ab}^{ij}$%
, therefore in a generic theory $F_{0i}^{a}=C^{j}F_{ji}^{a}$. In
consequence, the time-like diffeomorphisms can be obtained from spatial
diffeomorphisms \emph{on-shell} through the redefinition of local parameter $%
\varepsilon ^{i}=\varepsilon ^{0}C^{i}$.

\subparagraph{Non-generic theories.}

The above analysis can be easily generalized to degenerate (non-generic)
theories, in which the symplectic matrix $\Omega _{ab}^{ij}$ has $K$
independent zero modes $\left( \mathbf{U}_{\rho }\right) _{i}^{a}=U_{\rho
i}^{a}$,
\begin{equation}
\Omega _{ab}^{ij}U_{\rho j}^{b}\approx 0\,,\qquad \left( \rho =1,\ldots
,K\right) \;\left( 2n<K\leq 2nN\right) \,,
\end{equation}%
where $K=2n$ corresponds to the generic case. Then the first class
constraints are
\begin{equation}
\mathcal{H}_{\rho }\equiv U_{\rho \,j}^{a}\phi _{a}^{i}\approx 0\,,
\end{equation}%
which, apart from $2n$ \emph{improved spatial diffeomorphisms} also generate
an \emph{additional }($K-$ $2n$)\emph{-parameter symmetry}. The
number of physical degrees of freedom is given by
\begin{equation}
f_{2n+1}(N)=nN-N-\frac{K}{2}\,,\quad \quad (n,N>1)\,.  \label{K}
\end{equation}%
This formula is a generalization of (\ref{degrees of freedom}) obtained for
generic theories, and it enables the counting of the degrees of freedom in
the sectors of phase space which have more than minimal number of local
symmetries. However, during its evolution, the degenerate system can change
the number of degrees the freedom (loosing them) if it reaches the point
where the symplectic form has lower rank
\cite{Saavedra-Troncoso-Zanelli,Saavedra-Troncoso-Zanelli2}.
Therefore, the formula (\ref{K}) is valid only on an open set not containing
degenerate points.

\section{Regularity conditions}

In the previous chapter it was implicitly supposed that all constraints were
regular. Now the regularity conditions in CS theories are going to be
analyzed.

First, consider the original set of constraints obtained from Dirac-Bergman
algorithm, $\phi _{a}^{i}\approx 0$ and$\;\chi _{a}\approx 0$. Constraints $%
\phi _{a}^{i}$ are regular since they are linear in momenta. Thus, the
regularity of CS theories is determined by momentum-independent constraints $%
\chi _{a}$. It is convenient to write $\chi _{a}$ in the basis of spatial
1-forms $dx^{i}$ as
\begin{equation}
K_{a}\equiv d^{2n}x\,\chi _{a}=-k\left( n+1\right) \,g_{aa_{1}\cdots
a_{n}}F^{a_{1}}\cdots F^{a_{n}}\approx 0\,.  \label{constraint K}
\end{equation}
Their small variations evaluated at $K_{a}=0$ are
\begin{equation}
\delta K_{a}=\mathbf{J}_{ab}D\delta A^{b}\,,
\end{equation}
and they define the $(2n-2)$-form $\mathbf{J}_{ab}$, which can be identified
as the Jacobian,
\begin{equation}
\mathbf{J}_{ab}\equiv -kn\left( n+1\right) \,g_{aba_{1}\cdots
a_{n-1}}\,\left. F^{a_{1}}\cdots F^{a_{n-1}}\right| _{K=0}\,.
\label{CS Jacobian}
\end{equation}
According to Dirac's definition, a sufficient and necessary condition for $%
K_{a}$ to be regular is
\begin{equation}
\Re \left( \mathbf{J}_{ab}\right) =N\,.  \label{rc}
\end{equation}
Since $\mathbf{J}_{ab}$ is field dependent, its rank may change in phase
space. In particular, for a pure gauge configuration $F^{a}=0$, the Jacobian
$\mathbf{J}_{ab}$ has rank zero. For other configurations, the rank of
Jacobian can range from zero to $N$, and the irregularities are always of
\emph{multilinear} type because in the expression (\ref{constraint K}),
thanks to the antisymmetric tensor $\varepsilon ^{i_{1}j_{1}\cdots
i_{n}j_{n}},$ the phase space coordinate $A_{i}^{a},$ for one particular
choice of $i$, occurs only linearly.

\subparagraph{Regularity conditions of first and second class constraints.}

Suppose that the set of constraints $\left( \phi _{a}^{i},\chi _{a}\right) $
is regular. A new equivalent set, with separated first and second class
constraints, is introduced via the transformation
\begin{equation}
\mathcal{T}:\left( \phi _{b}^{j},\chi _{c}\right) \rightarrow \left(
\mathcal{H}_{i},\theta _{\alpha },\mathcal{G}_{a}\right) \,,
\end{equation}%
where the matrix of transformation is given by
\begin{equation}
\mathcal{T}=\left[
\begin{array}{cc}
\left(
\begin{array}{ll}
F_{ij}^{b} & 0 \\
0 & S_{\alpha j}^{b}%
\end{array}%
\right) & 0 \\
\delta _{a}^{b}D_{j} & -\delta _{a}^{c}%
\end{array}%
\right] \,.  \label{transformation}
\end{equation}%
This transformation preserves the regularity of the original constraints
only if it is invertible at $\Sigma $, \emph{i.e.}, its rank is equal to the
number of constraints, $\Re \left( \mathcal{T}\right) =2n(N+1)$. For the
rank of $\mathcal{T}$, one obtains\footnote{%
If a matrix $M=\left(
\begin{array}{ll}
A & B \\
C & D%
\end{array}%
\right) $ has an invertible submatrix $A$, then $\Re \left( M\right) =\Re
\left( A\right) +\Re \left( D-CA^{-1}B\right) .$ This is a consequence of
the fact that the zero modes of $M$\ have the form $\ \left(
\begin{array}{c}
-A^{-1}B\mathcal{\,}v \\
v%
\end{array}%
\right) $, where $v$ is a zero mode of $D-CA^{-1}B$.}

\begin{equation}
\Re \left( \mathcal{T}\right) =\Re \left( F_{ij}^{b}\right) +\Re \left(
S_{\alpha j}^{b}\right) +N\,,  \label{sum}
\end{equation}%
implying that $N$ constraints $\mathcal{G}_{a}$ are always regular (if $\chi
_{a}$\ are regular), while the regularity of $\mathcal{H}_{i}$ and $\theta
_{\alpha }$ requires two extra conditions:
\begin{equation}
\Re \left( F_{ij}^{b}\right) =2n\,,\qquad \qquad \Re \left( S_{\alpha
j}^{b}\right) =2n(N-1)\,,
\end{equation}%
which mean that $C^{i}F_{ij}^{b}=0$ and $C^{\alpha }S_{\alpha j}^{b}=0$ have
to have the unique solution $C^{i}=C^{\alpha }=0.$

\subparagraph{Generic condition and the regularity.}

The study of dynamics in CS theories requires not only the analysis of
regularity, but also of genericity. A generic configuration has the
symplectic form $\Omega _{ab}^{ij}$ with maximal rank, while a regular
configuration has the invertible Jacobian $\mathbf{J}_{ab}$. The relation
between $\Omega _{ab}^{ij}$ and $\mathbf{J}_{ab}$, Eqs. (\ref{sympl. m.})
and (\ref{CS Jacobian}), is given by
\begin{equation}
2\,d^{2n}x\,\Omega _{ab}^{ij}\equiv dx^{i}dx^{j}\mathbf{J}_{ab}\,.
\end{equation}
In spite of the fact that both conditions are expressed in terms of the same
matrix $\Omega _{ab}^{ij}$, they are independent.

This is illustrated with four different examples.

\begin{enumerate}
\item \emph{An irregular and non-generic configuration} is the \emph{pure
gauge,} $F=0,$ which occurs in any non-Abelian CS theory in $D\geq 5$.

\item \emph{Irregular and generic configurations} can be found in
five-dimensional \emph{AdS}$_5$-CS supergravity, see Ref. \cite%
{Chandia-Troncoso-Zanelli}.

\item \emph{Regular and generic configurations }can also be found in
five-dimensional \emph{AdS}$_{5}$-CS supergravity, as it is discussed in
Chapter \ref{SCS}.

\item \emph{Regular and non-generic configurations} occur in a
five-dimensional CS theory based on a direct product $G_{1}\otimes G_{2},$
for a particular choice of the invariant tensor.
\end{enumerate}

In order to demonstrate the last example, the group indices can be taken as $%
a=(r,\alpha )$ corresponding to $G_1$ and $G_2$ respectively, and the
invariant tensor chosen to have non vanishing components $g_{rs1}=g_{rs}$ $%
\left( r,s=1,2,\ldots \right) $ and $g_{\alpha \beta \bar{1}}=g_{\alpha
\beta }$ $\left( \alpha ,\beta =\bar{1},\bar{2},\ldots \right) ,$ with
tensors $g_{rs}$ and $g_{\alpha \beta }$ invertible. Then the configuration
\begin{equation}
F^a=\left( f^1dx^1dx^2,\;h^{\bar{1}}dx^3dx^4\right)  \label{conf}
\end{equation}
is regular and non-generic. Indeed, the Jacobian evaluated at (\ref{conf})
has maximal rank
\begin{equation}
\mathbf{J}_{ab}=\left(
\begin{array}{cc}
-6k\,g_{rs}\,f^1dx^1dx^2 & 0 \\
0 & -6k\,g_{\alpha \beta }\,h^{\bar{1}}dx^3dx^4%
\end{array}
\right) \,,
\end{equation}
while $\Omega _{ab}^{ij}$ with non-vanishing components
\begin{equation}
\Omega _{rs}^{34}=-3k\,g_{rs}\,f^1,\qquad \qquad \Omega _{\alpha \beta
}^{12}=-3k\,g_{\alpha \beta }\,h^{\bar{1}},
\end{equation}
has $2N$ zero modes
\begin{equation}
\mathbf{V}_i^a=\left(
\begin{array}{c}
u^r\delta _i^1+v^r\delta _i^2 \\
u^\alpha \delta _i^3+v^\alpha \delta _i^4%
\end{array}
\right) \,,
\end{equation}
with $2N$ arbitrary functions $u^a\left( x\right) $ and $v^a\left( x\right) $%
, and is therefore degenerate. This particular CS theory has no physical
degrees of freedom, according to the formula (\ref{K}) applicable to regular
non-generic theories, for $n=2$ and $K=2N$.

As a consequence of the existence of both regularity and genericity issues,
the regularization problem is much more delicate in CS theories. For
example, in a \emph{pure gauge} sector, $F^{a}=0$, the constraint surface is
defined by the \emph{pure gauge} configurations $A_{i}^{a}=-D\lambda
_{i}^{a} $, and the multilinear constraints $\chi _{a}\approx 0$ can be
exchanged by the equivalent regular set $\mathcal{A}_{i}^{a}\equiv
A_{i}^{a}+D\lambda _{i}^{a}\approx 0$. In that case, all constraints $%
\left\{ \mathcal{A}_{i}^{a}\approx 0,\text{ }\phi _{a}^{i}\approx 0\right\} $
are second class and there are no physical degrees of freedom in the \emph{%
pure gauge} sector , as expected.

A more general situation occurs around the background $F^{a}=\left(
F^{r},F^{\alpha }\right) $, where only one block of the field-strength
vanishes, $F^{\alpha }=0,$ leading to the irregular constraints $\chi
_{\alpha }$. It is supposed that the rest of constraints $\chi _{r}$ are
regular. In that case, $\chi _{\alpha }\approx 0$ are exchanged by $\mathcal{%
A}_{i}^{\alpha }\equiv A_{i}^{\alpha }+D\lambda _{i}^{\alpha }\approx 0$,
and the constraints $\left\{ \mathcal{A}_{i}^{\alpha }\approx 0,\text{ }\phi
_{\alpha }^{i}\approx 0\right\} $ are second class, so that the variables $%
\left( A_{i}^{\alpha },\text{ }\pi _{\alpha }^{i}\right) $ can be eliminated
from the corresponding reduced phase space. In consequence, the dynamics of
this sector is effectively determined by the regular constraints $\left\{
\chi _{r}\approx 0,\text{ }\phi _{r}^{i}\approx 0\right\} ,$ \emph{i.e.}, by%
\emph{\ }the submatrix of the symplectic form $\Omega _{rs}^{ij}=\left\{
\phi _{r}^{i},\phi _{s}^{j}\right\} ^{\ast }$. Therefore, the problems of
irregularity and genericity are decoupled, and the whole dynamics follows
only from $\Omega _{rs}^{ij}$. The underlying reason for the decoupling of
two problems, is that regular and irregular sectors do not intersect in the
phase space, and that there is an effective symplectic form determining the
dynamics of the each sector.

Although a wide class of irregular CS sectors are of the type described
above (when the irregularity is a consequence of $F^{\alpha }=0$), there may
be another \textquotedblleft accidental\textquotedblright\ irregular
configurations, specific only for a certain CS theory, where an independent
analysis is required.

\section{Conclusions: the phase space of CS\ theories}

The dynamical structure of higher-dimensional CS theories invariant under a
Lie group with more than one generator, is complex and crucially depends on
the symplectic form $\Omega $. Since the rank of this matrix, which
determines the number and character of constraints in a theory, changes
throughout the phase space, CS theories have the following general features:

\begin{itemize}
\item There exist \emph{regular} and \emph{irregular }sectors of phase
space, where the constraints $\chi _a$ are functionally independent or not,
respectively. A system cannot spontaneously, in finite time, evolve from one
sector to another.

\item If the rank of the symplectic form is maximal, the theory is \emph{%
generic}, otherwise, it is \emph{degenerate.} This classification is
independent from the regularity, although both conditions are expressed in
terms of the same matrix $\Omega _{ab}^{ij}$.

\item During its evolution a regular system can reach a point in \emph{%
configurational} space where it is irregular, but it passes this point
without any effect. The reason is that these two sectors do not have
intersections in the \emph{phase space}.

\item During its evolution a regular system can reach the point where it is
degenerate, \emph{i.e.,} the symplectic form has lower rank. Then the system
cannot leave this sector of lower rank since it gains additional local
symmetry and looses physical degrees of freedom there.

\item In Chern-Simons theories, it is not possible to separate first from
second class constraints explicitly, for all configurations.
\end{itemize}

The features mentioned above make the dynamics of CS theories more
complicated and with rich structure.

\chapter{\emph{AdS}-Chern-Simons supergravity\label{SCS}}

It is known that gravity in 2+1\ dimensions, described by the
Einstein-Hilbert action with or without cosmological constant, is exactly
soluble and quantizable \cite{WittenCS}. This is possible because it has no
local degrees of freedom, and because its action is a CS form and can be
seen as a gauge theory for the (\emph{A})\emph{dS }or Poincar\'{e} groups.
This is an insight that motivates the study of similar theories, represented
by CS forms, in all odd dimensions \cite%
{Chamseddine'89,Chamseddine'90,Troncoso-Zanelli '98,Troncoso-Zanelli '99}%
\textbf{\ }(see Appendix \ref{AdS5}). In these theories, gravity becomes a
truly gauge theory, where the vielbein ($e^{a}$) and the spin-connection ($%
\omega ^{ab}$) are components of the same connection field $\mathbf{A},$ for
the (\emph{A})\emph{dS} or Poincar\'{e} algebra. In higher dimensions ($%
D\geq 5$), the CS actions are not equivalent to the Einstein-Hilbert
actions, as they contain terms nonlinear in curvature, and torsion is a
dynamical field \cite{Troncoso-Zanelli '00}.

In this chapter, CS supergravities based on the supersymmetric extensions of
the \emph{AdS} group, are studied. These theories contain, apart from the
gravitational fields ($e,\omega $) and the gravitini ($\psi $), a number of
the additional bosonic gauge fields. The supersymmetry algebra closes \emph{%
off shell}, without the need to introduce auxiliary fields.\ The number of
boson and fermion components are not equal \cite{Troncoso-Zanelli '99a}.

The simplest higher-dimensional CS supergravity with propagating degrees of
freedom occurs in five dimensions. An additional simplification is that the
Lagrangian of the five-dimensional theory does not contain torsion
explicitly in the purely gravitational sector, which is just a polynomial in
the curvature and vielbein.

\section{$D=5$ supergravity}

\subsubsection{\emph{a}) Algebra}

Five-dimensional \emph{AdS}-Chern-Simons supergravity is based on the
supersymmetric extension of the \emph{AdS\ }group, $SU(2,2\left\vert
N\right. )$. The superalgebra $su(2,2\left\vert N\right. )$ is generated by
the following anti-Hermitean generators \cite{Chandia-Troncoso-Zanelli}:
\begin{equation}
\begin{array}{llll}
so(2,4):\quad & \mathbf{J}_{AB}\,, & \left( A,B=0,\ldots ,5\right) \,, & (15%
\text{ generators})\,, \\
su(N): & \mathbf{T}_{\Lambda }\,, & \left( \Lambda =1,\ldots ,N^{2}-1\right)
\,, & (N^{2}-1)\,, \\
\text{SUSY}: & \mathbf{Q}_{r}^{\alpha },\;\mathbf{\bar{Q}}_{\alpha }^{r}\,,\;
& \left( \alpha =1,\ldots ,4;\;r=1,\ldots ,N\right) \,, & (8N)\,, \\
u(1): & \mathbf{G}_{1}\,, &  & (1)\mathbf{\,,}%
\end{array}
\label{super ads}
\end{equation}%
where $\eta _{AB}=$ diag $\left( -,+,+,+,+,-\right) .$ Lorentz rotations and
\emph{AdS} translations are generated by $\mathbf{J}_{ab}$ and$\;\mathbf{J}%
_{a}\equiv \mathbf{J}_{a5}\quad (a,b=0,\ldots ,4)$ respectively, and
supersymmetry (SUSY) generators transform as Dirac spinors in a vector
representation of $SU(N)$. The dimension of this superalgebra is
\begin{equation}
\mathcal{N}\left( SU(2,2\left\vert N\right. )\right) =N^{2}+8N+15\,.
\end{equation}

\subsubsection{\emph{b}) Field content}

The fundamental field is the Lie-algebra-valued connection 1-form, $\mathbf{A%
}=A^{M}\mathbf{G}_{M}$, with components
\begin{equation}
\mathbf{A}=\frac{1}{\ell }\,e^{a}\mathbf{J}_{a}+\frac{1}{2}\,\omega ^{ab}%
\mathbf{J}_{ab}+a^{\Lambda }\mathbf{T}_{\Lambda }+\left( \bar{\psi}^{r}%
\mathbf{Q}_{r}-\mathbf{\bar{Q}}^{r}\psi _{r}\right) +\phi \,\mathbf{G}_{1}%
\mathbf{\,.}  \label{connection}
\end{equation}%
Apart from the purely gravitational part $\left( \frac{1}{\ell }%
\,e^{a},\omega ^{ab}\right) $, where $\ell $ is the \emph{AdS} radius, this
theory contains a fermionic sector $\left( \psi _{r}^{\alpha },\;\bar{\psi}%
_{\alpha }^{r}\right) $ (gravitini), and bosonic fields $a^{\Lambda }$ and $%
\phi $ demanded by supersymmetry. The components of the field-strength $%
\mathbf{F}=d\mathbf{A}+\mathbf{A}^{2}=F^{M}\mathbf{G}_{M}$ along the bosonic
generators are
\begin{equation}
\begin{array}{lll}
F^{a5} & = & \frac{1}{\ell }\,T^{a}+\frac{1}{2}\,\bar{\psi}^{r}\Gamma
^{a}\psi _{r}\,,\medskip \\
F^{ab} & = & R^{ab}+\frac{1}{\ell ^{2}}\,e^{a}e^{b}-\frac{1}{2}\bar{\psi}%
^{r}\Gamma ^{ab}\psi _{r}\,,\medskip \\
F^{\Lambda } & = & \mathcal{F}^{\Lambda }+\bar{\psi}^{s}\left( \tau
^{\Lambda }\right) _{s}^{r}\psi _{r}\,,\medskip \\
F^{\phi } & = & d\phi -i\bar{\psi}^{r}\psi _{r}\,,%
\end{array}%
\end{equation}%
where $N\times N$ matrices $\tau ^{\Lambda }$ are generators of $su(N),$ and
$a\equiv a^{\Lambda }\tau _{\Lambda }$ and $\mathcal{F}=da+a^{2}$ are the
corresponding field and its field-strength, $d\phi $ is the $u(1)$
field-strength and the torsion ($T^{a}$) and curvature ($R^{ab}$) two-forms
are given in Appendix \ref{AdS5}. The component of the field-strength along
the fermionic generator $\left( -\mathbf{\bar{Q}}^{r}F_{r}\text{ term}%
\right) $, is
\begin{equation}
F_{r}=D\psi _{r}\,,
\end{equation}%
where the super covariant derivative of a $p$-form $\mathbf{X}$ is $D\mathbf{%
X}=\mathbf{X}+\left[ \mathbf{A},\mathbf{X}\right] $ (see Appendix \ref{sG}).
In particular, for the spinor $\psi _{r},$ it is
\begin{equation}
D\psi _{r}=\left( \nabla +e\hspace{-0.18cm}/\right) \psi _{r}-a_{r}^{s}\psi
_{s}+i\left( \frac{1}{4}-\frac{1}{N}\right) \phi \text{\thinspace }\psi _{r}%
\text{\thinspace },  \label{D psi}
\end{equation}%
where $\nabla \psi _{r}=\left( d+\omega \hspace{-0.26cm}/\right) \psi _{r}$
is the Lorentz covariant derivative, with $\omega \hspace{-0.26cm}/\equiv
\frac{1}{4}\,\omega ^{ab}\Gamma _{ab}$ and $e\hspace{-0.18cm}/\equiv \frac{1%
}{2\ell }\,e^{a}\Gamma _{a}.$

\subsubsection{\emph{c}) Action}

The CS Lagrangian, defined by (\ref{dL_CS}), in five dimensions becomes
\begin{equation}
dL_{\text{CS}}=ik\left\langle \mathbf{F}^{3}\right\rangle
=ik\,g_{MNK}\,F^{M}F^{N}F^{K}\,,  \label{dL 5}
\end{equation}%
where $\left\langle \mathbf{\ldots }\right\rangle $ stands for a symmetrized
invariant supertrace (symmetric in bosonic and antisymmetric in fermionic
indices) which is explicitly given in Appendix \ref{SAdS}.\footnote{%
The symmetrized supertrace $\left\langle \;\;\;\right\rangle \equiv
\left\langle \;,\;,\;\right\rangle $ has three Lie-algebra-valued entries.
For example, $\left\langle \mathbf{A}^{3}d\mathbf{A}\right\rangle \equiv
\left\langle \mathbf{A}^{2},\mathbf{A},d\mathbf{A}\right\rangle
=\left\langle \mathbf{A,A}^{2},d\mathbf{A}\right\rangle .$} The constant $k$
is dimensionless and real, and antisymmetric wedge product acts between the
forms. The CS action can be explicitly written as
\begin{equation}
I_{\text{CS}}\left[ A\right] =\int\limits_{\mathcal{M}}L_{\text{CS}}(\mathbf{%
A})=ik\int\limits_{\mathcal{M}}\left\langle \mathbf{A}\left( d\mathbf{A}%
\right) ^{2}+\frac{3}{2}\,\mathbf{A}^{3}d\mathbf{A}+\frac{3}{5}\,\mathbf{A}%
^{5}\right\rangle +B\left[ A\right] \,,  \label{I5}
\end{equation}%
where $B\left[ A\right] $ is a boundary term which must be added so that the
action is stationary on the classical orbits. In components, the CS\
supergravity action takes the form originally obtained by Chamseddine \cite%
{Chamseddine'90},%
\begin{equation}
L_{\text{CS}}=L_{\text{g}}\left( \omega ,e\right) +L_{\mathtt{SU(N)}}\left(
a\right) +L_{\mathtt{U(1)}}\left( \omega ,e,\phi \right) +L_{\mathtt{f}%
}\left( \omega ,e,a,\phi ,\psi \right) \,,
\end{equation}%
where
\begin{equation}
\begin{array}{lll}
L_{\text{g}} & = & \frac{k}{8}\,\varepsilon _{abcde}\,\left( \frac{1}{\ell }%
\,R^{ab}R^{cd}e^{e}+\frac{2}{3\ell ^{3}}\,R^{ab}e^{c}e^{d}e^{e}+\frac{1}{%
5\ell ^{5}}\,e^{a}e^{b}e^{c}e^{d}e^{e}\right) \,,\medskip \\
L_{\mathtt{SU(N)}} & = & -ik\,\text{Tr}_{N}\left( \,a\mathcal{F}^{2}-\frac{1%
}{2}\,a^{3}\mathcal{F}+\frac{1}{10}\,a^{5}\right) \,,\medskip \\
L_{\mathtt{U(1)}} & = & k\,\left( \frac{1}{4^{2}}-\frac{1}{N^{2}}\right)
\phi \left( d\phi \right) ^{2}+\frac{3k}{4\ell ^{2}}\,\left( T^{a}T_{a}-%
\frac{\ell ^{2}}{2}\,R^{ab}R_{ab}-R^{ab}e_{a}e_{b}\right) \phi +\frac{3k}{N}%
\,\mathcal{F}^{\Lambda }\mathcal{F}_{\Lambda }\phi \,,\medskip \\
L_{\mathtt{f}} & = & -\frac{3ik}{4}\,\bar{\psi}^{r}\left[ \frac{1}{\ell }%
\,T^{a}\Gamma _{a}+\frac{1}{2}\,\left( R^{ab}+e^{a}e^{b}\right) \Gamma
_{ab}+2i\left( \frac{1}{N}+\frac{1}{4}\right) \,d\phi -\bar{\psi}^{s}\psi
_{s}\right] D\psi _{r}\medskip \\
&  & -\frac{3ik}{2}\bar{\psi}^{r}\left( \mathcal{F}_{r}^{s}-\frac{1}{2}\,%
\bar{\psi}^{s}\psi _{r}\right) D\psi _{s}+\text{c.c.}\,,%
\end{array}%
\end{equation}%
and $\mathcal{F}_{r}^{s}=\mathcal{F}^{\Lambda }\left( \tau _{\Lambda
}\right) _{r}^{s}.$ On the basis of the form of $D\psi _{r}$ given by (\ref%
{D psi}), one can see that the fermions carry a $U(1)$ charge $q=\frac{1}{4}-%
\frac{1}{N}$. The pure gravitational Lagrangian $L_{\text{g}}$ contains the
standard Einstein-Hilbert Lagrangian with negative cosmological constant,
plus an additional term quadratic in curvature which is a dimensional
continuation of the Gauss-Bonnet density from four to five dimensions.

From the Lagrangian $L_{\mathtt{U(1)}}$ and the relation (\ref{D psi}) it
can be seen that the case $N=4$ is exceptional since the dynamics of $\phi $
changes (it looses the cubic term in the Lagrangian) and fermions become
neutral ($q=0$). It is shown in the Appendix \ref{SAdS} that for $N=4$, the $%
U(1)$ generator becomes a central extension.

\subsubsection{\emph{d}) Local symmetries}

As all CS theories, the \emph{AdS}-CS supergravity action is invariant under
diffeomorphisms, $\delta x^{\mu }=\xi ^{\mu }\left( x\right) $, $\delta
_{\xi }\mathbf{A}=-\pounds _{\xi }\mathbf{A}\,,\,$ and infinitesimal gauge
transformations,\emph{\ }$\delta _{\lambda }\mathbf{A}=-D\mathbf{\lambda }$
(see the Section \ref{CSI}). In particular, under local supersymmetry
transformations with parameter $\epsilon _{r}^{\alpha }\left( x\right) ,$
from $\mathbf{\lambda }=\bar{\epsilon}^{r}\mathbf{Q}_{r}-\mathbf{\bar{Q}}%
^{r}\epsilon _{r}$ one obtains
\begin{equation}
\begin{array}{llllll}
\delta _{\epsilon }e^{a} & = & -\frac{1}{2}\,\left( \bar{\psi}^{r}\Gamma
^{a}\epsilon _{r}-\bar{\epsilon}^{r}\Gamma ^{a}\psi _{r}\right) \,,\qquad
\qquad \medskip & \delta _{\epsilon }\psi _{\alpha }^{r} & = & -D\epsilon
_{\alpha }^{r}\,, \\
\delta _{\epsilon }\omega ^{ab} & = & \frac{1}{4}\,\left( \bar{\psi}%
^{r}\Gamma ^{ab}\epsilon _{r}-\bar{\epsilon}^{r}\Gamma ^{ab}\psi _{r}\right)
\,,\medskip & \delta _{\epsilon }\bar{\psi}_{r}^{\alpha } & = & -D\bar{%
\epsilon}_{r}^{\alpha }\,, \\
\delta _{\epsilon }a^{\Lambda } & = & \bar{\psi}^{r}\left( \tau ^{\Lambda
}\right) _{r}^{s}\epsilon _{s}-\bar{\epsilon}\left( \tau ^{\Lambda }\right)
_{r}^{s}\psi _{s}\,, & \delta _{\epsilon }\phi & = & -i\left( \bar{\psi}%
^{r}\epsilon _{r}-\bar{\epsilon}\psi _{r}\right) \,.%
\end{array}%
\end{equation}%
Unlike standard supergravities, here the supersymmetry algebra closes \emph{%
off shell }by construction, without requiring auxiliary fields \cite%
{Troncoso-Zanelli '99a}.

The CS action is not invariant under \emph{large} gauge transformations
\begin{equation}
\mathbf{A}^{g}=g\left( \mathbf{A}+d\right) g^{-1}\,,\qquad \qquad g\in
SU\left( 2,2\left\vert N\right. \right) \,,
\end{equation}%
where $g=e^{\mathbf{\lambda }}$. Using the fact that the field-strength
transforms homogenously, $\mathbf{F}^{g}=g\mathbf{F}g^{-1}$, leading to $dL_{%
\mathtt{CS}}^{g}=dL_{\mathtt{CS}},$ one obtains that the Lagrangian changes
as $L_{\mathtt{CS}}^{g}=L_{\mathtt{CS}}+\omega $, where $\omega $ is a
closed form ($d\omega =0$) which is not exact for nontrivial topology.

\subsubsection{\emph{e}) Field equations}

Varying (\ref{I5}) with respect to the connection, yields
\begin{equation}
\delta I_{\text{CS}}=3ik\int\limits_{\mathcal{M}}\left\langle \mathbf{F}%
^{2}\,\delta \mathbf{A}\right\rangle \,,  \label{var I}
\end{equation}%
provided $\delta B$ and the boundary conditions are chosen so that the
boundary term in $\delta I_{\text{CS}}$ vanishes. Then the action has an
extremum if the equations of motion are satisfied,
\begin{equation}
\left\langle \mathbf{G}_{M}\mathbf{F}^{2}\right\rangle
=g_{MNK}F^{N}F^{K}=0\,.
\end{equation}%
One solution of these equations is a connection which is locally flat, a
\emph{pure gauge} field, $\mathbf{A}=gdg^{-1}$. The theory with
configurations in the sector around a \emph{pure gauge} field has no
propagating degrees of freedom and the entire dynamics is contained in the
nontrivial topology, but in that case the theory is irregular. In general,
there are physical degrees of freedom in the bulk, as well as at the
boundary.

The asymptotic dynamics is sensitive to the choice of boundary conditions
and the topology of $\partial \mathcal{M}$. The boundary conditions must be
chosen so that $\delta B$ can be integrated to get $B$, or, if $B=0$, so
that $\delta I_{\text{CS}}$ does not contain additional boundary terms in (%
\ref{var I}). Here it is supposed that the adequate boundary conditions
exist, and they will be discussed later, in connection with the conserved
charges.

\section{Conserved charges}

In Chapter 5, the Hamiltonian dynamics of CS theories in $D\geq 5$ was
analyzed. It was shown that, in regular sectors of phase space, gauge
invariance is expressed by the presence of first class constraints $\mathcal{%
G}_{M}$ which generate gauge transformations. There is also diffeomorphism
invariance in these theories. Spatial diffeomorphisms are generated by first
class constraints $\mathcal{H}_{i}$, while time-like diffeomorphisms are not
independent, but realized as \emph{on-shell} symmetries. Generic sectors of
these theories do not have additional independent local symmetries. The
symplectic form, which defines the kinetic term in CS theories, was also
discussed. It was emphasized that it can change its rank throughout
configuration space. Depending on the rank of the symplectic form, which in
general depends on the background, CS theories can be either regular or
irregular, generic or degenerate.

The study of the boundary dynamics has been intentionally left out of the
analysis. In 2+1 dimensions, where there are no locally propagating degrees
of freedom, the boundary dynamics is purely topological. In higher
dimensions, besides purely topological degrees of freedom, there also exist
local ones. In this chapter, the boundary dynamics is analyzed (see \cite%
{Banados-Garay-Henneaux2} for a review of the bosonic case). From now on,
only \emph{regular} and \emph{generic }CS theories are considered, and they
in general have local degrees of freedom.

In the Hamiltonian formalism it is assumed that space-time is $\mathcal{M}%
\simeq \mathbb{R}\times \sigma $, where $\mathbb{\sigma }$ is an Euclidean
manifold. The PB of the canonical fields $\left( A_{i}^{M},\pi
_{M}^{i}\right) $ on the phase space $\Gamma $ is given by
\begin{equation}
\left\{ \pi _{M}^{i},A_{j}^{N}\right\} =-\delta _{j}^{i}\delta
_{M}^{N}\,\delta =-\left( -\right) ^{\varepsilon _{M}}\left\{ A_{j}^{N},\pi
_{M}^{i}\right\} \,,
\end{equation}%
where $\delta $ denotes the Dirac's $\delta $-function at the spatial
section, $A_{0}^{M}$ is a Lagrange multiplier, and the number
$\varepsilon_{M}=0,1$ (mod $2$) is the Grassmann parity of
$A_{\mu }^{M}$ and $\pi
_{M}^{\mu }$ (see Appendix \ref{sG} for the conventions). The Hamiltonian
for the action (\ref{I5}) is given by
\begin{equation}
H_{T}=\int d^{4}x\,\left( A_{0}^{M}\mathcal{G}_{M}+u_{i}^{M}\phi
_{M}^{i}\right) \,,
\end{equation}%
where boundary terms have been neglected for the moment. The constraints are
given by
\begin{eqnarray}
\phi _{M}^{i} &=&\pi _{M}^{i}-\mathcal{L}_{M}^{i}\approx 0\,,  \notag \\
\mathcal{G}_{M} &=&-\chi _{M}+D_{i}\phi _{M}^{i}\approx 0\,,
\end{eqnarray}%
where the covariant derivative acts on $\phi _{M}^{i}$ as $D_{i}\phi
_{M}^{i}=\partial _{i}\phi _{M}^{i}+f_{MN}^{\quad \;K}A_{i}^{K}\phi _{M}^{i}$
and, for $D=5$,
\begin{eqnarray}
\mathcal{L}_{M}^{i} &=&ik\,\varepsilon ^{ijkl}\,\left(
g_{MNK}\,F_{jk}^{N}A_{l}^{K}-\frac{1}{4}\,g_{MNL}\,f_{KS}^{\quad
\;L}A_{j}^{N}A_{k}^{K}A_{l}^{S}\right) \,,  \notag \\
\chi _{M} &=&-\frac{3ik}{4}\,\varepsilon
^{ijkl}g_{MNK}\,F_{ij}^{N}F_{kl}^{K}\approx 0\,.
\end{eqnarray}%
The symplectic form defining the kinetic term in the action is a function on
phase space
\begin{equation}
\Omega _{MN}^{ij}=-3ik\,\varepsilon ^{ijkl}g_{MNK}\,F_{kl}^{K}\,,
\end{equation}%
whose rank can vary throughout $\Gamma $. In the regular and generic
sectors, the action does not have other independent symmetries apart from
spatial diffeomorphisms and gauge transformations, and the constraints
satisfy the Poisson brackets algebra
\begin{eqnarray}
\left\{ \mathcal{G}_{M},\mathcal{G}_{N}\right\} &=&f_{MN}^{\quad \;K}\,%
\mathcal{G}_{K}\,\delta \,,  \notag \\
\left\{ \mathcal{G}_{M},\phi _{N}^{i}\right\} &=&f_{MN}^{\quad \;K}\,\phi
_{K}^{i}\,\delta \,, \\
\left\{ \phi _{M}^{i},\phi _{N}^{j}\right\} &=&\Omega _{MN}^{ij}\,\delta \,,
\notag
\end{eqnarray}%
where $\mathcal{G}_{M}$ are first class constraints (generators of gauge
transformations), while among $\phi $'s there are four first class
constraints (generators of spatial diffeomorphisms) and the rest are second
class constraints. The number of locally propagating degrees of freedom in
this theory, according to (\ref{degrees of freedom}), is
\begin{equation}
f_{5}\left( N\right) =N^{2}+8N+13\,.
\end{equation}

\subparagraph{Regular generic background.}

In the following, a class of backgrounds (solutions of the constraints) is
chosen so that they provide a regular and generic theory. They also allow
separating first and second class constraints among $\phi $'s, which is in
general a difficult task. However, in the case of $N=4,$ the invariant
tensor takes the same form as invariant tensor of $g\otimes u\left( 1\right)
$. Explicitly, the supersymmetric algebra $su(2,2\left\vert 4\right. )$ is
generated by the $u(1)$ generator $\mathbf{G}_{1}$ and the $%
psu(2,2\left\vert 4\right. )$ generators $\mathbf{G}_{M^{\prime }}=\left(
\mathbf{J}_{AB},\mathbf{T}_{\Lambda },\mathbf{Q}_{r}^{\alpha },\mathbf{\bar{Q%
}}_{\alpha }^{r}\right) ,$ so that the invariant tensor $g_{MNK}\,,$
decomposed as $\mathbf{G}_{M}\rightarrow \left( \mathbf{G}_{M^{\prime }},%
\mathbf{G}_{1}\right) ,$ takes the simpler form (see Appendix \ref{SAdS}):
\begin{equation}
g_{MNK}\quad \rightarrow \quad \left\{ g_{M^{\prime }N^{\prime }K^{\prime
}}\,,\quad g_{M^{\prime }N^{\prime }1}=-\frac{i}{4}\,\gamma _{M^{\prime
}N^{\prime }}\,,\quad g_{M^{\prime }11}=0\,,\quad g_{111}=0\right\} \,.
\end{equation}%
Here $\gamma _{M^{\prime }N^{\prime }}$ is the invertible Killing metric of $%
PSU(2,2\left\vert 4\right. )$.$\medskip $

\subparagraph{(\emph{i})\emph{\ Regular background. }}

The background has to be such that the Jacobian matrix
\begin{equation}
J_{MN}=-6ik\,g_{MNK}\,F^{K}=\left(
\begin{array}{cc}
J_{M^{\prime }N^{\prime }} & \frac{3}{2}\,k\,\gamma _{N^{\prime }K^{\prime
}}\,F^{K^{\prime }} \\
\frac{3}{2}\,k\,\gamma _{M^{\prime }K^{\prime }}\,F^{K^{\prime }} & 0%
\end{array}%
\right)
\end{equation}%
is invertible. (From now on, the forms are defined on the spatial section.)
When the submatrix $J_{M^{\prime }N^{\prime }}$ is invertible, the
regularity conditions require that $F^{K^{\prime }}$ be non-zero, for at
least one $K^{\prime }$. The simplest solution occurs when the fermionic
field vanishes ($\psi _{r}^{\alpha }=0$) and the space-time is locally \emph{%
AdS} ($F^{AB}=0$), while the $su(4)$ field-strength $\mathcal{F}^{\Lambda }$
has a component only along $dx^{1}dx^{2}$,
\begin{equation}
\mathcal{F}^{\Lambda }=\mathcal{F}_{12}^{\Lambda }\,dx^{1}dx^{2}\neq 0\,.
\label{solution 1}
\end{equation}%
This configuration is on the constraint surface if the $u(1)$ field $\phi $
has a field-strength satisfying $f_{34}=0,\ $while the remaining components $%
f_{ij}=\partial _{i}\phi _{j}-\partial _{j}\phi _{i}$ are arbitrary.$%
\medskip $

\subparagraph{\textbf{(}\emph{ii})\emph{\ Generic background. }}

The background is generic if the symplectic form
\begin{equation}
\mathbf{\Omega }=\Omega _{MN}^{ij}=\left(
\begin{array}{cc}
\Omega _{M^{\prime }N^{\prime }}^{ij} & \Omega _{M^{\prime }1}^{ij} \\
\Omega _{N^{\prime }1}^{ij} & 0%
\end{array}%
\right)
\end{equation}%
has maximal rank. Since $\mathbf{\Omega }$ has always four zero modes
\begin{equation}
\Omega _{MN}^{ij}F_{jk}^{N}=-\delta _{k}^{i}\chi _{M}\approx 0\,,
\label{4 modes}
\end{equation}%
the maximal rank is $\Re \left( \mathbf{\Omega }\right) =$ $4\left( \mathcal{%
N}-1\right) $. This is satisfied if the following two conditions are
fulfilled:
\begin{equation}
\begin{array}{llll}
\text{(I)}\qquad & \Omega _{M^{\prime }N^{\prime }}^{ij} & = & \text{%
non-degenerate,\smallskip } \\
\text{(II)} & \det f_{ij} & \neq & 0\qquad (f_{34}=0)\,.%
\end{array}
\label{solution 2}
\end{equation}%
Consequently, the inverse $\Delta _{ij}^{M^{\prime }N^{\prime }}$ and $%
f^{ij} $ exist,
\begin{equation}
\Delta _{ik}^{M^{\prime }K^{\prime }}\,\Omega _{K^{\prime }N^{\prime
}}^{kj}=\delta _{i}^{j}\delta _{N^{\prime }}^{M^{\prime }}\,,\qquad \qquad
f_{ij}\,f^{jk}=\delta _{k}^{i}\,,
\end{equation}%
and the rank of the symplectic form is:\footnote{%
If the symplectic matrix $\Omega _{MN}^{ij}$ has invertible submatrix $%
\Omega _{M^{\prime }N^{\prime }}^{ij}$, then its rank is $\Re \left( \Omega
_{MN}^{ij}\right) =\Re \left( \Omega _{M^{\prime }N^{\prime }}^{ij}\right)
+\Re \left( M^{ij}\right) $, where $M^{ij}\equiv \Omega _{M^{\prime
}1}^{ik}\,\Delta _{kl}^{M^{\prime }N^{\prime }}\Omega _{N^{\prime }1}^{lj}$
(see the footnote at p. \pageref{sum}). However, on the basis of the
identity (\ref{4 modes}), the matrix $M^{ij}$ weakly vanishes,
\begin{equation*}
M^{ij}=f^{ij}\chi _{1}-\Omega _{M^{\prime }1}^{ik}\,\Delta _{kl}^{M^{\prime
}N^{\prime }}\,f^{lj}\chi _{N^{\prime }}\approx 0\,,
\end{equation*}%
and therefore it has zero rank on the constraint surface.}
\begin{equation}
\Re \left( \Omega _{MN}^{ij}\right) =\Re \left( \Omega _{M^{\prime
}N^{\prime }}^{ij}\right) =4\left( \mathcal{N}-1\right) \,.
\label{rank omega}
\end{equation}

\subparagraph{\textbf{(}\emph{iii})\emph{\ First and second class
constraints. }}

Since the submatrix $\Omega _{M^{\prime }N^{\prime }}^{ij}$ is invertible,
it is possible to separate the first and second class constraints. The
second class constraints are $\phi _{M^{\prime }}^{i}$, while
\begin{equation}
\tilde{\phi}_{1}^{i}=\phi _{1}^{i}-\Delta _{jk}^{M^{\prime }N^{\prime
}}\Omega _{N^{\prime }1}^{ki}\phi _{M^{\prime }}^{j}\approx 0
\end{equation}%
are first class constraints related to the generators of spatial
diffeomorphisms,
\begin{equation}
\mathcal{H}_{i}\equiv f_{ij}\tilde{\phi}_{1}^{j}=F_{ij}^{M}\phi
_{M}^{j}\approx 0\,.
\end{equation}

In general, one could introduce Dirac brackets and eliminate unphysical
degrees of freedom coming from the second class constraints. But it is more
convenient not to do so in order to maintain explicitly covariant
expressions.

\subparagraph{Improved generators.}

The generators of gauge symmetry are given by
\begin{equation}
G\left[ \lambda \right] =\int d^{4}x\,\lambda ^{M}\mathcal{G}_{M}=\int
d^{4}x\,\lambda ^{M}\left( -\chi _{M}+D_{i}\phi _{M}^{i}\right) \,,
\label{G}
\end{equation}%
and their action on phase space functions $F$ is
\begin{equation}
\delta _{\lambda }F=\left\{ F,G\left[ \lambda \right] \right\} =\left(
-\right) ^{\varepsilon _{F}\varepsilon _{M}}\int d^{4}x\,\lambda ^{M}\left\{
F,\mathcal{G}_{M}\right\} \,.
\end{equation}%
The generators can be made to have local functional derivatives. This means
that the variation of the generator $G\left[ \lambda ,z\right] $ takes the
form $\int dx\,\delta z^{A}\frac{\delta G}{\delta z^{A}},$ without
derivatives $\partial _{j}\left( \delta z^{A}\right) $ which would give rise
to boundary terms. The generators (\ref{G}), however, vary as
\begin{equation}
\delta G\left[ \lambda \right] =\delta G_{Q}\left[ \lambda \right] -\delta Q%
\left[ \lambda \right] \,,  \label{delta G}
\end{equation}%
where $\delta G_{Q}\left[ \lambda \right] $ is the bulk term and $\delta Q%
\left[ \lambda \right] $ is a boundary term. The generators of local
transformations should be the so-called \emph{improved generators},\emph{\ }%
which differ from the original ones by boundary terms (the Regge-Teitelboim
approach \cite{Regge-Teitelboim}),
\begin{equation}
G_{Q}\left[ \lambda \right] \equiv G\left[ \lambda \right] +Q\left[ \lambda %
\right] \,,
\end{equation}%
such that their functional derivatives are local functions. In order find
the explicit expressions for the improved generators, it is more convenient
to rewrite the original generators (\ref{G}) as
\begin{equation}
G\left[ \lambda \right] =\int\limits_{\sigma }\left\langle \mathbf{\lambda }%
\left( -\mathbf{K}+D\mathbf{\Phi }\right) \right\rangle \,,  \label{G_}
\end{equation}%
where the 4-form $\mathbf{K}$ ($K_{Mijkl}=\chi _{M}\,\varepsilon _{ijkl}$),
and the 3-form $\mathbf{\Phi }$ ($\Phi _{Mjkl}=\phi _{M}^{i}\,\varepsilon
_{ijkl}$) are defined on the 4-dimensional spatial manifold $\sigma $ by
\begin{eqnarray}
\mathbf{K} &\equiv &-3ik\,\mathbf{F}^{2}\approx 0\,,  \notag \\
\mathbf{\Phi } &\equiv &\mathbf{\Pi }-ik\mathbf{\,}\left( \left\{ \mathbf{%
\mathbf{A,F}}\right\} -\frac{1}{2}\mathbf{\,\mathbf{A}}^{3}\right) \approx
0\,.
\end{eqnarray}%
Here $\Pi _{Mjkl}=\pi _{M}^{i}\,\varepsilon _{ijkl}.$ Then, the generators (%
\ref{G_}) vary as
\begin{equation}
\delta G\left[ \lambda \right] =\int\limits_{\sigma }\left[ -\left\langle
\mathbf{\lambda }\delta \mathbf{K}\right\rangle +\left\langle \mathbf{%
\lambda }D\delta \mathbf{\Phi }\right\rangle +\left\langle \left[ \mathbf{%
\lambda ,\Phi }\right] \delta \mathbf{A}\right\rangle \right] \,,
\label{var G}
\end{equation}%
with the following variations of the constraints:
\begin{eqnarray}
\delta \mathbf{K} &=&-3ik\,\left\{ \mathbf{F,}D\delta \mathbf{A}\right\} \,,
\notag \\
\delta \mathbf{\Phi } &=&\delta \mathbf{\Pi }-ik\,\left\{ \mathbf{F,}\delta
\mathbf{\mathbf{A}}\right\} -ik\,\left\{ \mathbf{A,}D\delta \mathbf{A}%
\right\} +\frac{ik}{2}\,\delta \mathbf{A}^{3}.
\end{eqnarray}%
Therefore, the bulk term in (\ref{var G}) has the form
\begin{equation}
\delta G_{Q}\left[ \lambda \right] =\int\limits_{\sigma }\left\langle \delta
\mathbf{AX}\left( \lambda \right) +\delta \mathbf{\Pi }D\mathbf{\mathbf{%
\lambda }}\right\rangle \,,  \label{delta GQ}
\end{equation}%
where the 3-form $\mathbf{X}$ is given by the expression
\begin{equation}
\mathbf{X}\left( \lambda \right) =ik\left\{ D\mathbf{\lambda ,F}\right\} -%
\frac{3ik}{2}\left\{ D\mathbf{\lambda ,A}^{2}\right\} +ik\left\{ \left[
\mathbf{F},\mathbf{\lambda }\right] ,\mathbf{A}\right\} -\left[ \mathbf{%
\lambda },\mathbf{\Phi }\right] \,.  \label{X}
\end{equation}%
From Eqs. (\ref{delta GQ}) and (\ref{X}), the functional derivatives of $%
G_{Q}$ are
\begin{eqnarray}
\frac{\delta G_{Q}\left[ \lambda \right] }{\delta A_{i}^{M}} &=&\frac{1}{3!}%
\,\varepsilon ^{ijkl}X_{Mjkl}\left( \lambda \right) \,,  \notag \\
\frac{\delta G_{Q}\left[ \lambda \right] }{\delta \pi _{M}^{i}} &=&-\left(
-\right) ^{\varepsilon _{M}}D_{i}\lambda ^{M}\,.  \label{f derivatives}
\end{eqnarray}%
Thus, the improved generators indeed generate local gauge transformations,%
\begin{equation}
\delta _{\lambda }A_{i}^{M}=\left\{ A_{i}^{M},G_{Q}\left[ \lambda \right]
\right\} =\left( -\right) ^{\varepsilon _{M}}\frac{\delta G_{Q}\left[
\lambda \right] }{\delta \pi _{M}^{i}}=-D_{i}\lambda ^{M}\,,
\end{equation}%
and it can also be shown that momenta transform as%
\begin{equation}
\delta _{\lambda }\pi _{M}^{i}=\left\{ \pi _{M}^{i},G_{Q}\left[ \lambda %
\right] \right\} =-\frac{\delta G_{Q}\left[ \lambda \right] }{\delta
A_{i}^{M}}=-\frac{1}{3!}\,\varepsilon ^{ijkl}X_{Mjkl}\left( \lambda \right)
\,.
\end{equation}%
The improved generators, however, are not constraints but take the value $%
G_{Q}\approx Q$ on the constraint surface. $Q\left[ \lambda \right] $
generates gauge transformations at the boundary and is called the \emph{%
charge}.

\subparagraph{Conserved charges.}

The boundary term in (\ref{var G}), denoted by $-\delta Q\left[ \lambda %
\right] $, is
\begin{equation}
\delta Q\left[ \lambda \right] =-6ik\int\limits_{\partial \sigma
}\left\langle \mathbf{\lambda F}\delta \mathbf{A}\right\rangle
-2ik\int\limits_{\partial \sigma }\left\langle D\mathbf{\lambda A}\delta
\mathbf{\mathbf{A}}\right\rangle -\int\limits_{\partial \sigma }\left\langle
\mathbf{\lambda }\delta \mathbf{\Phi }\right\rangle \,.  \label{delta Q}
\end{equation}%
This expression can be integrated out provided the connection is fixed at
the boundary,
\begin{equation}
\mathbf{A\;\longrightarrow \;\bar{A}}\,,\qquad \text{at }\mathbf{\;}\partial
\sigma ,  \label{ac}
\end{equation}%
where $\mathbf{\bar{A}}$ is a regular generic configuration. The choice of
boundary conditions is not unique, and (\ref{ac}) is the simplest one which
still gives a non trivial asymptotic dynamics. More general possibility is
to solve $Q$ from (\ref{delta Q}) without fixing of all components of the
connection at the boundary.

Using (\ref{delta Q}) and the boundary conditions (\ref{ac}), the charge is
obtained as
\begin{equation}
Q\left[ \lambda \right] =-6ik\int\limits_{\partial \sigma }\left\langle
\mathbf{\lambda \bar{F}A}\right\rangle -2ik\int\limits_{\partial \sigma
}\left\langle \bar{D}\mathbf{\lambda \bar{A}\mathbf{A}}\right\rangle
-\int\limits_{\partial \sigma }\left\langle \mathbf{\lambda \Phi }%
\right\rangle \,.  \label{Q}
\end{equation}%
The charge can be explicitly written as
\begin{equation}
Q\left[ \lambda \right] \approx -2ik\int\limits_{\partial \sigma
}g_{MNK}\,\left( 3\lambda ^{M}\bar{F}^{N}+\bar{D}\lambda ^{M}\bar{A}%
^{N}\right) \,A^{K},  \label{Q expl}
\end{equation}%
where the term proportional to the constraints $\mathbf{\Phi }$ vanishes
\emph{on-shell,} and it does not give contributions to the charge. The
second term at the \emph{r.h.s. }of (\ref{Q}), comes from the variation of
the constraints $\phi _{M}^{i}=\left( \phi _{1}^{i},\phi _{M^{\prime
}}^{i}\right) $, where $\phi _{M^{\prime }}^{i}$ are second class. On the
other hand, one should not expect to have a contribution of the second class
constraints, which naturally do not appear in the gauge generator (\ref{G_})
on the reduced phase space, where $\phi _{M^{\prime }}^{i}=0$, what can be
explicitly provided by introducing the appropriate Dirac brackets. Both
approaches should be equivalent, because the second class constraints do not
generate gauge transformations, and there are no conserved charges
associated to them.

One can see that the term $\left\langle \bar{D}\mathbf{\alpha ,\bar{A},%
\mathbf{A}}\right\rangle $ vanishes for the local $PSU\left( 2,2\left\vert
4\right. \right) $ parameter $\mathbf{\alpha }\equiv \alpha ^{M^{\prime }}%
\mathbf{G}_{M^{\prime }}$, once the asymptotic conditions are used.
Evaluated at the background ($\mathbf{A}=\mathbf{\bar{A}}$), this term is
indeed zero, due to the identity $g_{M^{\prime }NK}\,\bar{A}^{N}\bar{A}%
^{K}\equiv 0$. The fields asymptotically tend to the background, so that the
connection behaves as $\mathbf{A}\sim \mathbf{\bar{A}}+\Delta \mathbf{A}$,
while the local parameter is $\mathbf{\alpha \sim \bar{\alpha}+}\Delta
\mathbf{\alpha }$, where $\mathbf{\bar{\alpha}}$ is a covariantly constant
vector ($\bar{D}\mathbf{\bar{\alpha}}=0$) which describes symmetries of the
vacuum at the boundary. Then the term
\begin{equation}
\left\langle \bar{D}\mathbf{\alpha ,\bar{A},\mathbf{A}}\right\rangle \sim
\left\langle \bar{D}\Delta \mathbf{\alpha },\mathbf{\bar{A},}\Delta \mathbf{%
\mathbf{A}}\right\rangle \,,
\end{equation}%
gives a contribution of second order, since $\Delta \mathbf{A}$ and $\Delta
\mathbf{\alpha }$ are subleading compared to $\mathbf{\bar{A}}$ and $\mathbf{%
\bar{\alpha}}$, in the limit in which the boundary is taken to infinity.

\subparagraph{Charge algebra.}

The PB algebra of improved gauge generators can be found directly from the
definition of the functional derivatives (\ref{f derivatives}), and it has
the form
\begin{equation}
\left\{ G_{Q}\left[ \lambda \right] ,G_{Q}\left[ \eta \right] \right\}
=\int\limits_{\sigma }\left\langle \mathbf{X}\left( \lambda \right) D\mathbf{%
\eta }-X\left( \eta \right) D\mathbf{\lambda }\right\rangle \,,  \label{GG}
\end{equation}%
where the expression for $\mathbf{X}$ is given by (\ref{X}). This algebra
closes, and it has the general form%
\begin{equation}
\left\{ G_{Q}\left[ \lambda \right] ,G_{Q}\left[ \eta \right] \right\} =G_{Q}%
\left[ \left[ \lambda ,\eta \right] \right] +C\left[ \lambda ,\eta \right]
\,,  \label{GG=G+C}
\end{equation}%
where $\left[ \lambda ,\eta \right] ^{M}=-f_{KN}^{\quad \;M}\lambda ^{N}\eta
^{K}$, and $C\left[ \lambda ,\eta \right] $ is a boundary term. Thus, the
classical algebra acquires a central extension $C$, called the \emph{central
charge}, which emerges as a consequence of working with the improved
generators. In order to calculate $C$, rather than starting from (\ref{GG}),
it is more convenient to find it from the gauge transformation of the charge
$Q$,
\begin{equation}
\delta _{\eta }Q\left[ \lambda \right] =\left\{ Q\left[ \lambda \right] ,Q%
\left[ \eta \right] \right\} =Q\left[ \left[ \lambda ,\eta \right] \right] +C%
\left[ \lambda ,\eta \right] \,.  \label{algebra Q}
\end{equation}%
On the basis of the Brown-Henneaux theorem \cite{Brown-HenneauxQ}, the
central extension obtained from the gauge transformation of the charge, Eq. (%
\ref{algebra Q}), is the same as the central charge in the algebra of
improved generators, Eq. (\ref{GG=G+C}), evaluated on the background. This
follows from the fact that the charge algebra (\ref{algebra Q}) is valid
only on the reduced phase space, after gauge fixing of all first class
constraints $\mathcal{G}_{M}$, so that the original generators $G\left[
\lambda \right] \ $are all strongly zero.

The central charge can be evaluated as follows. In the class of generic
regular configurations $\bar{A}$ given by (\ref{solution 1}) and (\ref%
{solution 2}), the vacuum $\bar{A}_v$ is the one for which the charges
vanish, $\bar{Q}\left[ \lambda \right] \equiv Q\left[ \lambda \right] _{\bar{%
A}_v}=0$. Then, the central charges are obtained as $C\left[ \lambda ,\eta %
\right] =\delta _\eta Q\left[ \lambda \right] _{\bar{A}_v}.$ This will be
calculated in the next section.

\section{Killing spinors and BPS states}

In supergravity theories, the anticommutation relation are of the form $%
\left\{ Q,Q^{\dagger }\right\} \sim P+J+\cdots $, giving that the sum of the
total charges is bounded from below, since it is proportional to $\sum
\left( QQ^{\dagger }+QQ^{\dagger }\right) \geq 0$ (Bogomol'nyi bound \cite%
{Bogomolnyi}). In standard supergravity theories it leads to the positivity
of energy \cite{Witten(E),Deser-Teitelboim,Grisaru(E)}. When the bound is
saturated, the corresponding states, the so-called BPS states, have some
unbroken supersymmetries. Therefore, the existence of BPS\ states is
important for the stability of the theory. In what follows, a BPS state will
be constructed which is also regular and generic.

For the bosonic BPS states $\psi _r^\alpha =0$, and the supersymmetry
transformations $\delta _\epsilon \psi _r^\alpha =-D\epsilon _r^\alpha $
leave this condition invariant if the local fermionic parameter satisfies
\begin{equation}
D\epsilon _r^\alpha =0\,.  \label{Killing eq.}
\end{equation}
This is the Killing equation for the spinor $\epsilon _r^\alpha ,$ and its
solutions are Killing spinors.

Assuming a generic regular configuration, conditions (\ref{solution 1}) and (%
\ref{solution 2}) are satisfied if:
\begin{equation}
\begin{array}{llllll}
\psi _{r} & = & 0\,,\qquad \qquad & \mathcal{F}^{\Lambda } & = & \mathcal{F}%
_{12}^{\Lambda }\,drd\varphi ^{2}\neq 0\,, \\
F^{AB} & = & 0\,, & \det \,f_{ij} & \neq & 0\,,\qquad f_{34}=0\,.%
\end{array}
\label{Gen Reg}
\end{equation}%
The local coordinates on $\mathcal{M}$ are chosen as $x^{\mu }=\left(
t,r,x^{n}\right) $, where $x^{n}$ $\left( n=2,3,4\right) $ parametrize the
boundary $\partial \sigma ,$ placed at the infinity of the radial coordinate
$r$.

\subparagraph{Locally \emph{AdS} space-time.}

The \emph{AdS} space-time ($F^{AB}=0$) can be described by the metric
\begin{equation}
ds_{\text{AdS}}^{2}=\ell ^{2}\,\left( dr^{2}+e^{2r}\eta _{\bar{n}\bar{m}%
}\,dx^{\bar{n}}dx^{\bar{m}}\right) \,,  \label{AdS metric}
\end{equation}%
where $x^{\bar{n}}=\left( t,x^{n}\right) $ and $\eta _{\bar{n}\bar{m}}=$
diag $\left( -,+,+,+,+\right) $. The vielbein ($e^{a}$) and the spin
connection ($\omega ^{ab}$) are given by
\begin{equation}
\begin{array}{ll}
e^{1}=\ell dr\,, & \omega ^{\bar{n}1}=\frac{1}{\ell }\,e^{\bar{n}}=e^{r}dx^{%
\bar{n}}\,, \\
e^{\bar{n}}=\ell e^{r}dx^{\bar{n}}\,,\qquad \qquad & \omega ^{\bar{n}\bar{m}%
}=0\,.%
\end{array}
\label{ads connection}
\end{equation}%
It is easy to check that this space-time is indeed torsionless ($T^{a}=0$)
and with constant negative curvature ($R^{ab}=-\frac{1}{\ell ^{2}}%
\,e^{a}e^{b}$). Then the \emph{AdS} connection is
\begin{equation}
\mathbf{W}=\frac{1}{4}\,W^{AB}\Gamma _{AB}=\frac{1}{2\ell }\,\left[
e^{1}\Gamma _{1}+e^{\bar{n}}\Gamma _{\bar{n}}\left( 1+\Gamma _{1}\right) %
\right] \,.  \label{W}
\end{equation}%
The Killing spinors $\varepsilon ^{\alpha }$ for the metric (\ref{AdS metric}%
), which are solutions of
\begin{equation}
\left( d+\mathbf{W}\right) \varepsilon =0\,,  \label{AdS Killing eq.}
\end{equation}%
and they have the form \cite{Lu-Pope-Townsend}
\begin{equation}
\varepsilon =e^{-\frac{r}{2}\,\Gamma _{1}}\,\left[ 1-x^{\bar{n}}\Gamma _{%
\bar{n}}\left( 1+\Gamma _{1}\right) \right] \,\varepsilon _{0}\,,
\label{AdS Killing}
\end{equation}
where $\varepsilon _{0}$ is a constant spinor. The derivation of
(\ref{AdS Killing}) is given in the Appendix \ref{Kill}.
Changing the topology of $\partial \sigma $
by the identification of the coordinates $x^{n}$ gives a
locally \emph{AdS} space-time, and it eliminates the
$x^{\bar{n}}$-dependence of $\varepsilon $, which therefore
must be (anti)chiral under $\Gamma _{1}$. Some examples of
Killing spinors in the locally \emph{AdS} space-time
are given in \cite{Aros-Martinez-Troncoso-Zanelli}.

\subparagraph{Consistency of the Killing equation.}

The Killing equation (\ref{Killing eq.}) for the $su(4)$ spinor $\epsilon
_{r}^{\alpha }$ can be written in components as
\begin{equation}
\left[ \delta _{r}^{s}\left( d+\mathbf{W}\right) -a_{r}^{s}\right] \epsilon
_{s}=0\,,  \label{Killing 2}
\end{equation}%
where $W^{AB}=\left( \frac{1}{\ell }\,e^{a},\omega ^{ab}\right) $ is the
\emph{AdS} connection, $a$ is the $su(4)$ connection, and (for $N=4$) the $%
u(1)$ field is decoupled and it does not appear in the covariant derivative.
The consistency of this equation requires $DD\epsilon =\left[ \mathbf{F}%
,\epsilon \right] =0,$ or in components,
\begin{equation}
\left( \frac{1}{4}\,\delta _{r}^{s}\,F^{AB}\Gamma _{AB}-\mathcal{F}%
_{r}^{s}\right) \epsilon _{s}=0\,.  \label{DDE}
\end{equation}%
For the configurations (\ref{Gen Reg}), this equation reduces to
\begin{equation}
\mathcal{F}_{r}^{s}\,\epsilon _{s}=0\,.  \label{consistency}
\end{equation}%
This equation has nonvanishing solution provided the matrix $\mathcal{F}%
_{r}^{s}$ has zero eigenvalues and $\epsilon _{r}$ are its zero modes. In
order to have a nontrivial $su(4)$ curvature, $\mathcal{F}^{\Lambda }$ must
be nonvanishing for more than one value of the index $\Lambda $, so that the
contributions of all components cancel. Using the local isomorphism $%
su(4)\simeq so(6)$, it is more convenient to represent the $su(4)$ curvature
as $\mathcal{F}_{r}^{s}=\frac{1}{2}\mathcal{F}^{IJ}\left( \tau _{IJ}\right)
_{r}^{s}$, where
\begin{equation}
\tau _{IJ}=\frac{1}{2}\,\hat{\Gamma}_{IJ}\,,\qquad \left( I,J=1,\ldots
,6\right)
\end{equation}%
are the $so(6)$ generators, $\hat{\Gamma}_{IJ}=\frac{1}{2}\,\left[ \hat{%
\Gamma}_{I},\hat{\Gamma}_{J}\right] $ and $\hat{\Gamma}_{I}$ are the
Euclidean $\hat{\Gamma}$-matrices. Furthermore, the commuting matrices $\tau
_{12}$ and $\tau _{34}$ generate the $u(1)\otimes u(1)$ subalgebra of $so(6)$%
. Since $\left( \tau _{12}\right) ^{2}=\left( \tau _{34}\right) ^{2}=-\frac{1%
}{4}$, the eigenvalues of $\tau _{12}$ and $\tau _{34}$ are $\pm \frac{i}{2}$%
. Considering the \textquotedblleft twisted\textquotedblright\
configuration, \emph{i.e.}, assuming
\begin{eqnarray}
\left( \tau _{12}\right) _{r}^{s}\epsilon _{s} &=&\frac{i}{2}\,\epsilon
_{r}\,,  \notag \\
\left( \tau _{34}\right) _{r}^{s}\epsilon _{s} &=&-\frac{i}{2}\,\epsilon
_{r}\,,  \label{twist}
\end{eqnarray}%
and that the only $u(1)$ curvature components are $\mathcal{F}^{12}=da^{12}$
and $\mathcal{F}^{34}=da^{34},$ then Eq. (\ref{consistency}) becomes $\frac{i%
}{2}\,\left( \mathcal{F}^{12}-\mathcal{F}^{34}\right) \,\epsilon _{r}=0\,,$
whose solution is
\begin{equation}
\mathcal{F}^{12}=\mathcal{F}^{34}.
\end{equation}%
In terms of the connection, this implies
\begin{equation}
a^{12}=a^{34}+d\eta \,,  \label{1234}
\end{equation}%
where $\eta \left( r,x^{n}\right) $ is an arbitrary function. Since $%
\mathcal{F}_{12}^{12}\left( r,x^{2}\right) \;$can only depend on two
coordinates by virtue of the Bianchi identity, the simplest solution reads
\begin{equation}
\begin{array}{lll}
a^{12}\left( x^{2}\right) & = & \rho \left( x^{2}\right) \,dr\,, \\
\mathcal{F}^{12}\left( x^{2}\right) & = & -\rho ^{\prime }\left(
x^{2}\right) \,drdx^{2}\,.%
\end{array}
\label{su(4)}
\end{equation}

\subparagraph{The Killing spinors.}

Since $a_{r}^{s}\,\epsilon _{s}=\frac{i}{2}\,d\eta \,\epsilon _{r}$, the $%
su(2,2\left\vert 4\right. )$ Killing equation (\ref{Killing 2}) reduces to
\begin{equation}
\left( d+\mathbf{W}-\frac{i}{2}\,d\eta \right) \,\epsilon _{s}=0\,,
\end{equation}%
where $\mathbf{W}$ is given by (\ref{W}). The spinor $\epsilon _{s}$ can be
factorized as
\begin{equation}
\epsilon _{s}^{\alpha }=u_{s}\,\varepsilon ^{\alpha }\,,
\end{equation}%
where the spinor $\varepsilon ^{\alpha }$ is the \emph{AdS} Killing spinor (%
\ref{AdS Killing}), while the $su(4)$ vector $u_{s}$ is a solution of the
equation
\begin{equation}
\left( d-\frac{i}{2}\,d\eta \right) \,u_{s}=0\,,
\end{equation}%
and it has the form $u_{s}=e^{\frac{i}{2}\,\eta }u_{0s}$. Therefore, The
spinor $\epsilon _{s}$ is
\begin{equation}
\epsilon _{s}=e^{\frac{i}{2}\,\eta }e^{-\frac{r}{2}\,\Gamma _{1}}\,\left[
1-x^{\bar{n}}\Gamma _{\bar{n}}\left( 1+\Gamma _{1}\right) \right] \,\epsilon
_{0s}\,,  \label{Killing spinor}
\end{equation}%
where the constant spinor $\epsilon _{0s}^{\alpha }$ satisfies the
conditions (\ref{twist})
\begin{equation}
\left( \tau _{12}\right) _{r}^{s}\epsilon _{0s}=\frac{i}{2}\,\epsilon
_{0r}\,,\qquad \qquad \left( \tau _{34}\right) _{r}^{s}\epsilon _{0s}=-\frac{%
i}{2}\,\epsilon _{0r}\,.
\end{equation}%
The norm $\left\Vert \epsilon \right\Vert ^{2}=\bar{\epsilon}\epsilon $ is
constant and positive,
\begin{equation}
\left\Vert \epsilon \right\Vert ^{2}=\left\Vert \epsilon _{0}\right\Vert
^{2}>0\,,
\end{equation}%
where $\bar{\epsilon}_{\alpha }^{r}=$ $\epsilon _{\beta }^{r\dagger }\left(
\Gamma _{0}\right) _{\alpha }^{\beta }$ (see Appendix \ref{Kill}).

Therefore, the existence of configurations with some unbroken
supersymmetries, which saturate the Bogomol'nyi bound, is important for the
stability of the theory. Among them, there is the ground state $\bar{A}_{v}.$

\subparagraph{The central charge.}

A pending issue is to find the explicit expression of the central charge, $%
C, $ for the charge algebra (\ref{algebra Q}). This can be done for the
configurations that asymptotically tend to the background solution which is
a BPS state. Since $\bar{F}^{M^{\prime }}=0$ at the boundary, the only
nonvanishing field-strength is $\bar{f},$ and the charge (\ref{Q}) becomes
\begin{equation}
\bar{Q}\left[ \lambda \right] =-\frac{3k}4\int\limits_{\partial \sigma }\bar{%
f}\,\left( \lambda _{IJ}\bar{a}^{IJ}+\lambda _{AB}\bar{W}^{AB}\right) \,.
\label{Q vacuum}
\end{equation}
The goal is to find the vacuum, for which all charges vanish, $\bar{Q}\left[
\lambda \right] =0.$ The $su(4)$ fields are given by (\ref{1234}) and (\ref%
{su(4)}), so that the first term in the expression for the charge (\ref{Q
vacuum}) is proportional to the \emph{pure gauge} $d\eta $. Therefore, for
the vacuum state, $\eta $ can be chosen so that $d\eta =0$. On the basis of (%
\ref{ads connection}), the second term in the charge (\ref{Q vacuum})
vanishes if the \emph{AdS} parameters $\lambda _{n5},$ $\lambda _{n1}$ obey
the asymptotic conditions:
\begin{equation}
\varepsilon ^{nmk}\bar{f}_{nm}\,\left( \lambda _{k5}+\lambda _{k1}\right)
e^r\rightarrow 0\,,\qquad \left( r\rightarrow \infty \right) \,.
\end{equation}

Supposing that all conditions are fulfilled and $\bar{Q}=0$, the charge $Q%
\left[ \lambda \right] $ in (\ref{Q}) is found to change under the gauge
transformations $\delta _{\eta }\mathbf{A}=-D\mathbf{\eta }$, as
\begin{equation}
C\left[ \lambda ,\eta \right] =2ik\int\limits_{\partial \sigma }\left\langle
\left( 3\mathbf{\lambda \bar{F}+}\bar{D}\mathbf{\lambda \bar{A}}\right) \bar{%
D}\mathbf{\eta }\right\rangle \,.
\end{equation}%
Finally, after substitution of the BPS background, the charge becomes
\begin{equation}
C\left[ \lambda ,\eta \right] =k\int\limits_{\partial \sigma }\bar{f}\,\eta
^{M^{\prime }}d\lambda _{M^{\prime }}\,.  \label{C ext}
\end{equation}%
The algebra (\ref{GG=G+C}), with the central charge (\ref{C ext}), is a
supersymmetric extension of the WZW$_{4}$ algebra
\cite{Nair-Schiff'90,Nair-Schiff'92,%
Banados-Garay-Henneaux1,Losev-Moore-Nekrasov-Shatashvili}.
This algebra has a nontrivial central extension for $PSU(2,2\left\vert
4\right. ),$ which depends on the $2$-form $\bar{f}$. In the Ref. \cite%
{Banados-Garay-Henneaux1}, the WZW$_{4}$ algebra describing the asymptotic
symmetries of a CS theory based on $g\otimes u(1)$ algebra was studied. The
background chosen there was irregular, and it was not possible to define the
$u(1)$ charge, so that the WZW$_{4}$ algebra was associated only to the
subalgebra $g$. In the case of the superalgebra (\ref{GG=G+C}), all
generators are well defined because the chosen background is regular and
generic. In contrast, the super WZW$_{4}$ algebra obtained here is
associated to the full gauge group $SU(2,2\left\vert 4\right. )$.

\section{Conclusions}

In this chapter, CS supergravities based on the supersymmetric extension of
the \emph{AdS}$_5$ algebra, $su(2,2\left| N\right. )$ were analyzed. This
supergravity theories contain the gravitational fields, $2N$ gravitini, $%
su(N)$ bosonic fields, and $u(1)$ field. The action, apart from the
Einsten-Hilbert term, contains terms nonlinear in curvature and is
torsionless. The sypersymmetry algebra closes \emph{off-shell}, without
bringing in the auxiliary fields.

In the case of $N=4,$ the invariant tensor of $su(2,2\left\vert 4\right. )$
algebra has the same form as the invariant tensor of $g\otimes u(1)$. The
theory has rich local dynamics, with $61$ locally propagating degrees of
freedom in the regular generic sector, as well as nontrivial asymptotic
dynamics.

The asymptotic dynamics depends on the choice of boundary conditions and is
determined by the subset of the gauge transformations which preserve these
conditions at the boundary. A result of this analysis was to show that
adequate boundary conditions exist, and to find the corresponding symmetries
and the charges at the boundary.

The following results are obtained:

\begin{itemize}
\item A class of backgrounds is found, which provide a regular and generic
theory. They are locally \emph{AdS} space-times with bosonic $su(4)$ and $%
u(1)$ matter. It is impossible to have a regular and generic \emph{AdS}
space-time without both types of bosonic matter fields. On the other hand,
if \emph{AdS} space-time is not required, then the $su(4)$ field-strength
can vanish (\emph{pure-gauge}).

\item Around the chosen backgrounds, the first and second class constraints
are explicitly separated. This, in general, is an extremely difficult task
for an arbitrary CS theory.

\item The charges corresponding to the complete gauge symmetry are obtained
for the simplest choice of the asymptotic conditions $A\rightarrow \bar{A}.$
This problem is not trivial since higher-dimensional CS theories are
irregular systems and there are sectors in the phase space where it is not
possible to define some generators.

\item BPS states exist among the considered backgrounds, which is important
for the stability of the theory.

\item The supersymmetric extension of the classical WZW$_4$ algebra,
associated to $su(2,2\left| 4\right. ),$ is obtained. The nontrivial central
extension occurs only for $psu(2,2\left| 4\right. )$ subalgebra.
\end{itemize}

In addition to these results, work on the asymptotics of the CS supergravity
is in progress, because there are still many questions to be answered. One
of them is related to the finding of the asymptotic symmetries and the
physical interpretation of the conserved charges. For example, is the mass
associated to the generator of the local time translations $Q\left[ \lambda
^{a5}\right] $? In Ref. \cite{Banados}, the energy and angular momentum of
the black hole embedded in this supergravity theory were found, but the
considered solution belonged to the irregular sector. There are
also other black hole solutions in the five-dimensional CS
supergravity which may be considered
\cite{Banados-Teitelboim-Zanelli}--\cite{Aros-Troncoso-Zanelli}.
Then, the natural
question is to analyze which charges (or their combinations) are bounded
from below by the Bogomol'nyi bound. For example, in $3D$ supergravity, the
black hole solution with the zero mass and zero angular momentum is a BPS
state, as well as the extreme case $M=\ell \left\vert J\right\vert $
\cite{Coussaert-Henneaux}.

Furthermore, for the BPS states obtained in the last section, it is
straightforward to make the mode expansion of the super WZW$_{4}$ algebra,
for instance when a boundary has a topology $S^{1}\times S^{1}\times S^{1},$
or $S^{1}\times S^{2}$. It is interesting to see which are the implications
of different topologies to the asymptotic symmetry.

One of the interesting questions which remains to be investigated, is
whether there is a regular and generic background (for example, when the $%
su(4)$ field does not vanish at the boundary) which leads to the super WZW$%
_{4}$\ with the $u(1)$ central extension, as well. It seems, in general,
possible.

\chapter{List of main results and open problems}

\begin{center}
{\large \textbf{Main results}}\textbf{\medskip }
\end{center}

\begin{itemize}
\item The action for the two-dimensional super Wess-Zumino-Witten model
coupled to supergravity is obtained by canonical methods, so that it is
invariant under local supersymmetry transformations by construction.

\item Standard Dirac's procedure for dealing with constrained systems is
generalized to the cases when the constraints are functionally dependent
(irregular). These irregular systems are classified, and regularized when
possible, for classical theories with finite number of degrees of freedom.

\item Higher-dimensional Chern-Simons theories are analyzed in the context
of irregular systems, and the criterion which recognizes irregular sectors
in their phase space is presented.

\item The dynamical content of \emph{AdS}$_{5}$-Chern-Simons supergravity
theory in five dimensions, based on the group $SU(2,2\left\vert 4\right. )$,
is studied and, among them, a class of BPS states is found.

\item The classical super WZW$_{4}$ algebra associated to $su(2,2\left\vert
4\right. )$, with nontrivial central extension, is obtained as the charge
algebra for \emph{AdS}$_{5}$-CS supergravity.
\end{itemize}

\newpage

\begin{center}
\textbf{{\large Open problems}\medskip }
\end{center}

\begin{itemize}
\item The method of construction of a super WZW action coupled to
supergravity can be generalized to the cases of $N>1$ supersymmetry, in $D=2$%
.

\item It is not clear whether all types of irregular systems can be
consistently quantized, and how irregular sectors will present themselves
after the quantization.

\item The dynamic of irregular and degenerate Chern-Simons theories, in
which the irregular sectors intersect with the surfaces of phase space with
lower rank of the symplectic form, is not well understood.

\item The pending questions in the \emph{AdS}$_{5}$-Chern-Simons
supergravity are the identification of the asymptotic symmetries and
conserved charges in terms of observables (the mass, angular momentum,
electric charge, etc.), as well as analysis of its boundary dynamics for
different choices of topology of the space-time.

\item One of more general problems is to obtain an action for a super WZW
model in higher dimensions. It can be done, for example, solving the CS
theory in $D=5$ based on the supersymmetric extension of $U(1)\otimes U(1)$,
and finding its theory at the boundary.
\end{itemize}

\appendix

\chapter{Hamiltonian formalism\label{HF}}

In this chapter, the Hamiltonian formalism for the systems with bosonic
degrees of freedom is reviewed \cite{DiracCJM,Castellani}, \cite%
{DiracFormalism1}--\cite{Bergman}. A generalization of the formalism to the
systems with fermionic degrees of freedom can be found in Refs. \cite%
{Berezin-Marinov}--\cite{Gervais-Sakita}, while for the reviews see Refs.
\cite{Dirac,Henneaux-Teitelboim,Blagojevic}, \cite{Hanson-Regge-Teitelboim}--%
\cite{Wipf}.

Consider a \emph{classical} system with \emph{finite} number of degrees of
freedom, described by the action
\begin{equation}
I[q,\dot{q}]=\int_{t_{0}}^{t_{1}}dt\,L(q^{i},\dot{q}^{i})\,,\qquad
(i=1,\cdots ,N)\,,  \label{action}
\end{equation}%
which depends at most on first derivatives of the local coordinates $%
q^{i}(t) $ (up to divergence terms) and it does not depend on time
explicitly. The classical dynamics is derived from the Hamilton's
variational principle, as a stationary point of the action under variations $%
\delta q(t)$ with fixed endpoints $\delta q(t_{0})=\delta q(t_{1})=0$.

\subsubsection{\emph{a}) Dirac-Bergman algorithm}

In order to pass to the Hamiltonian formalism, momenta are defined in the
usual way as
\begin{equation}
p_{i}=\frac{\delta I}{\delta \dot{q}^{i}}\,,\qquad (i=1,\cdots ,N)\,,
\label{momenta}
\end{equation}%
and Hamiltonian dynamics happens on $2N$-dimensional phase space
\begin{equation}
\Gamma =\left\{ z^{n}\mid n=1,\ldots ,2N\right\} \,,
\end{equation}%
with local coordinates $z^{n}=\left( q^{i},p_{j}\right) $. When all
velocities can be solved from the equations (\ref{momenta}) in terms of
canonical variables, the evolution of a system is uniquely determined from
its initial configuration by means of Hamilton equations. When all
velocities $\dot{q}^{i}$ cannot be uniquely solved from equations (\ref%
{momenta}) in terms of canonical variables, all momenta are not independent.
In consequence, there are \emph{primary constraints,}
\begin{equation}
\varphi _{\alpha }\left( z\right) =0\,,\qquad (\alpha =1,\cdots ,N_{P})\,,
\end{equation}%
which define the \emph{primary constraint surface}
\begin{equation}
\Sigma _{P}=\left\{ \bar{z}\in \Gamma \mid \varphi _{\alpha }(\bar{z}%
)=0\;\left( \alpha =1,\ldots ,N_{P}\right) \;\left( N_{P}\leq 2N\right)
\right\} \,.
\end{equation}

Although the primary constraints vanish on $\Sigma _{P}$, their derivatives
do not, thus it is useful to make difference between the concepts of weak
and strong equalities. A function $F(z)$, defined and differentiable in a
neighborhood of $z\in \Gamma $, is \emph{weakly }equal to zero if it
vanishes on $\Sigma _{P}$\emph{, }
\begin{equation}
F(z)\approx 0\quad \Longleftrightarrow \quad \left. F(z)\right\vert _{\Sigma
_{P}}=0\,,
\end{equation}%
and it is \emph{strongly} equal to zero if the function $F$ and its first
derivatives vanish on $\Sigma _{P}$,
\begin{equation}
F(z)=0\quad \Longleftrightarrow \quad \left. F,\;\frac{\partial F}{\partial
z^{n}}\right\vert _{\Sigma _{P}}=0\,.
\end{equation}%
With this conventions, the primary constraints are
\begin{equation}
\varphi _{\alpha }\left( z\right) \approx 0\,,\qquad (\alpha =1,\cdots
,N_{P})\,.  \label{primary}
\end{equation}%
Primary constraints are functionally independent if the \emph{regularity
conditions} (RCs) \cite{DiracCJM} are fulfilled: the constraints $\varphi
_{\alpha }\approx 0$\ are regular\ if and only if their small variations $%
\delta \varphi _{\alpha }\ $evaluated on $\Sigma _{p}$\ define $N_{P}$\
linearly independent functions of $\delta z^{n}$.

When the RCs are satisfied, there is relation between the strong and weak
equalities. If a phase space function $F$ is weakly equal to zero, then it
is strongly equal to a linear combination of constraints,
\begin{equation}
F\approx 0\quad \Longleftrightarrow \quad F=u^{\alpha }\varphi _{\alpha }\,.
\label{weak-strong}
\end{equation}
Therefore, the existence of primary constraints naturally leads to
appearance of arbitrary functions of time $u^{\alpha }(t)$ in a theory.

\subparagraph{Total Hamiltonian and time evolution.}

The canonical Hamiltonian, obtained by Legendre transformation of the
Lagrangian,
\begin{equation}
H_{0}(q,p)=p_{i}\dot{q}^{i}-L(q,\dot{q})\,,
\end{equation}%
depends only on canonical variables. Due to presence of primary constraints,
canonical variables are not independent and $H_{0}$ is not unique, therefore
the Legendre transformation is not invertible. Using the relation between
the weak and strong equalities (\ref{weak-strong}), one introduces \emph{%
total Hamiltonian} as
\begin{equation}
H=H_{0}+u^{\alpha }\varphi _{\alpha }\,.  \label{total H}
\end{equation}%
It gives an invertible Legendre transformation, but the dynamics following
from (\ref{total H}) depends on $N_{P}$ arbitrary functions $u^{\alpha }(t)$%
. Introducing a Poisson bracket (PB) on the phase space $\Gamma $ as
\begin{equation}
\{F,G\}=\frac{\partial F}{\partial z^{n}}\,\omega ^{nm}\,\frac{\partial G}{%
\partial z^{m}}\,,  \label{PB omega}
\end{equation}%
where $\omega ^{nm}$ is the antisymmetric symplectic form which determines
the basic PB $\{z^{n},z^{m}\}=\omega ^{nm}$, Hamilton equations can be
written as
\begin{equation}
\dot{z}^{n}=\left\{ z^{n},H_{0}\right\} +u^{\alpha }\left\{ z^{n},\varphi
_{\alpha }\right\} \approx \left\{ z^{n},H\right\} \,.  \label{H.eqs.}
\end{equation}%
Time evolution of phase space functions $F$ is determined by
\begin{equation}
\dot{F}=\left\{ F,H_{0}\right\} +u^{\alpha }\left\{ F,\varphi _{\alpha
}\right\} \approx \left\{ F,H\right\} \,,  \label{evolution}
\end{equation}%
and in general it is not unique for given initial conditions $F\left(
t_{0}\right) $.

\subparagraph{Consistency conditions.}

Consistency of the theory during its time evolution requires that the
primary constraints are preserved in time,
\begin{equation}
\dot{\varphi}^{\alpha }=\left\{ \varphi ^{\alpha },H_{0}\right\} +u^{\beta
}\left\{ \varphi _{\alpha },\varphi _{\beta }\right\} \approx 0\,.
\label{cc}
\end{equation}
These \emph{consistency conditions} reduce to one of the following
possibilities\footnote{%
The systems in which the consistency conditions are not satisfied are
excluded because, in such inconsistent models, the action has no stationary
points.}:

\begin{itemize}
\item If $\left\{ \varphi _\alpha ,\varphi _\beta \right\} \approx 0$ and $%
\left\{ \varphi _\alpha ,H_0\right\} \approx 0$, then Eq. (\ref{cc}) is
automatically satisfied (it reduces to $0\approx 0$).

\item If $\left\{ \varphi _\alpha ,\varphi _\beta \right\} \approx 0$ and $%
\left\{ \varphi _\alpha ,H_0\right\} \neq 0$, Eq. (\ref{cc}) is independent
on multipliers and gives a \emph{secondary constraint}.

\item If $\left\{ \varphi _\alpha ,\varphi _\beta \right\} \neq 0$, Eq. (\ref%
{cc}) becomes an algebraic equation in multipliers, leading to restrictions
on some of them.
\end{itemize}

Therefore, consistency conditions can give \emph{secondary constraints}, the
evolution of secondary constraints can give a new generation of constraints,
and so on, until it stops after finite number of generations (because the
dimension of $\Gamma $ is finite). This procedure results with the complete
set of constraints
\begin{equation}
\phi _{r}(z)\approx 0\,,\qquad \left( r=1,\ldots ,R\right) \,,
\label{constraints}
\end{equation}%
and a number of determined multipliers $u$.

A system is \emph{regular} if all constraints satisfy the RCs; otherwise,
the system is \emph{irregular}. Here only regular systems are considered.
Then the conditions (\ref{constraints}) define ($2n-R$)-dimensional \emph{%
constraint surface}
\begin{equation}
\Sigma =\left\{ \bar{z}\in \Gamma \mid \phi _{r}(\bar{z})=0\;\left(
r=1,\ldots ,R\right) \;\left( R\leq 2N\right) \right\} \,,
\end{equation}%
and weak and strong equalities are defined with respect to $\Sigma $.
Dirac-Bergman procedure guarantees that the system remains on the constraint
surface during its evolution.

\subparagraph{First and second class functions.}

A function $F(z)$ is said to be \emph{first class} if it has a PB with all
constraints weakly vanished,
\begin{equation}
\left\{ F,\phi _{r}\right\} \approx 0\,,\qquad (r=1,\ldots ,R)\,.
\end{equation}%
A function that is not first class, is called \emph{second class}. As a
consequence of Dirac-Bergman procedure, the total Hamiltonian $H$ is a first
class function.

In particular, the constraints (\ref{constraints}) can be divided into first
and second class constraints. While distinction of primary and secondary
constraints is of minor importance in the final form of the Hamiltonian
theory, the classification on first and second class constraints has
important dynamical consequences.

First class constraints commute with all other constraints, $\left\{ \phi
\,_{\text{I class}},\phi _{r}\right\} \approx 0$, thus their consistency
conditions give no restrictions on multipliers. Consequently, the final
dynamics is not uniquely determined by initial conditions, and unphysical
difference is related to the existence of\emph{\ local symmetries} in a
theory.

Second class constraints have $\left\{ \phi \,_{\text{II class}},\phi
_{r}\right\} \neq 0$, and their consistency conditions solve a number
multipliers. The total Hamiltonian with solved multipliers is
\begin{equation}
H^{\prime }=H_{0}^{\prime }+v^{a}\phi _{a}\,,  \label{H'}
\end{equation}
where $\phi _{a}$ are \emph{primary first class} constraints. Both
Hamiltonians $H^{\prime }$ and $H_{0}^{\prime }$ are first class functions.

\subsubsection{\emph{b}) Dirac brackets}

Consider a set of all second class constraints,
\begin{equation}
\theta _{m}(z)\approx 0\,,\qquad (m=1,\ldots ,N_{2})\,,
\end{equation}%
where the corresponding PB matrix, or \emph{Dirac matrix}, is
non-degenerate,
\begin{equation}
\{\theta _{m},\theta _{k}\}=\Delta _{mk}\,,
\end{equation}%
with an inverse $\Delta ^{mk}$ $\left( \Delta _{mn}\Delta ^{nk}=\delta
_{m}^{k}\right) $. The rank of Dirac matrix is even, as it is antisymmetric,
thus there are always even number of second class constraints.

The \emph{Dirac bracket} (DB) of two phase space functions is defined by
\begin{equation}
\left\{ F,G\right\} ^{\ast }=\left\{ F,G\right\} -\left\{ F,\theta
_{m}\right\} \Delta ^{mk}\left\{ \theta _{k},G\right\} \,,  \label{DB}
\end{equation}%
and it has all properties of a PB: antisymmetry, bilinearity and it obeys
product rule and Jacobi identity. By construction, DB of second class
constraints with any function $F$ vanishes,
\begin{equation}
\left\{ \theta _{m},F\right\} ^{\ast }=0\,.
\end{equation}%
Therefore, using DB\ instead of PB, the second class constraints can be
exchanged by strong equalities (set to zero before evaluating DB) on the
\emph{reduced phase space} $\tilde{\Gamma}\subset \Gamma ,$ where $\theta
_{m}=0.$

The construction of DB has \emph{iterative property}: a subset of second
class constraints can be used to define a preliminary DB $\left\{
\;\;\right\} ^{*}$. The next set of secondary constraints define a new DB $%
\left\{ \;\;\right\} ^{**}$, where a preliminary DB is used instead of a PB
in the definition (\ref{DB}); and so on, until all second class constraints
are exhausted.

\subparagraph{Gauge conditions.}

Unobservable degrees of freedom can be eliminated by imposing \emph{gauge
conditions}
\begin{equation}
\Psi _{a}\left( z\right) \approx 0\,.  \label{gauge fix}
\end{equation}%
The choice of functions $\Psi _{a}$ has to be such that: (\emph{a}) gauge
conditions are \emph{accessible}, or that the equation $\Psi _{a}\left(
z\right) $ $+$ $\delta _{\varepsilon }\Psi _{a}\left( z\right) =0$ has a
solution in $\varepsilon $, and (\emph{b}) this solution is \emph{unique},
\emph{i.e.}, there is no a residual gauge symmetry which preserves gauge
conditions (\ref{gauge fix}). When (\emph{a}) and (\emph{b}) are fulfilled,
then the number of the gauge conditions is equal to the number of first
class constraints
\begin{equation}
\chi _{a}\left( z\right) \approx 0\,,\qquad (a=1,\ldots ,N_{1})\,,
\end{equation}%
and the matrix
\begin{equation}
\left\{ \chi _{a},\Psi _{b}\right\} =\sigma _{ab}
\end{equation}%
is invertible. The gauge conditions must be preserved in time and their
consistency conditions determine all multipliers. $\Psi _{a}$ are treated as
any other constraints in a theory and, together with $\chi _{a}$, they form
a set of second class constraints. One can define DB which treat all
constraints and gauge conditions as strong equalities at reduced phase space
$\Gamma ^{\ast },$ containing only physical degrees of freedom which number
is given by the formula%
\begin{equation}
N^{\ast }=2N-\left( 2N_{1}+N_{2}\right) \,.
\end{equation}%
This formula applies only to the regular systems.

\subsubsection{\emph{c}) Local symmetries}

Due to the presence of arbitrary multipliers $v^{a}$ in the Hamiltonian (\ref%
{H'}), the evolution of a variable $F\left( t\right) \ $cannot be uniquely
determined from the given initial values $F(t_{0})$. After a short interval $%
\delta t=t-t_{0}$, the difference of two $F$'s, for two different choices of
multipliers $v_{1}^{a}$ and $v_{2}^{a}$, has the form
\begin{equation}
\Delta F\left( \delta t\right) =\varepsilon ^{a}\left\{ F,\phi _{a}\right\}
\,,  \label{symm}
\end{equation}%
where $\varepsilon ^{a}=\left( v_{2}^{a}-v_{1}^{a}\right) \,\delta t$. This
difference is unphysical and corresponds to a \emph{gauge transformation}
generated by \emph{primary first class} constraints $\phi _{a}$. Two
successive gauge transformations of type (\ref{symm}), with parameters $%
\varepsilon _{1}^{a}$ and $\varepsilon _{2}^{a}$, gives
\begin{equation}
\left( \Delta _{1}\Delta _{2}-\Delta _{2}\Delta _{1}\right) F\left( \delta
t\right) =\varepsilon _{1}^{a}\varepsilon _{2}^{b}\left\{ F,\left\{ \phi
_{a},\phi _{b}\right\} \right\} \,,
\end{equation}%
thus $\left\{ \phi _{a},\phi _{b}\right\} $, containing \emph{secondary}
first class constraints, also generates gauge transformations. Dirac
conjectured that \emph{all} first class constraints are generators of gauge
symmetries, what will be discussed below.

Physical \emph{observables} are quantities independent on arbitrary
multipliers. These are gauge invariant objects.

\subparagraph{Generator of local symmetries.}

The Hamiltonian formalism provides an algorithm to construct the generators
of all gauge symmetries of the equations of motion (\ref{H.eqs.}). If the
gauge transformations have canonical form
\begin{equation}
\delta _{\varepsilon }z^{n}=\left\{ z^{n},G\left[ \varepsilon \right]
\right\} \,,  \label{dz}
\end{equation}%
where $\varepsilon (t)$ is an infinitesimal local parameter, and the
generator is
\begin{equation}
G\left[ \varepsilon \right] =\dot{\varepsilon}G_{1}+\varepsilon G_{0}\,,
\end{equation}%
then necessary and sufficient conditions that Hamilton equations (\ref%
{H.eqs.}) are invariant under (\ref{dz}) are
\begin{eqnarray}
G_{1} &=&\text{\textsc{pfc}}\,,  \notag \\
G_{0}+\left\{ G_{1},H\right\} &=&\text{\textsc{pfc}}\,, \\
\left\{ G_{0},H\right\} &=&\text{\textsc{pfc}}\,,  \notag
\end{eqnarray}%
where `\textsc{pfc}' stands for a primary first class constraint. Thus, $%
G_{0}$ and $G_{1}$ are determined only up to primary first class
constraints. This is Castellani's method and it can be generalized to the
systems containing higher derivatives of a local parameter \cite%
{Castellani,Blagojevic}.

\subparagraph{Dirac conjecture.}

Dirac conjectured that all first class constraints are generators of the
gauge symmetries \cite{DiracCJM}. From Castellani's method it can be seen
that the generators contain, apart from the primary first class constraints $%
G_{1}$, also secondary constraints appearing in $\left\{ G_{1},H\right\} $.
Since any higher power of constraints can be treated as strong equality,
Castellani's algorithm stops if one obtains $\left\{ G_{1},H\right\} =\phi
^{K}=0$ $\left( K\geq 2\right) $. Therefore, the Dirac conjecture is
replaced by the statement that all first class constraints generate gauge
symmetry, apart from those appearing in the consistency conditions as a
higher power of a constraint $\phi $, and those following from the
consistency conditions of $\phi $ \cite{Blagojevic}.

\subsubsection{\emph{d}) Extended action\label{ET}}

The total Hamiltonian (\ref{total H}) contains only primary constraints.
Because the separation on primary and secondary constraints has no physical
implications, it is natural to define a Hamiltonian which contains all
constraints in a theory, or \emph{extended Hamiltonian}
\begin{equation}
H_{E}=H_{0}+u^{r}\phi _{r}\,.
\end{equation}
This Hamiltonian contains more dynamical variables than $H_{T},$ thus the
dynamics derived from the extended Hamiltonian is not equivalent to the
Lagrangian one, but the difference is unphysical. The introduction of $H_{E}$
is a new feature of the Hamiltonian scheme, which extends the Lagrangian
formalism by making manifest all gauge freedom.

The action with the dynamics equivalent to that obtained from $H_{E}$ is a
canonical \emph{extended action}
\begin{equation}
I_{E}\left[ q,p,u\right] =\int\limits_{t_{0}}^{t_{1}}dt\,\left( \dot{q}%
^{i}\,p_{i}-H_{0}-u^{r}\phi _{r}\right) \,,  \label{canonical action}
\end{equation}%
and it can always be reduced to the original action by gauge fixing of all
extra multipliers.

In a theory with only first class constraints, the following PB are
satisfied
\begin{equation}
\left\{ \phi _{r},\phi _{s}\right\} =C_{rs}^{\;\;\;p}\phi _{p}\,,\qquad
\qquad \left\{ \phi _{r},H_{0}\right\} =C_{r}^{\;s}\phi _{s}\,,
\label{brackets}
\end{equation}%
where $C_{rs}^{\;\;\;p}\left( z\right) $ and $C_{r}^{\;s}\left( z\right) $
are structure functions. Then the action $I_{E}\left[ q,p,u\right] $ is
invariant under the following gauge transformations \cite%
{Henneaux-Teitelboim}:
\begin{eqnarray}
\delta _{\varepsilon }z^{n} &=&\left\{ z^{n},\varepsilon ^{r}\phi
_{r}\right\} \,,  \label{tr1} \\
\delta _{\varepsilon }u^{r} &=&\dot{\varepsilon}^{r}+C_{ps}^{\;\;\;r}\,u^{s}%
\varepsilon ^{p}+C_{s}^{\;r}\,\varepsilon ^{s}\,.  \label{tr2}
\end{eqnarray}

\chapter{Superspace notation in $D=2$ \label{super}}

Two-dimensional space-time manifold $\mathcal{M}$ with signature $\left(
-,+\right) $ is parametrized by the local coordinates $x^\mu =\left( \tau
,\sigma \right) ,$ where $\mu =0,1$.

In the tangent Minkowski space, the local coordinates $x^{m}$ $\left(
m=0,1\right) $ are exchanged by the \emph{light-cone} coordinates $x^{a}$ $%
(a=+,-),$ defined by $x^{\pm }\equiv \frac{1}{2}\,(x^{0}\pm x^{1})$. In the
\emph{light-cone} basis, the Minkowski metric $\eta _{mn}=$ diag $\left(
-1,1\right) ,$ and its inverse $\eta ^{mn},$ become%
\begin{equation}
\eta _{ab}=\left(
\begin{array}{cc}
0 & -2 \\
-2 & 0%
\end{array}%
\right) \,,\qquad \qquad \eta ^{ab}=\left(
\begin{array}{cc}
0 & -\frac{1}{2} \\
-\frac{1}{2} & 0%
\end{array}%
\right) \,,
\end{equation}%
and therefore the raising and lowering of the tangent space indices is
performed as $V_{\pm }=-2V^{\mp }$. The Levi-Civita tensor $\varepsilon
^{mn} $ $\left( \varepsilon ^{01}=1\right) $ in the \emph{light-cone} basis
takes the form%
\begin{equation}
\varepsilon ^{ab}=\left(
\begin{array}{cc}
0 & -\frac{1}{2} \\
\frac{1}{2} & 0%
\end{array}%
\right) \,,\qquad \qquad \varepsilon _{ab}=\left(
\begin{array}{cc}
0 & 2 \\
-2 & 0%
\end{array}%
\right) \,.
\end{equation}

\subparagraph{Representation of $\protect\gamma $-matrices.}

Dirac matrices, defined in the tangent space, satisfy the \emph{Clifford
algebra}
\begin{equation}
\{\gamma ^{m},\gamma ^{n}\}=2\eta ^{mn}\,.  \label{Clifford}
\end{equation}%
The following representation is used:
\begin{equation}
\gamma ^{0}=\left(
\begin{array}{cc}
0 & 1 \\
-1 & 0%
\end{array}%
\right) \,,\quad \gamma ^{1}=\left(
\begin{array}{cc}
0 & 1 \\
1 & 0%
\end{array}%
\right) \,,\quad \gamma \equiv \gamma ^{0}\gamma ^{1}=\left(
\begin{array}{cc}
1 & 0 \\
0 & -1%
\end{array}%
\right) \,,  \label{gamma}
\end{equation}%
so that there is the identity Tr $\left( \gamma ^{m}\gamma ^{n}\gamma
\right) =2\varepsilon ^{mn}$. The projective $\gamma $-matrices $\gamma
^{\pm }=\frac{1}{2}\,(\gamma ^{0}\pm \gamma ^{1})$ are represented as:
\begin{equation}
\gamma ^{+}=\left(
\begin{array}{cc}
0 & 1 \\
0 & 0%
\end{array}%
\right) \,,\qquad \qquad \gamma ^{-}=\left(
\begin{array}{cc}
0 & 0 \\
-1 & 0%
\end{array}%
\right) \,.
\end{equation}

\subparagraph{Spinors.}

A Majorana spinor is a Dirac spinor $\theta _{\alpha }\equiv \left(
\begin{array}{c}
\theta _{+} \\
\theta _{-}%
\end{array}
\right) $ which obeys the Majorana condition $\theta =C\bar{\theta}^{T}$,
with $\bar{\theta}\equiv \theta ^{\dagger }\gamma ^{0}$, and $C_{\alpha
\beta }$ is the charge conjugation matrix ($C^{-1}\gamma _{\mu }C=-\gamma
_{\mu }^{T}$). In the representation (\ref{gamma}) and with $C=\gamma ^{0}$,
the Majorana spinors are real, $\theta _{\alpha }^{*}=\theta _{\alpha }$.
The inverse tensor $\left( C^{-1}\right) ^{\alpha \beta }\equiv C^{\alpha
\beta }$ performs the raising of spinor indices ($\theta ^{\alpha
}=C^{\alpha \beta }\theta _{\beta }$), while $C_{\alpha \beta }$ performs
their lowering ($\theta _{\alpha }=C_{\alpha \beta }\theta ^{\beta }$). In
components, it gives $\theta _{\pm }=\pm \theta ^{\mp }$. Spinor contraction
is denoted by $\theta \xi \equiv \theta ^{\alpha }\xi _{\alpha }=-\theta
_{\alpha }\xi ^{\alpha }$.

\subparagraph{Super covariant derivative.}

$\left( 1,1\right) $ superspace is parametrized by four real coordinates $%
z^A=(x^a,\theta _\alpha )$, where $x^a$ are the \emph{\emph{light-cone}}
coordinates and $\theta _\alpha $ is a Majorana spinor. The supercovariant
derivative is
\begin{equation}
D_\alpha =\bar{\partial}_\alpha +i\left( \gamma ^m\theta \right) _\alpha
\partial _m\,,  \label{sD}
\end{equation}
where $\partial _m\equiv \partial /\partial x^m$ and $\bar{\partial}_\alpha
\equiv \partial /\partial \bar{\theta}^\alpha $. More explicitly, the
derivative (\ref{sD}) in the representation (\ref{gamma}) is
\begin{equation}
D^{\pm }\equiv \partial _{\theta _{\mp }}-i\theta _{\mp }\partial _{\pm }\,,
\end{equation}
where $D_\alpha \equiv \left(
\begin{array}{c}
-D^{+} \\
D^{-}%
\end{array}
\right) $.

\subparagraph{Super $\protect\delta $-function.}

A generalization of the $\delta $-function to the super $\delta $-function
is
\begin{equation}
\delta _{\pm 12}\equiv \theta _{\mp 12}\,\delta (x_{1}^{\pm }-x_{2}^{\pm
})\,,
\end{equation}%
where $\theta _{12}=\theta _{1}-\theta _{2}$. It has the following
properties:
\begin{equation}
\begin{array}{llllll}
\int d^{4}z_{1}\,\delta _{\pm 12} & = & 1\,, & \delta _{\pm 21} & = &
-\delta _{\pm 12}\,, \\
F(z_{1})\,\delta _{\pm 12} & = & F(z_{2})\,\delta _{\pm 12}\,,\qquad \qquad
& D_{1}^{\pm }\delta _{\pm 12} & = & -D_{2}^{\pm }\delta _{\pm 12}\,,%
\end{array}%
\end{equation}%
where $d^{4}z\equiv d^{2}x\,d^{2}\theta $ and basic integrals for Grassman
odd numbers are $\int d\theta =0$ and $\int d\theta \,\theta =1$.

\chapter{Components of the vielbein and metric in the \emph{light-cone}
basis \label{cone}}

At each point of the curved space-time $\mathcal{M}$, there is a \emph{\emph{%
light-cone}} basis of 1-forms $e^{a}\equiv {e^{a}}_{\mu }\,dx^{\mu }$. The
vielbein ${e^{a}}_{\mu }$ $(a=+,-;~~\mu =0,1)$ is expressed in terms of
variables $(h^{-},h^{+},F,f)$ as
\begin{equation}
{e^{\pm }}_{\mu }=e^{F\pm f}{{\hat{e}}^{\pm }}_{\;\;\mu }\,,\quad \qquad {{%
\hat{e}}^{a}}_{\;\;\mu }=\frac{1}{2}\,\left(
\begin{array}{cc}
-h^{+} & 1 \\
h^{-} & -1%
\end{array}%
\right) \,.  \label{vielbein}
\end{equation}%
The inverse vielbein ${e^{\mu }}_{a}$ (${e_{a}}^{\mu }{e_{\mu }}^{b}=\delta
_{a}^{b}$ and ${e_{\mu }}^{a}{e_{a}}^{\nu }=\delta _{\mu }^{\nu }$) is
\begin{equation}
{e^{\mu }}_{\pm }=e^{-(F\pm f)}{{\hat{e}}^{\mu }}_{\;\;\pm }\,,\qquad \quad {%
{\hat{e}}^{\mu }}_{\;\;a}=\frac{2}{h^{-}-h^{+}}\,\left(
\begin{array}{cc}
1 & 1 \\
h^{-} & h^{+}%
\end{array}%
\right) \,.  \label{invers}
\end{equation}%
The related basis of tangent vectors $\partial _{a}\equiv {e^{\mu }}%
_{a}\partial _{\mu }$ can be written as
\begin{equation}
\partial _{\pm }=e^{-(F\pm f)}\hat{\partial}_{\pm }\,,\qquad \quad \hat{%
\partial}_{\pm }=\frac{2}{h^{-}-h^{+}}\,(\partial _{0}+h^{\mp }\partial
_{1})\,.  \label{partial}
\end{equation}%
It follows from (\ref{vielbein}) that the components of the metric tensor $%
g_{\mu \nu }=\eta _{ab}\,{e^{a}}_{\mu }{e^{b}}_{\nu }$ are
\begin{equation}
g_{\mu \nu }=e^{2F}\hat{g}_{\mu \nu }\,,\qquad \qquad \hat{g}_{\mu \nu
}\equiv -\frac{1}{2}\,\left(
\begin{array}{cc}
-2h^{+}h^{-} & h^{+}+h^{-} \\
h^{+}+h^{-} & -2%
\end{array}%
\right) \,,
\end{equation}%
while the inverse metric $g^{\mu \nu }$ has components
\begin{equation}
g^{\mu \nu }=e^{-2F}\hat{g}^{\mu \nu }\,,\qquad \,\qquad \hat{g}^{\mu \nu
}\equiv -\frac{2}{\left( h^{-}-h^{+}\right) ^{2}}\,\left(
\begin{array}{cc}
2 & h^{+}+h^{-} \\
h^{+}+h^{-} & 2h^{-}h^{+}%
\end{array}%
\right) \,.
\end{equation}%
Here $\sqrt{-g}=e^{2F}\sqrt{-\hat{g}}\equiv e^{2F}\frac{h^{-}-h^{+}}{2}$.

The Riemannian connection on $\mathcal{M}$ is defined by $T^{a}=0$ as
\begin{equation}
\omega _{a}=\varepsilon ^{bc}\,{e^{\mu }}_{b}\,\partial _{c}e_{a\mu }\,,
\end{equation}%
where $\varepsilon ^{mn}\,(\varepsilon ^{01}=1)$ is the constant totally
antisymmetric tensor in the Minkowski space or, in \emph{\emph{light-cone}}
basis, $\varepsilon ^{-+}=-\varepsilon ^{+-}=\frac{1}{2}$. Written in terms
of variables (\ref{invers}) and (\ref{partial}), the connection becomes
\begin{equation}
\omega _{\pm }=e^{-(F\pm f)}\left[ \hat{\omega}_{\pm }\mp \hat{\partial}%
_{\pm }(F\mp f)\right] \,,\qquad \qquad \hat{\omega}_{\pm }\equiv \mp \frac{%
2\partial _{1}h^{\mp }}{h^{-}-h^{+}}\,.
\end{equation}

Covariant derivative in the conformally flat (\emph{light-cone})
coordinates, acting on a Weyl field with the weight $n$, is
\begin{equation}
\triangledown _{a}=\partial _{a}+\frac{n}{2}\,\omega _{a}\,,
\end{equation}%
or explicitly
\begin{equation}
\triangledown _{\pm }=e^{(\pm n-1)F-(n\pm 1)f}\,\hat{\triangledown}_{\pm
}\,e^{\mp nF+nf}\,,
\end{equation}%
where $\hat{\triangledown}_{a}=\hat{\partial}_{a}+\frac{n}{2}\,\hat{\omega}%
_{a}$.

\chapter{Symplectic form in Chern-Simons theories\label{sm}}

Consider a Chern-Simons (CS) theory in $D=2n+1\geq 5$ dimensions described
by the action in Hamiltonian form (first order formalism)
\begin{eqnarray}
I_{\text{CS}}[A_{i}^{a}] &=&\int\limits_{\mathbb{R}\times \sigma }L_{\text{CS%
}}(A)=\int\limits_{t_{0}}^{t_{1}}dt\int\limits_{\sigma }d^{2n}x\,\left(
\mathcal{L}_{a}^{i}\,\dot{A}_{i}^{a}-A_{0}^{a}\chi _{a}\right) \,,  \label{i}
\\
dL_{\text{CS}} &=&k\,\,g_{a_{1}\cdots a_{n+1}}\,F^{a_{1}}\cdots
F^{a_{n+1}}\,,  \label{dl}
\end{eqnarray}%
where $A_{i}^{a}$ define a space of gauge fields $\mathcal{A}$ and $%
A_{0}^{a} $ is Lagrange multiplier. Similarly to the basis of 1-forms $%
dx^{\mu }$ on the space-time manifold $\mathcal{M}$, one can define a basis
of 1-forms $\underline{\delta }A_{i}^{a}\left( x\right) $ on the manifold of
gauge fields $\mathcal{A}$. The \emph{symplectic form} $\hat{\Omega}$ is a
quadratic 2-form defining the kinetic term in (\ref{i}),
\begin{eqnarray}
I_{\text{CS}} &=&\int dt\,\hat{\Omega}\,\underline{\delta }\dot{A}\underline{%
\delta }A  \notag \\
&=&\int dt\,\int d^{2n}x\,\int d^{2n}x^{\prime }\,\Omega _{ab}^{ij}\left(
x,x^{\prime }\right) \,\underline{\delta }\dot{A}_{i}^{a}\left( x\right) \,%
\underline{\delta }A_{j}^{b}\left( x^{\prime }\right) \,,
\end{eqnarray}%
and it can be expressed as
\begin{equation}
\hat{\Omega}_{ab}^{ij}\left( x,x^{\prime }\right) =\frac{\delta \mathcal{L}%
_{b}^{j}\left( x^{\prime }\right) }{\delta A_{i}^{a}\left( x\right) }-\frac{%
\delta \mathcal{L}_{a}^{i}\left( x\right) }{\delta A_{j}^{b}\left( x^{\prime
}\right) }\,.  \label{omega}
\end{equation}%
The symplectic form can be calculated using
\begin{equation}
\frac{\delta ^{2}I_{\text{CS}}}{\delta \dot{A}_{i}^{a}\left( x\right) \delta
A_{j}^{b}\left( x^{\prime }\right) }=\frac{\delta \mathcal{L}_{a}^{i}\left(
x\right) }{\delta A_{j}^{b}\left( x^{\prime }\right) }\,,
\end{equation}%
and varying $dL_{\mathtt{CS}}$ in (\ref{dl}), so that one obtains
\begin{eqnarray}
\delta L_{\text{CS}} &=&k\left( n+1\right) \,g_{aa_{1}\cdots
a_{n}}\,F^{a_{1}}\cdots F^{a_{n}}\delta A^{a}\,,  \notag \\
\delta ^{2}L_{\text{CS}} &=&kn\left( n+1\right) \,g_{aba_{1}\cdots
a_{n-1}}\,F^{a_{1}}\cdots F^{a_{n-1}}D\delta A^{a}\delta A^{b}  \notag \\
&=&dt\,d^{2n}x\,\frac{kn}{2^{n-1}}\,\left( n+1\right) \,\varepsilon
^{iji_{1}j_{1}\cdots i_{n-1}j_{n-1}}\,g_{aba_{1}\cdots
a_{n-1}}\,F_{i_{1}j_{1}}^{a_{1}}\cdots F_{i_{n-1}j_{n-1}}^{a_{n-1}}\delta
\dot{A}_{i}^{a}\delta A_{j}^{b}  \notag \\
&&+\quad \dot{A}\;\text{-\ independent part}\,.
\end{eqnarray}%
Therefore, one finds that the symplectic form is diagonal in continual
indices (without non-local operators),
\begin{equation}
\hat{\Omega}_{ab}^{ij}\left( x,x^{\prime }\right) =\Omega _{ab}^{ij}\left(
x\right) \delta \left( x-x^{\prime }\right) \,,
\end{equation}%
with the symplectic matrix $\Omega _{ab}^{ij}$ given by
\begin{equation}
\Omega _{ab}^{ij}\equiv -\frac{kn}{2^{n-1}}\,\left( n+1\right) \,\varepsilon
^{iji_{2}j_{2}\cdots i_{n}j_{n}}g_{aba_{2}\cdots
a_{n}}F_{i_{2}j_{2}}^{a_{1}}\cdots F_{i_{n}j_{n}}^{a_{n}}\,.
\end{equation}

In Hamiltonian approach, there is always a primary constraint $\phi
_{a}^{i}\equiv \pi _{a}^{i}-\mathcal{L}_{a}^{i}\approx 0$ such that, by
definition of PB and (\ref{omega}), gives
\begin{equation}
\left\{ \phi _{a}^{i},\phi _{b}^{j}\right\} =\Omega _{ab}^{ij}\delta \,.
\end{equation}%
Therefore, the explicit expression for $\mathcal{L}_{a}^{i}\left( A\right) $
is not necessary, since the symplectic matrix determines the dynamics of the
theory.

The matrix $\Omega _{ab}^{ij}$ is degenerate because it always has at least $%
2n$ zero modes $\mathbf{V}_{i}$, solutions of the matrix equation $\Omega
_{ab}^{ik}\left( V_{j}\right) _{k}^{b}=0.$ It follows from the identity
\begin{equation}
\Omega _{ab}^{ik}F_{kj}^{b}=-\delta _{j}^{i}\chi _{a}\approx 0\,,
\label{identity}
\end{equation}%
where $\left( V_{j}\right) _{k}^{b}=F_{kj}^{b}$. The above identity can be
shown using the fact that the tensor $g_{aa_{1}\cdots
a_{n}}F_{i_{1[}j_{1}}^{a_{1}}\cdots F_{i_{n}j_{n}]}^{a_{n}}$ (the definition
of antisymmetrization includes the factor $\frac{1}{\left( 2n-1\right) !}$)
is \emph{totally antisymmetric} in indices $\left[ i_{1}j_{1}\cdots
i_{n}j_{n}\right] $, and it is therefore proportional to the Levi-Civita
tensor,
\begin{equation}
g_{aa_{1}\cdots a_{n}}F_{i_{1[}j_{1}}^{a_{1}}\cdots
F_{i_{n}j_{n}]}^{a_{n}}=C_{a}\,\varepsilon _{i_{1}j_{1}\cdots i_{n}j_{n}}\,,
\end{equation}%
with the factor of proportionality
\begin{equation}
C_{a}=\frac{1}{\left( 2n\right) !}\,g_{aa_{1}\cdots a_{n}}\varepsilon
^{i_{1}j_{1}\cdots i_{n}j_{n}}F_{i_{1[}j_{1}}^{a_{1}}\cdots
F_{i_{n}j_{n}]}^{a_{n}}\,\text{.}
\end{equation}%
Then the identity
\begin{equation}
g_{aa_{1}\cdots a_{n}}\varepsilon ^{sj_{1}i_{2}j_{2}\cdots
i_{n}j_{n}}F_{k[j_{1}}^{a_{1}}\cdots F_{i_{n}j_{n}]}^{a_{n}}=\left(
2n-1\right) !\,C_{a}\,\delta _{k}^{s}\,
\end{equation}%
is equivalent to (\ref{identity}).

\chapter{Anti-de Sitter group, \emph{AdS}$_{D}$\label{AdS5}}

The $D$-dimensional \emph{AdS}$_{D}$ group, $SO(D-1,2)$, is the isometry
group of the $D$-dimensional hyperboloid
\begin{equation}
H_{D}:-x_{0}^{2}+x_{1}^{2}+\cdots +x_{D-1}^{2}-x_{D}^{2}=-\ell ^{2}
\end{equation}%
defined in a $(D+1)$-dimensional space-time with signature
\begin{equation}
\eta _{AB}=(-,+,\;\cdots \;,+,-)\,,\qquad \left( A,B=0,\ldots ,D\right) \,.
\end{equation}%
The group has $D\left( D+1\right) /2$ generators represented by $\mathbf{J}%
_{AB}=-\mathbf{J}_{BA},$ which satisfy the Lie-algebra
\begin{equation}
\left[ \mathbf{J}_{AB},\mathbf{J}_{CD}\right] =\eta _{AD}\,\mathbf{J}%
_{BC}-\eta _{BD}\,\mathbf{J}_{AC}-\eta _{AC}\,\mathbf{J}_{BD}+\eta _{BC}\,%
\mathbf{J}_{AD}\,.
\end{equation}%
The generators $\mathbf{J}_{AB}$ can be decomposed into
\begin{equation}
\mathbf{J}_{AB}:\quad \left\{
\begin{array}{l}
\mathbf{J}_{a}\equiv \mathbf{J}_{aD}\,, \\
\mathbf{J}_{ab}\,,%
\end{array}%
\right. \qquad \left( a,b=0,\ldots ,D-1\right) \,,
\end{equation}%
leading to the \emph{AdS}$_{D}$ algebra in the form
\begin{equation}
\begin{array}{lll}
\left[ \mathbf{J}_{ab},\mathbf{J}_{cd}\right] & = & \eta _{ad}\,\mathbf{J}%
_{bc}-\eta _{bd}\,\mathbf{J}_{ac}-\eta _{ac}\,\mathbf{J}_{bd}+\eta _{bc}\,%
\mathbf{J}_{ad}\,,\smallskip \\
\left[ \mathbf{J}_{ab},\mathbf{J}_{c}\right] & = & \eta _{bc}\,\mathbf{J}%
_{a}-\eta _{ac}\,\mathbf{J}_{b}\,,\smallskip \\
\left[ \mathbf{J}_{a},\mathbf{J}_{b}\right] & = & \mathbf{J}_{ab}\,,%
\end{array}%
\end{equation}%
where the metric $\eta _{ab}$ is $(-,+,\ldots ,+)$.

The \emph{AdS}$_{D}$\emph{\ }group\emph{\ }is related to the Poincar\'{e}
group via the \emph{Wigner-In\"{o}n\"{u} contraction} \cite{Inonu}, as
follows. After defining $\mathbf{P}_{a}=\frac{1}{\ell }\,\mathbf{J}_{a}$,
the commutator $\left[ \mathbf{P}_{a},\mathbf{P}_{b}\right] =\frac{1}{\ell
^{2}}\,\mathbf{J}_{ab}$ vanishes in the flat space limit, $\ell \rightarrow
\infty $. Since $\mathbf{J}_{ab}$ become the generators of Lorentz
transformations and $\mathbf{P}_{a}$ generators of translations in $a$-th
direction, the \emph{AdS}$_{D}$ algebra reduces to the $D$-dimensional
Poincar\'{e} group $ISO(D-1,1)$.

This motivates to construct a connection associated to \emph{AdS}$_{D}$
whose components are the \emph{vielbein} $e^{a},$ and the \emph{%
spin-connection} $\omega ^{ab}$,
\begin{equation}
\mathbf{A}\equiv \frac{1}{2}\,W^{AB}\mathbf{J}_{AB}=\frac{1}{\ell }\,e^{a}%
\mathbf{J}_{a}+\frac{1}{2}\,\omega ^{ab}\mathbf{J}_{ab}\,\text{.}
\end{equation}%
The corresponding field-strength, $\mathbf{F=}d\mathbf{A+A}^{2}\mathbf{,}$
has the form
\begin{equation}
\begin{array}{lll}
\mathbf{F} & \mathbf{=} & \frac{1}{\ell }\,T^{a}\mathbf{J}_{a}+\frac{1}{2}%
\,F^{ab}\mathbf{J}_{ab}\,,\smallskip \\
F^{ab} & \equiv & R^{ab}+\frac{1}{\ell ^{2}}\,e^{a}e^{b}\text{\thinspace },%
\end{array}%
\end{equation}%
where the \emph{torsion} ($T^{a}$) and \emph{Ricci curvature} ($R^{ab}$) are
\begin{eqnarray}
R^{ab} &=&d\omega ^{ab}+\omega _{\;\,c}^{a}\omega ^{cb}\,,  \notag \\
T^{a} &=&de^{a}+\omega _{\;\,b}^{a}e^{b}.
\end{eqnarray}%
\emph{AdS} space-time, which has a constant curvature $R^{ab}=-\frac{1}{\ell
^{2}}\,e^{a}e^{b}$, can be interpreted as a \emph{pure gauge} solution $%
F^{ab}=0,$ $T^{a}=0$.

\chapter{Supersymmetric extension of \emph{AdS}$_{5}$, $SU(2,2\left\vert
N\right. )$\label{SAdS}}

\subsubsection{\emph{a}) Generators}

The supersymmetric extension of the \emph{AdS} group in five dimensions is
the super unitary group $SU(2,2\left\vert N\right. )$ \cite%
{Chandia-Troncoso-Zanelli,Nahm,Strathdee}, containing supermatrices of unit
superdeterminant which leave invariant the (real) quadratic form
\begin{equation}
q=\theta ^{\ast \alpha }G_{\alpha \beta }\theta ^{\beta }+z^{\ast
r}g_{rs}z^{s}\,,\qquad (\alpha =1,\ldots ,4;\;\;r=1,\ldots ,N)\,.
\label{q form}
\end{equation}%
Here $\theta ^{\alpha }$ are complex Grassman numbers (with complex
conjugation defined as $\left( \theta ^{\alpha }\theta ^{\beta }\right)
^{\ast }=\theta ^{\ast \beta }\;\theta ^{\ast \alpha }$), and $G_{\alpha
\beta }$ and $g_{rs}$ are Hermitean matrices, antisymmetric and symmetric
respectively, which can be chosen as
\begin{equation}
G_{\alpha \beta }=i\left( \Gamma _{0}\right) _{\alpha \beta }\,,\qquad
\qquad g_{rs}=\delta _{rs}\,.
\end{equation}%
The bosonic sector of this supergroup is
\begin{equation}
SU(2,2)\otimes SU(N)\otimes U(1)\subset SU(2,2\left\vert N\right. )\,,
\end{equation}%
where the \emph{AdS} group is present on the basis of the isomorphism $%
SU(2,2)\simeq SO(2,4).$ Therefore, the generators of $su(2,2\left\vert
N\right. )$ algebra are
\begin{equation}
\begin{array}{lll}
so(2,4):\quad & \mathbf{J}_{AB}=\left( \mathbf{J}_{ab,}\mathbf{J}_{a}\right)
\,,\qquad \medskip & \left( A,B=0,\ldots ,5\right) \,, \\
su(N): & \mathbf{T}_{\Lambda }\,,\medskip & \left( \Lambda =1,\ldots
,N^{2}-1\right) \,, \\
\text{SUSY}: & \mathbf{Q}_{r}^{\alpha },\;\mathbf{\bar{Q}}_{\alpha
}^{r}\,,\medskip & \left( \alpha =1,\ldots ,4;\;r=1,\ldots ,N\right) \,, \\
u(1): & \mathbf{G}_{1}\,, &
\end{array}%
\end{equation}%
where $\eta _{AB}=$ diag $\left( -,+,+,+,+,-\right) ,$ and \emph{AdS}
rotations and translations are $\mathbf{J}_{ab}$ and$\;\mathbf{J}_{a}\equiv
\mathbf{J}_{a5}\;(a,b=0,\ldots ,4)$.

\subsubsection{\emph{b}) Representation of generators}

A representation of the superalgebra acting in $\left( 4+N\right) $%
-dimensional superspace $\left( \theta ^{\alpha },y^{r}\right) $ is given by
the $\left( 4+N\right) \times \left( 4+N\right) $ supermatrices as follows.

\begin{itemize}
\item \textbf{\emph{AdS} generators}\emph{\ }
\begin{equation}
\mathbf{J}_{AB}=\left(
\begin{array}{cc}
\frac{1}{2}\,\left( \Gamma _{AB}\right) _{\alpha }^{\beta } & 0 \\
0 & 0%
\end{array}%
\right) \,,
\end{equation}%
with the $4\times 4$ matrices $\Gamma _{AB}$ defined by
\begin{equation}
\Gamma _{AB}:\qquad \;\left\{
\begin{array}{l}
\Gamma _{ab}=\frac{1}{2}\,\left[ \Gamma _{a},\Gamma _{b}\right] \,,\smallskip
\\
\Gamma _{a5}=\Gamma _{a}\,,%
\end{array}%
\right.
\end{equation}%
where $\Gamma _{a}$ are the Dirac matrices in five dimensions with the
signature $\left( -,+,+,+,+\right) $;

\item $su(N)$\textbf{\ generators}
\begin{equation*}
\mathbf{T}_\Lambda =\left(
\begin{array}{cc}
0 & 0 \\
0 & \left( \tau _\Lambda \right) _r^s%
\end{array}
\right) \,,
\end{equation*}
where $\tau _\Lambda $ are anti-Hermitean generators of $su(N)$ acting in $N$%
-dimensional space $y^r$,
\begin{equation}
\left[ \tau _{\Lambda _1},\tau _{\Lambda _2}\right] =f_{\Lambda _1\Lambda
_2}^{\quad \;\Lambda _3}\,\tau _{\Lambda _3}\,;
\end{equation}

\item \textbf{Supersymmetry generators}
\begin{equation}
\mathbf{Q}_{q}^{\gamma }=\left(
\begin{array}{cc}
0 & 0 \\
-\delta _{q}^{s}\delta _{\alpha }^{\gamma } & 0%
\end{array}%
\right) \,,\qquad \qquad \mathbf{\bar{Q}}_{\gamma }^{q}=\left(
\begin{array}{cc}
0 & \delta _{r}^{q}\delta _{\gamma }^{\beta } \\
0 & 0%
\end{array}%
\right) \,;
\end{equation}

\item $u(1)$\textbf{\ generator}
\begin{equation}
\mathbf{G}_{1}=\left(
\begin{array}{cc}
\frac{i}{4}\,\delta _{\alpha }^{\beta } & 0 \\
0 & \frac{i}{N}\,\delta _{r}^{s}%
\end{array}%
\right) \,.
\end{equation}
\end{itemize}

\subsubsection{\emph{c}) The algebra}

From the given representation of the supermatrices, it is straightforward to
find the explicit form of the corresponding Lie algebra. The commutators of
the bosonic generators $\mathbf{J}_{AB}$, $\mathbf{T}_{\Lambda }$ and $%
\mathbf{G}_{1}$ close the algebra $su(2,2)\otimes su(N)\otimes u(1),$%
\begin{eqnarray}
\left[ \mathbf{J}_{AB},\mathbf{J}_{CD}\right] &=&\eta _{AD\,}\mathbf{J}%
_{BC}-\eta _{BD}\,\mathbf{J}_{AC}-\eta _{AC}\,\mathbf{J}_{BD}+\eta _{BC}\,%
\mathbf{J}_{AD}\,,  \notag \\
\left[ \mathbf{T}_{\Lambda _{1}},\mathbf{T}_{\Lambda _{2}}\right]
&=&f_{\Lambda _{1}\Lambda _{2}}^{\quad \;\;\Lambda _{3}}\,\mathbf{T}%
_{\Lambda _{3}}\,.
\end{eqnarray}%
The supersymmetry generators transform as spinors under \emph{AdS} and as
vectors under $su(N)$,
\begin{equation}
\begin{array}{ll}
\left[ \mathbf{J}_{AB},\mathbf{Q}_{r}^{\alpha }\right] =-\frac{1}{2}\,\left(
\Gamma _{AB}\right) _{\beta }^{\alpha }\,\mathbf{Q}_{r}^{\beta }\,,\qquad
\qquad \medskip & \left[ \mathbf{T}_{\Lambda },\mathbf{Q}_{r}^{\alpha }%
\right] =\left( \tau _{\Lambda }\right) _{r}^{s}\,\mathbf{Q}_{s}^{\alpha }\,,
\\
\left[ \mathbf{J}_{AB},\mathbf{\bar{Q}}_{\alpha }^{r}\right] =\frac{1}{2}\,%
\mathbf{\bar{Q}}_{\beta }^{r}\,\left( \Gamma _{AB}\right) _{\alpha }^{\beta
}\,, & \left[ \mathbf{T}_{\Lambda },\mathbf{\bar{Q}}_{\alpha }^{r}\right] =-%
\mathbf{\bar{Q}}_{\alpha }^{s}\,\left( \tau _{\Lambda }\right) _{s}^{r}\,,%
\end{array}%
\end{equation}%
and they carry $u(1)$ charge:
\begin{eqnarray}
\left[ \mathbf{G}_{1},\mathbf{Q}_{r}^{\alpha }\right] &=&-i\,\left( \frac{1}{%
4}-\frac{1}{N}\right) \,\mathbf{Q}_{r}^{\alpha }\,,  \notag \\
\left[ \mathbf{G}_{1},\mathbf{\bar{Q}}_{\alpha }^{r}\right] &=&i\,\left(
\frac{1}{4}-\frac{1}{N}\right) \,\mathbf{\bar{Q}}_{\alpha }^{r}\,.
\end{eqnarray}%
The anticommutator of the supersymmetry generators has the following form:
\begin{equation}
\left\{ \mathbf{Q}_{r}^{\alpha }\mathbf{,\bar{Q}}_{\beta }^{s}\right\} =%
\frac{1}{4}\,\delta _{r}^{s}\,\left( \Gamma ^{AB}\right) _{\beta }^{\alpha
}\,\mathbf{J}_{AB}-\delta _{\beta }^{\alpha }\,\left( \tau ^{\Lambda
}\right) _{r}^{s}\,\mathbf{T}_{\Lambda }+i\,\delta _{\beta }^{\alpha
}\,\delta _{r}^{s}\,\mathbf{G}_{1}\mathbf{\,,}
\end{equation}%
what can be shown using the orthogonality relations for $\Gamma $- and $\tau
$-matrices,
\begin{equation}
\begin{array}{lll}
\frac{1}{2}\,\left( \Gamma ^{AB}\right) _{\beta }^{\alpha }\,\left( \Gamma
_{AB}\right) _{\lambda }^{\rho } & = & \delta _{\beta }^{\alpha }\delta
_{\lambda }^{\rho }-4\delta _{\lambda }^{\alpha }\delta _{\beta }^{\rho
}\,,\smallskip \\
\left( \tau ^{\Lambda }\right) _{s}^{r}\,\left( \tau _{\Lambda }\right)
_{q}^{p} & = & \delta _{q}^{r}\delta _{s}^{p}-\frac{1}{N}\,\delta
_{s}^{r}\delta _{q}^{p}\,.%
\end{array}%
\end{equation}

\subsubsection{\emph{d}) Killing metric}

Denote all generators as $\mathbf{G}_{M}=\left( \mathbf{G}_{M^{\prime }},%
\mathbf{G}_{1}\right) $, where $\mathbf{G}_{M^{\prime }}$ are the generators
of $PSU(2,2\left\vert N\right. )$ (closing the algebra without $U(1)$
generator). The components of the Killing metric of $SU(2,2\left\vert
N\right. )$ is an invariant tensor of rank two which is symmetric in the
bosonic and antisymmetric in the fermionic indices, and has the form
\begin{equation}
g_{MN}=\left\langle \mathbf{G}_{M}\mathbf{G}_{N}\right\rangle =-\left(
\begin{array}{cc}
\gamma _{M^{\prime }N^{\prime }} & 0 \\
0 & 0%
\end{array}%
\right) \,,
\end{equation}%
where $\gamma _{M^{\prime }N^{\prime }}$ is the Killing metric of $%
PSU(2,2\left\vert N\right. )$, and $\left\langle \mathbf{\cdots }%
\right\rangle $ stands for the supertrace Str\thinspace $\left( \mathbf{%
\cdots }\right) $, which is the difference between the trace of the upper
and lower diagonal blocks. The components of the invertible Killing metric $%
\gamma _{M^{\prime }N^{\prime }}$ are
\begin{equation}
\begin{array}{lll}
\gamma _{\left[ AB\right] \left[ CD\right] } & = & \eta _{\left[ AB\right] %
\left[ CD\right] }\,,\smallskip \\
\gamma _{\Lambda _{1}\Lambda _{2}} & = & \text{Tr}_{N}\left( \tau _{\Lambda
_{1}}\tau _{\Lambda _{2}}\right) \,,\smallskip \\
\gamma _{\binom{\alpha }{r}\binom{s}{\beta }} & = & -\delta _{\beta
}^{\alpha }\delta _{r}^{s}\,,%
\end{array}
\label{psu}
\end{equation}%
and it raises and lowers $PSU(2,2\left\vert N\right. )$ indices. Note that
the metric $g_{MN}$ is not invertible (the corresponding supergroup is not
semi-simple)\thinspace . Here $\eta _{\left[ AB\right] \left[ CD\right]
}\equiv \eta _{AD}\,\eta _{BC}-\eta _{AC}\,\eta _{BD}\,$.

\subsubsection{\emph{e}) Symmetric invariant tensor}

The invariant tensor of rank three, \emph{completely} symmetric in bosonic
and antisymmetric in fermionic indices, can be calculated from
\begin{equation}
g_{MNK}=\left\langle \mathbf{G}_{M}\mathbf{G}_{N}\mathbf{G}_{K}\right\rangle
=\frac{1}{2}\,\text{Str\thinspace }\left[ \left( \mathbf{G}_{M}\mathbf{G}%
_{N}+\left( -\right) ^{\varepsilon _{M}\varepsilon _{N}}\mathbf{G}_{N}%
\mathbf{G}_{M}\right) \mathbf{G}_{K}\right] \,.  \label{3}
\end{equation}%
Note that, due to the cyclic property of the supertrace,
(anti)symmetrization in first two indices in (\ref{3}) leads to the
completely (anti)symmetric tensor $g_{MNK}$. It has the following
non-vanishing components:
\begin{equation}
\begin{array}{lll}
g_{\left[ AB\right] \left[ CD\right] \left[ EF\right] } & = & -\frac{i}{2}%
\,\varepsilon _{ABCDEF}\,,\medskip \\
g_{\Lambda _{1}\Lambda _{2}\Lambda _{3}} & = & -\gamma _{\Lambda _{1}\Lambda
_{2}\Lambda _{3}}\,,\medskip \\
g_{\left[ AB\right] \left( _{r}^{\alpha }\right) \left( _{\beta }^{s}\right)
} & = & \frac{1}{4}\,\left( \Gamma _{AB}\right) _{\beta }^{\alpha }\delta
_{r}^{s}\,,\medskip \\
g_{\Lambda \left( _{r}^{\alpha }\right) \left( _{\beta }^{s}\right) } & = &
\frac{1}{2}\,\delta _{\beta }^{\alpha }\left( \tau _{\Lambda }\right)
_{r}^{s}\,,\medskip \\
g_{1\left[ AB\right] \left[ CD\right] } & = & -\frac{i}{4}\,\eta _{\left[ AB%
\right] \left[ CD\right] }\,,\medskip \\
g_{1\Lambda _{1}\Lambda _{2}} & = & -\frac{i}{N}\,\gamma _{\Lambda
_{1}\Lambda _{2}}\,,\medskip \\
g_{1\left( _{r}^{\alpha }\right) \left( _{\beta }^{s}\right) } & = & \frac{i%
}{2}\,\left( \frac{1}{4}+\frac{1}{N}\right) \delta _{\beta }^{\alpha }\delta
_{r}^{s}\,,\medskip \\
g_{111} & = & -i\,\left( \frac{1}{4^{2}}-\frac{1}{N^{2}}\right) \,,%
\end{array}%
\end{equation}%
where $\gamma _{\Lambda _{1}\Lambda _{2}\Lambda _{3}}\equiv \frac{1}{2}\,$Tr$%
_{N}\left( \left\{ \tau _{\Lambda _{1}},\tau _{\Lambda _{2}}\right\} \tau
_{\Lambda _{3}}\right) $ is the symmetric invariant tensor of rank three for
$su(N)$ and $\Gamma $-matrices are normalized so that
\begin{equation}
\text{Tr}_{4}\left( \Gamma _{a}\,\Gamma _{b}\,\Gamma _{c}\,\Gamma
_{d}\,\Gamma _{e}\right) =-4i\,\varepsilon _{abcde}\,,\qquad \left(
\varepsilon ^{abcde5}\equiv \varepsilon ^{abcde},\quad \varepsilon
^{012345}=1\right) \,.
\end{equation}

In the special case $N=4$, the invariant tensor $g_{MNK}$ of $%
SU(2,2\left\vert 4\right. )$ simplifies to:
\begin{equation}
\begin{array}{lll}
g_{\left[ AB\right] \left[ CD\right] \left[ EF\right] } & = & -\frac{i}{2}%
\,\varepsilon _{ABCDEF}\,,\medskip \\
g_{\Lambda _{1}\Lambda _{2}\Lambda _{3}} & = & -\gamma _{\Lambda _{1}\Lambda
_{2}\Lambda _{3}}\,,\medskip \\
g_{\left[ AB\right] \left( _{r}^{\alpha }\right) \left( _{\beta }^{s}\right)
} & = & \frac{1}{4}\,\left( \Gamma _{AB}\right) _{\beta }^{\alpha }\delta
_{r}^{s}\,,\medskip \\
g_{\Lambda \left( _{r}^{\alpha }\right) \left( _{\beta }^{s}\right) } & = &
\frac{1}{2}\,\delta _{\beta }^{\alpha }\left( \tau _{\Lambda }\right)
_{r}^{s}\,,\medskip \\
g_{1M^{\prime }N^{\prime }} & = & -\frac{i}{4}\,\gamma _{M^{\prime
}N^{\prime }}\,,%
\end{array}%
\end{equation}%
with $g_{111}=0$ and the $PSU(2,2\left\vert 4\right. )$ Killing metric $%
\gamma _{M^{\prime }N^{\prime }}$ given by (\ref{psu}).

\chapter{Supergroup conventions \label{sG}}

Let $\mathbf{G}_{M}$ are supermatrices representing the generators of a Lie
supergroup. They satisfy the superalgebra
\begin{equation}
\left[ \mathbf{G}_{M},\mathbf{G}_{N}\right] =f_{MN}^{\quad \;K}\mathbf{G}%
_{K}\,,
\end{equation}%
where the commutators defined by $\left[ \mathbf{G}_{M},\mathbf{G}_{N}\right]
\equiv \mathbf{G}_{M}\mathbf{G}_{N}-\left( -\right) ^{\varepsilon
_{M}\varepsilon _{N}}\mathbf{G}_{N}\mathbf{G}_{M}$ and the numbers $%
\varepsilon _{M}\equiv \varepsilon \left( \mathbf{G}_{M}\right) $ are $0$
for bosonic and $1$ for fermionic generators (modulo 2). Summation
convention does not apply to $\left( -\right) ^{\varepsilon }$ factors. The
generators satisfy the generalized Jacobi identity
\begin{equation}
\left( -\right) ^{\varepsilon _{M}\varepsilon _{K}}\left[ \left[ \mathbf{G}%
_{M,}\mathbf{G}_{N}\right] ,\mathbf{G}_{K}\right] +\left( -\right)
^{\varepsilon _{M}\varepsilon _{N}}\left[ \left[ \mathbf{G}_{N,}\mathbf{G}%
_{K}\right] ,\mathbf{G}_{M}\right] +\left( -\right) ^{\varepsilon
_{K}\varepsilon _{N}}\left[ \left[ \mathbf{G}_{K,}\mathbf{G}_{M}\right] ,%
\mathbf{G}_{N}\right] =0\,,  \label{s Jacobi}
\end{equation}%
what in terms of the structure constants stands for
\begin{equation}
\left( -\right) ^{\varepsilon _{M}\varepsilon _{K}}f_{MN}^{\;\quad
S}f_{SK}^{\;\quad L}+\left( -\right) ^{\varepsilon _{M}\varepsilon
_{N}}f_{NK}^{\;\quad S}f_{SM}^{\;\quad L}+\left( -\right) ^{\varepsilon
_{K}\varepsilon _{N}}f_{KM}^{\;\quad S}f_{SN}^{\;\quad L}=0\;.
\end{equation}

The associated connection 1-form is $\mathbf{A}=A^{M}\mathbf{G}_{M}$, where
the components $A^{M}$ are Grassmann even fields (bosons) if $\varepsilon
_{M}=0$ and Grassmann odd fields (fermions) if $\varepsilon _{M}=1$. Then
one says that the corresponding generators are bosonic and fermionic as
well. Covariant derivatives act on a Lie-valued form $\mathbf{\alpha }$ as $D%
\mathbf{\alpha }=d\mathbf{\alpha }+\left[ \mathbf{A},\mathbf{\alpha }\right]
,$ where the commutator of a $p$-form $\mathbf{\alpha }$ and a $q$-form $%
\mathbf{\beta }$ is generalized to $\left[ \mathbf{\alpha },\mathbf{\beta }%
\right] \equiv \mathbf{\alpha \beta }-\left( -\right) ^{pq}\left( -\right)
^{\varepsilon _{\alpha }\varepsilon _{\beta }}\,\mathbf{\beta \alpha }\,.$

Denote the invariant multilinear form (supertrace) by $\left\langle \cdots
\right\rangle $, which is antisymmetric for fermionic generators. Then the
Killing metric, $g_{MN},$ and the invariant tensor of rank three, $g_{MNK}$,
are
\begin{equation}
\begin{array}{lll}
g_{MN} & = & \left\langle \mathbf{G}_{M}\mathbf{G}_{N}\right\rangle =\text{%
Str}\left( \mathbf{G}_{M}\mathbf{G}_{N}\right) \,,\medskip \\
g_{MNK} & = & \left\langle \mathbf{G}_{M}\mathbf{G}_{N}\mathbf{G}%
_{K}\right\rangle =\frac{1}{2}\,\text{Str}\left( \left[ \mathbf{G}_{M}%
\mathbf{G}_{N}+\left( -\right) ^{\varepsilon _{M}\varepsilon _{N}}\mathbf{G}%
_{N}\mathbf{G}_{M}\right] \mathbf{G}_{K}\right) \,.%
\end{array}%
\end{equation}%
The invariant tensors $f_{MN}^{\;\quad K}$, $g_{MN}$ and $g_{MNK}$ are
Grassmann even variables, $\varepsilon \left( g_{MN}\right) =\varepsilon
_{M}+\varepsilon _{N}=0$, \emph{etc.}, and they satisfy the identities
\begin{equation}
\begin{array}{c}
f_{MK}^{\;\quad S}g_{SN}-g_{MS}\,f_{KN}^{\;\quad S}=0\,,\medskip \\
g_{MNS}\,f_{LK}^{\;\quad S}-g_{MSK}\,f_{NL}^{\;\quad S}-\left( -\right)
^{\varepsilon _{L}\varepsilon _{N}}g_{SNK}\,f_{ML}^{\;\quad S}=0\,.%
\end{array}
\label{id}
\end{equation}%
The above identities follow from the definition of commutator and the cyclic
property of supertrace.

A symmetric tensor $T_{M_{1}\cdots M_{n}}$ of rank $n$ is defined by an
element of Lie algebra $\mathbf{T}$, as
\begin{equation}
T_{M_{1}\cdots M_{n}}=\left\langle \mathbf{G}_{M_{1}}\cdots \mathbf{G}%
_{M_{n}}\mathbf{T}\right\rangle \,.
\end{equation}
The identities (\ref{id}) are equivalent to $Dg_{MN}=0$ and $Dg_{MNK}=0$.

\subparagraph{Hamiltonian formalism.}

Hamiltonian formalism can be generalized to the systems containing fermions
\cite{Henneaux-Teitelboim}, \cite{Berezin-Marinov}--\cite{Gervais-Sakita}.
If $z^{A}$ are local coordinates on the phase space, then the Poisson
bracket (PB) of functions $F\left( z\right) $ and $G\left( z\right) $ is
\begin{equation}
\left\{ F,G\right\} =\frac{\partial ^{R}F}{\partial z^{A}}\,\omega ^{AB}\,%
\frac{\partial ^{L}G}{\partial z^{B}}\,,  \label{super PB}
\end{equation}%
where $\partial ^{R}/\partial z^{A}$ and $\partial ^{L}/\partial z^{A}$
stand for right and left derivatives, respectively. The basic PB are $%
\left\{ z^{A},z^{B}\right\} =\omega ^{AB}.$ The convention that all
derivatives are left is adopted ($\partial /\partial z\equiv \partial
^{L}/\partial z$):
\begin{equation}
\delta F=\delta z^{A}\,\frac{\partial F}{\partial z^{A}}\,.
\end{equation}%
The PBs (\ref{super PB}) are antisymmetric for bosons and symmetric for
fermions, and they satisfy the generalized Jacobi identity (\ref{s Jacobi}).

Particularly, for canonical variables $z^{A}=\left( A_{\mu }^{M}(x),\pi
_{M}^{\mu }(x)\right) $, the PB become
\begin{equation}
\left\{ F(x),G\left( x^{\prime }\right) \right\} =\left( -\right)
^{\varepsilon _{F}\varepsilon _{M}}\int d^{4}y\,\left[ \frac{\partial F(x)}{%
\partial A_{\mu }^{M}\left( y\right) }\frac{\partial G(x^{\prime })}{%
\partial \pi _{M}^{\mu }\left( y\right) }\right. -\left( -\right)
^{\varepsilon _{M}}\left. \frac{\partial F(x)}{\partial \pi _{M}^{\mu
}\left( y\right) }\frac{\partial G(x^{\prime })}{\partial A_{\mu }^{M}\left(
y\right) }\right] \,,  \label{SPB}
\end{equation}%
with the basic PBs
\begin{equation}
\left\{ \pi _{M}^{\mu }(x),A_{\nu }^{N}\left( x^{\prime }\right) \right\}
=-\delta _{\nu }^{\mu }\delta _{M}^{N}\,\delta ^{(4)}\left( x-x^{\prime
}\right) =-\left( -\right) ^{\varepsilon _{M}}\left\{ A_{\mu }^{M}(x),\pi
_{N}^{\nu }\left( x^{\prime }\right) \right\} \,.
\end{equation}%
The canonical Hamiltonian has the form
\begin{equation}
H=\int dx\,\left( \pi _{M}^{\mu }\dot{A}_{\mu }^{M}-\mathcal{L}\right) \,,
\label{SH}
\end{equation}%
and the corresponding Hamilton equations are
\begin{eqnarray}
\dot{A}_{\mu }^{M} &=&\left( -\right) ^{\varepsilon _{M}}\frac{\delta H}{%
\delta \pi _{M}^{\mu }}\approx \left\{ A_{\mu }^{M},H\right\} \,,  \notag \\
\dot{\pi}_{M}^{\mu } &=&-\frac{\delta H}{\delta A_{\mu }^{M}}\approx \left\{
\pi _{M}^{\mu },H\right\} \,.
\end{eqnarray}%
Using definitions (\ref{SPB}) and (\ref{SH}), it is straightforward to find
the generalization of generators of local symmetries, as well as to
introduce Dirac brackets defining a reduced phase space.

\chapter{Killing spinors for the \emph{AdS}$_{5}$ space-time\label%
{Kill}}

The \emph{AdS} space-time can be given by the metric
\begin{equation}
ds_{\text{AdS}}^{2}=\ell ^{2}\,\left( dr^{2}+e^{2r}\eta _{\bar{n}\bar{m}%
}\,dx^{\bar{n}}dx^{\bar{m}}\right) \,,\qquad \eta _{\bar{n}\bar{m}}=\left(
-,+,+,+,+\right) \,,
\end{equation}%
with the local coordinates $x^{\mu }=\left( t,r,x^{n}\right) $, where $%
x^{1}=r$ and $x^{\bar{n}}=\left( t,x^{n}\right) $. The corresponding
vielbein ($e^{a}$) and the spin-connection ($\omega ^{ab}$) are given by
\begin{equation}
\begin{array}{ll}
e^{1}=\ell dr\,, & \omega ^{\bar{n}1}=\frac{1}{\ell }\,e^{\bar{n}}=e^{r}dx^{%
\bar{n}}\,, \\
e^{\bar{n}}=\ell e^{r}dx^{\bar{n}}\,,\qquad \qquad & \omega ^{\bar{n}\bar{m}%
}=0\,,%
\end{array}%
\end{equation}%
so that the \emph{AdS} connection takes the form
\begin{equation}
\mathbf{W}=\frac{1}{2\ell }\,e^{a}\Gamma _{a}+\frac{1}{4}\,\omega
^{ab}\Gamma _{ab}=\frac{1}{2\ell }\,\left[ e^{1}\Gamma _{1}+e^{\bar{n}%
}\Gamma _{\bar{n}}\left( 1+\Gamma _{1}\right) \right] \,.
\label{locally AdS}
\end{equation}%
The Killing spinors $\varepsilon ^{\alpha }$ are solutions of the Killing
equation
\begin{equation}
D\varepsilon =\left( d+\mathbf{W}\right) \varepsilon =0\,,
\end{equation}%
which for (\ref{locally AdS}) splits to the system of the following partial
differential equations
\begin{eqnarray}
\left( \partial _{r}+\frac{1}{2}\,\Gamma _{1}\right) \varepsilon &=&0\,,
\label{r} \\
\left[ \partial _{\bar{n}}+e^{r}\,\Gamma _{\bar{n}}\left( 1+\Gamma
_{1}\right) \right] \varepsilon &=&0\,.  \label{n}
\end{eqnarray}%
In the ansatz\footnote{%
For $\mathbf{F}=0$, the gauge field is a \emph{pure gauge}, $\mathbf{A}%
=g^{-1}dg$, and the solution of the Killing equation $D\epsilon =\left( d+%
\mathbf{A}\right) \epsilon =0$ is $\epsilon =g^{-1}\epsilon _{0}.$}
\begin{equation}
\varepsilon =h^{-1}(r)\,\Theta ^{-1}(x^{\bar{n}})\,\varepsilon _{0}\text{%
\thinspace ,}  \label{fact}
\end{equation}%
where $\varepsilon _{0}^{\alpha }$ is a constant spinor and $h,\Theta \in
SO(2,4)$, the equation (\ref{r}) has the solution
\begin{equation}
h^{-1}(r)=e^{-\frac{r}{2}\,\Gamma _{1}}.  \label{fact1}
\end{equation}%
Then the equations (\ref{n}) reduce to
\begin{equation}
\begin{array}{l}
\left[ \partial _{\bar{n}}\Theta ^{-1}+C_{\bar{n}}\,\Theta ^{-1}\right]
\,\epsilon _{0}=0\,,\smallskip \\
C_{\bar{n}}\equiv e^{r}e^{\frac{r}{2}\,\Gamma _{1}}\Gamma _{\bar{n}}\left(
1+\Gamma _{1}\right) \,e^{-\frac{r}{2}\,\Gamma _{1}}\,.%
\end{array}
\label{theta eq}
\end{equation}%
Using the identity
\begin{equation}
e^{\frac{r}{2}\,\left( 1+\Gamma _{1}\right) }=1+\frac{1}{2}\,\left( 1+\Gamma
_{1}\right) \left( e^{r}-1\right) \,,
\end{equation}%
the coefficients $C_{\bar{n}}$ become
\begin{equation}
C_{\bar{n}}=\Gamma _{\bar{n}}\left( 1+\Gamma _{1}\right) \,.
\end{equation}%
Therefore, the general solution of the equation (\ref{theta eq}) is
\begin{equation}
\Theta ^{-1}\left( x^{\bar{n}}\right) =e^{-x^{\bar{n}}\Gamma _{\bar{n}%
}\left( 1+\Gamma _{1}\right) }=1-x^{\bar{n}}\Gamma _{\bar{n}}\left( 1+\Gamma
_{1}\right) \,,  \label{fact2}
\end{equation}%
where it was used that the matrices $\Gamma _{\bar{n}}\left( 1+\Gamma
_{1}\right) $ are nilpotent. Therefore, from (\ref{fact}), (\ref{fact1}) and
(\ref{fact2}), the \emph{AdS} Killing spinor has the form \cite%
{Lu-Pope-Townsend}
\begin{equation}
\varepsilon =e^{-\frac{r}{2}\,\Gamma _{1}}\,\left[ 1-x^{\bar{n}}\Gamma _{%
\bar{n}}\left( 1+\Gamma _{1}\right) \right] \,\varepsilon _{0}\,.  \label{Ks}
\end{equation}%
The norm of $\varepsilon $ is defined by $\left\Vert \varepsilon \right\Vert
^{2}\equiv \bar{\varepsilon}\varepsilon $, where $\bar{\varepsilon}%
=\varepsilon ^{\dagger }\Gamma _{0}$. Using $\Gamma _{\bar{n}}^{\dagger
}=\Gamma _{0}\Gamma _{\bar{n}}\Gamma _{0}$ and $\Gamma _{1}^{\dagger
}=\Gamma _{1}^{\dagger }$, as well as the Clifford algebra of $\Gamma $%
-matrices, one obtains
\begin{equation}
\left\Vert \varepsilon \right\Vert ^{2}=\bar{\varepsilon}_{0}\,\left[ 1+x^{%
\bar{n}}\left( 1-\Gamma _{1}\right) \Gamma _{\bar{n}}\right] \left[ 1-x^{%
\bar{n}}\Gamma _{\bar{n}}\left( 1+\Gamma _{1}\right) \right] \varepsilon
_{0}=\left\Vert \varepsilon _{0}\right\Vert ^{2}.
\end{equation}%
Therefore, the spinor $\varepsilon $ given by (\ref{Ks}) has a constant
positive norm, $\left\Vert \varepsilon \right\Vert =\left\Vert \varepsilon
_{0}\right\Vert >0$.

\end{document}